\documentclass[a4paper,12pt]{article}
\pdfoutput=1
\usepackage{geometry}
\usepackage{amsmath,amssymb,amsfonts}
\usepackage{xcolor,graphicx,cite,soul}
\usepackage{array,booktabs,longtable}
\usepackage{caption,microtype}
\usepackage{subcaption}
\usepackage{hyperref}
\usepackage{datetime}
\usepackage{appendix}
\usepackage{ulem}

\usepackage{multirow}
\usepackage{cancel}

\makeatletter
\providecommand{\tabularnewline}{\\}

\makeatother

\newcommand{\lsim}
{\;\raisebox{-.3em}{$\stackrel{\displaystyle <}{\sim}$}\;}

\newcommand\al{\alpha}
\newcommand\be{\beta}
\newcommand\tb{\tan\beta}

\newcommand\CBA{c_{\beta - \alpha}}
\newcommand\SBA{s_{\beta - \alpha}}

\newcommand\ReDiag{\mathop{%
  \raise .5pt\hbox{[}%
  \widetilde{\mathrm{Re}}%
  \raise .5pt\hbox{]}}}
\newcommand\ReOffDiag{\mathop{%
  \raise .5pt\hbox{$\llbracket$}%
  \widetilde{\mathrm{Re}}%
  \raise .5pt\hbox{$\rrbracket$}}}

\newcommand\MW{m_W}

\newcommand\Mh{m_h}
\newcommand\MH{m_H}
\newcommand\MA{m_A}
\newcommand\MHp{m_{H^\pm}}
\newcommand\mbar{\bar{m}^2}

\newcommand\mt{m_t}
\newcommand\msq{m_{12}^{2}}

\newcommand\refeq[1]{Eq.~(\ref{#1})}

\newcommand\refta[1]{Tab.~\ref{#1}}
\newcommand\refse[1]{Sect.~\ref{#1}}

\newcommand\citere[1]{Ref.~\cite{#1}}
\newcommand\citeres[1]{Refs.~\cite{#1}}

\newcommand{\CP}{{\cal CP}}
\newcommand{\cp}{{\CP}}

\newcommand{\tev}{\,\, \mathrm{TeV}}
\newcommand{\gev}{\,\, \mathrm{GeV}}

\newcommand\HB{\texttt{HiggsBounds}}
\newcommand\HS{\texttt{HiggsSignals}}
\newcommand\hpair{\texttt{HPAIR}}

\newcommand{\De}{\Delta}

\newcommand{\sig}{\sigma}

\def\reffi#1{\mbox{Fig.~\ref{#1}}}
\def\reffis#1{\mbox{Figs.~\ref{#1}}}

\def\Ga{\Gamma}
\newcommand{\GaHtot}{\ensuremath{\Ga_H^{\rm tot}}}
\def\ga{\gamma}
\def\de{\delta}
\def\la{\lambda}
\newcommand\kala{\ensuremath{\kappa_{\lambda}}}
\newcommand\laSM{\ensuremath{\lambda_{\mathrm{SM}}}}
\newcommand{\lahhh}{\ensuremath{\la_{hhh}}}
\newcommand{\lahhH}{\ensuremath{\la_{hhH}}}

\newcommand{\mhh}{\ensuremath{m_{hh}}}

\definecolor{Orange}{named}{orange}
\definecolor{Purple}{named}{purple}
\definecolor{Lightblue}{cmyk}{0.9,0.1,0.1,0.3}
\definecolor{dgelborange}{cmyk}{0.,0.3,0.5, 0.}
\definecolor{Lila}{rgb}{0.5,0.,1}
\definecolor{Darkgreen}{rgb}{0.,.7,0.2}

\newcommand{\vev}{vev}

\graphicspath{{figs/}}
\allowdisplaybreaks
\begin{document}
\thispagestyle{empty}

\def\thefootnote{\fnsymbol{footnote}}

\begin{flushright}
\mbox{}
DESY-22-203\\
IFT--UAM/CSIC-22-073 \\
KA-TP-30-2022 
\end{flushright}

\vspace{0.5cm}

\begin{center}

{\large\sc 
{\bf Sensitivity to Triple Higgs Couplings via Di-Higgs
  Production\\[.5em]
  in the 2HDM at the (HL-)LHC}}  

\vspace{1cm}

{\sc
F.~Arco$^{1,2}$%
\footnote{emails:
Francisco.Arco@uam.es,
Sven.Heinemeyer@cern.ch,
margarete.muehlleitner@kit.edu,\\
\mbox{}\hspace{19mm}kateryna.radchenko@desy.de}%
, S.~Heinemeyer$^{2}$%
, M.~M\"uhlleitner$^{3}$%
~and K.~Radchenko$^{4}$%
}

\vspace*{.7cm}

{\sl
$^1$Departamento de F\'isica Te\'orica, 
Universidad Aut\'onoma de Madrid, \\ 
Cantoblanco, 28049, Madrid, Spain

\vspace*{0.1cm}

$^2$Instituto de F\'isica Te\'orica (UAM/CSIC), 
Universidad Aut\'onoma de Madrid, \\ 
Cantoblanco, 28049, Madrid, Spain

\vspace*{0.1cm}

$^3$Institute for Theoretical Physics,
Karlsruhe Institute of Technology, 76128 Karlsruhe, Germany

\vspace{0.1em}

$^4$Deutsches Elektronen-Synchrotron DESY, Notkestr.\ 85, 22607 Hamburg,
Germany  
 
}

\end{center}

\vspace*{0.1cm}

\begin{abstract}
\noindent
An important task of the LHC is the investigation of the
Higgs-boson sector. Of particular interest is the reconstruction of the
Higgs potential, i.e.\ the measurement of the Higgs self-couplings.
Based on previous analyses, within the 2-Higgs-Doublet Models (2HDMs)
type~I and~II, we analyze several  
two-dimensional benchmark planes
that are over large parts in agreement with all theoretical and
experimental constraints. 
For these planes we evaluate di-Higgs production cross sections
at the (HL-)LHC with a center-of-mass energy of $13 \tev$ at
next-to-leading order in the heavy top-quark limit with the code \hpair. 
We investige in particular the process $gg \to hh$, with $h$
being the Higgs boson discovered at the LHC with a mass of about $125 \gev$.
The top box diagram of the loop-mediated gluon fusion
process into Higgs pairs interferes with the $s$-channel 
exchange of the two CP-even 2HDM Higgs bosons $h$ and
$H$ involving the trilinear couplings $\lahhh$ and $\lahhH$, respectively.
Depending on the size of the involved top-Yukawa 
and triple Higgs couplings as well as on the mass of $H$, the
contribution of the $s$-channel $H$~diagram can be dominating or be highly
suppressed. 
We find regions of the allowed parameter space in which the di-Higgs
production cross section can differ by many standard deviations from its
SM prediction, indicating possible access to deviations
in $\lahhh$ from the SM value $\laSM$ and/or
contributions involving $\lahhH$. 
The sensitivity to $\lahhH$ is further analyzed employing the 
\mhh\ distributions. We demonstrate how a possible measurement of
$\lahhH$ depends on the various experimenal uncertainties.
Depending on the underlying parameter space, the HL-LHC may have the
option not only to detect beyond-the-Standard-Model triple Higgs
couplings, but also to provide a first rough measurement of their sizes.
\end{abstract}


\def\thefootnote{\arabic{footnote}}
\setcounter{page}{0}
\setcounter{footnote}{0}

\newpage


\section{Introduction}
\label{sec:intro}

The discovery of a new scalar particle with a mass of $\sim125\gev$
by ATLAS and CMS~\cite{Aad:2012tfa,Chatrchyan:2012xdj,Khachatryan:2016vau}
--- within the experimental and theoretical uncertainties ---
is in agreement
with the properties of the Standard Model (SM) Higgs boson.
On the other hand, no conclusive sign of Higgs bosons beyond the SM
(BSM) has been observed so far.
However, the experimental results about the state at $\sim 125 \gev$,
whose couplings are known up to now to an experimental precision of
roughly $\sim 10-20\%$, leave ample room for interpretations in BSM
models.  Many BSM models feature
extended Higgs-boson sectors with correspondingly extended Higgs
potentials. Consequently, one of the main tasks of present and future
colliders will be to determine whether  
the observed scalar boson forms part of the Higgs sector of an extended
model, or not. 

In contrast to the Higgs couplings to the SM fermions and gauge
bosons, the trilinear  
Higgs self-coupling $\lahhh$ remains to be determined. So far it has
been constrained by ATLAS~\cite{ATLAS:2022kbf} to be inside the
range $-0.4 < \lahhh/\laSM < 6.3$ at the 95\% C.L.\
and $-1.24 < \lahhh/\laSM < 6.49$ at the 95\% C.L.~by
CMS~\cite{CMS:2022dwd},
both assuming a SM-like top-Yukawa coupling of the light Higgs.
Many BSM models can still induce significant deviations in the
trilinear coupling $\lahhh$ of the SM-like Higgs boson with respect
to the SM value, see, e.g., \citere{Abouabid:2021yvw} for an up-to-date
investigation. 
For recent reviews on the measurement of the triple
Higgs couplings
at future colliders see for instance \citeres{deBlas:2019rxi, DiMicco:2019ngk}.
In case a BSM Higgs sector manifests itself, it will be
a prime task to measure also the BSM trilinear Higgs self-couplings.

One of the simplest extensions of the SM Higgs sector is the 
2-Higgs-Doublet Model
(2HDM)~\cite{TDLee,Gunion:1989we,Aoki:2009ha,Branco:2011iw},
where a second Higgs doublet is added to the SM Higgs sector. 
After electroweak symmetry breaking this leads to five physical Higgs
bosons, two $\CP$-even bosons $h$ and $H$,
where by convention $\Mh<\MH$, one $\CP$-odd boson $A$ and two charged
Higgs bosons $H^\pm$.
The ratio of the two vacuum expectation values of the neutral
components of the two Higgs doublets is defined as
$\tb \equiv  v_2/v_1$.  
To avoid flavor-changing neutral currents at 
tree level, a $Z_2$~symmetry is imposed~\cite{Glashow:1976nt},
possibly softly broken by the parameter $\msq$.
Depending on how this symmetry is extended to the fermion sector, four
types of the 2HDM can be realized: 
type~I and~II, flipped (type~III) and lepton specific
(type~IV)~\cite{Aoki:2009ha}.

In \citere{Arco:2020ucn} an analysis was presented of the possible size
of triple Higgs couplings (THCs) in the 2HDM type~I and~II taking into
account all relevant experimental and theoretical
constraints.%
\footnote{For an analysis of THCs in the CP-conserving and
CP-violating 2HDM, the Next-to-2HDM and the Next-to-Minimal
Supersymmetric extension of the SM (NMSSM), see
\citere{Abouabid:2021yvw}, where in addition the 
constraints from Higgs pair production measurements at the LHC were
taken into account.}%
~For that analysis it was assumed that the
lightest $\CP$-even Higgs-boson $h$ is SM-like with a mass of 
$\Mh \sim 125 \gev$. All other Higgs bosons were assumed to be heavier.
(An update and extension to type~III and~IV was presented in
\citere{Arco:2022xum}.)
Future $e^+e^-$ linear colliders, like the ILC~\cite{Bambade:2019fyw}
and CLIC~\cite{Charles:2018vfv}, will play a key role for the
measurement of the Higgs potential and in detecting possible deviations
from the SM with high 
precision~\cite{Djouadi:1999gv,Abramowicz:2016zbo,Strube:2016eje,Roloff:2019crr,deBlas:2019rxi,DiMicco:2019ngk}.
Employing the results of \citere{Arco:2020ucn},
in \citere{Arco:2021bvf} the sensitivity of the ILC and CLIC to various
2HDM THCs (including BSM THCs) was analyzed. Further analyses
of THCs at $e^+e^-$ colliders
were presented in \citeres{Kon:2018vmv,Sonmez:2018smv}.
Recent reviews on triple Higgs couplings at $e^+e^-$
colliders can be found
in~\citeres{deBlas:2019rxi,DiMicco:2019ngk,Strube:2016eje,Roloff:2019crr}.

In this paper, based on the results of \citere{Arco:2020ucn}, we complement
the above results with an analysis of the senstivity to BSM triple Higgs
couplings at the LHC, and in particular its high-luminosity phase, the
HL-LHC. Further analyzes involving BSM triple Higgs couplings can be found in
\citeres{Djouadi:1999rca,Basler:2017uxn,Basler:2019iuu,DiMicco:2019ngk,Abouabid:2021yvw}. 
However, while these papers took the effects of BSM THCs into
account, to our knowledge no analysis for the (HL-)LHC exists
attempting to quantify the potential sensitivity to BSM triple Higgs
couplings.

The main Higgs pair production process at the LHC is gluon fusion into
Higgs pairs~\cite{Baglio:2012np}. Here we investige in particular
$gg \to hh$ in the 2HDM type~I and~II, with $h$ corresponding to the state
discovered at the LHC at $\sim 125 \gev$. The process is loop-mediated
already at leading order and consists of a triangle and a box top-loop
contribution. For small values of $\tan\beta$ the bottom loop 
plays only a minor role. In the SM, the box diagram interferes
destructively with the triangle contribution. In the 2HDM, we have
both the $h$ and $H$ $s$-channel exchange in the triangle
contribution, where a resonantly produced $H$ with subsequent decay
into $hh$ can lead to a significantly enhanced cross section. For our
analysis, we take into account the next-to-leading order QCD corrections to
the process in the heavy top-quark limit~\cite{Dawson:1998py} by making
use of the 
accordingly modified~\cite{Abouabid:2021yvw,Grober:2017gut} program \hpair. 
We find regions of the allowed parameter space in which the di-Higgs
production cross section can differ by several standard deviations from its
SM prediction, indicating possible access to deviations
in $\lahhh$ from $\laSM$ and/or contributions involving $\lahhH$.
The sensitivity to $\lahhH$ is further analyzed employing the 
\mhh\ distributions in the $gg \to hh$ production cross section.
We investigate how a possible measurement of $\lahhH$
depends on the assumed experimental uncertainties in \mhh, such as
smearing, bin width, as well as on the position of
the bins. We demonstrate that, 
depending on the underlying parameter space, the HL-LHC may have the
potential not only to detect BSM triple Higgs couplings, but also to 
provide a first rough measurement of their size.

Our paper is organized as follows. In \refse{sec:model} we briefly review
the 2HDM, fix our notation, define the benchmark planes used
later for our investigation and summarize the constraints that we
apply (which are the same as in \citeres{Arco:2020ucn,Arco:2022xum}).
The di-Higgs production cross sections in the
benchmark planes are presented in \refse{sec:xs} and analyzed
w.r.t.\ their dependence on the triple Higgs couplings in the
contribution from the $s$-channel $H$~exchange. 
In \refse{sec:sensitivity} we analyze a possible sensitivity of the
di-Higgs production cross section at the (HL-)LHC to $\lahhh$ and in
particular to $\lahhH$. 
Finally, in \refse{sec:mhh} we present the possible HL-LHC sensitivity
to $\lahhH$ via the \mhh\ distribution, and in \refse{sec:expuncert} also
assess its
dependence on smearing, bin width and position of the bins. 
Our conclusions are given in \refse{sec:conclusions}. 


\section{The Model and the constraints}
\label{sec:model}

In this section we give a short description of the 2HDM to fix our
notation. We briefly review the theoretical and experimental constraints,
which are taken over from 
\citeres{Arco:2020ucn,Arco:2022xum}.
Finally we will define the benchmark planes for our analysis of the
$gg \to hh$ production cross section.


\subsection{The 2HDM}
\label{sec:2hdm}

We assume the $\cp$-conserving
2HDM~\cite{TDLee,Gunion:1989we,Aoki:2009ha,Branco:2011iw}, where the
potential can be written as, 
\begin{eqnarray}
V &=& m_{11}^2 (\Phi_1^\dagger\Phi_1) + m_{22}^2 (\Phi_2^\dagger\Phi_2)
- \msq (\Phi_1^\dagger \Phi_2 + \Phi_2^\dagger\Phi_1)
+ \frac{\la_1}{2} (\Phi_1^\dagger \Phi_1)^2 +
\frac{\la_2}{2} (\Phi_2^\dagger \Phi_2)^2 \nonumber \\
&& + \la_3
(\Phi_1^\dagger \Phi_1) (\Phi_2^\dagger \Phi_2) + \la_4
(\Phi_1^\dagger \Phi_2) (\Phi_2^\dagger \Phi_1) + \frac{\la_5}{2}
[(\Phi_1^\dagger \Phi_2)^2 +(\Phi_2^\dagger \Phi_1)^2]  \;.
\label{eq:scalarpot}
\end{eqnarray}
After electroweak symmetry breaking (EWSB) the two $SU(2)_L$ doublets
$\Phi_1$ and $\Phi_2$ can be expanded around their two vacuum
expectation values (\vev s) $v_1$ and $v_2$, respectively, as
\begin{eqnarray}
\Phi_1 = \left( \begin{array}{c} \phi_1^+ \\ \frac{1}{\sqrt{2}} (v_1 +
    \rho_1 + i \eta_1) \end{array} \right) \;, \quad
\Phi_2 = \left( \begin{array}{c} \phi_2^+ \\ \frac{1}{\sqrt{2}} (v_2 +
    \rho_2 + i \eta_2) \end{array} \right) \;,
\label{eq:2hdmvevs}
\end{eqnarray}
with the \vev\ ratio given by $\tb \equiv v_2/v_1$. The vevs satisfy the 
relation $v = \sqrt{(v_1^2 +v_2^2)}$ where $v\simeq246\gev$ is the SM vev.
The eight (scalar) degrees of freedom, $\phi_{1,2}^\pm$, $\rho_{1,2}$ and
$\eta_{1,2}$, give rise to three Goldstone bosons,  $G^0$ and $G^\pm$,
and five physical scalar fields, two $\cp$-even scalar fields,
$h$ and $H$, where by convention $m_h < m_H$, one $\cp$-odd field,
$A$, and one charged Higgs pair, $H^\pm$. 
The mixing angles $\al$ and $\be$ diagonalize the $\CP$-even and
CP-odd/charged Higgs mass matrices, respectively.

The occurrence of tree-level flavor-changing neutral currents (FCNC) is
avoided by imposing a $Z_2$ symmetry,
only softly broken by the $\msq$ term in
the Lagrangian. The extension of the $Z_2$ symmetry to the Yukawa
sector prohibits tree-level FCNCs. 
This results in four variants of the 2HDM, 
depending on the $Z_2$ parities of the 
fermion types. In this article we focus on the Yukawa type~I and~II.
The assignment of the fermions to the Higgs doublets are
listed in \refta{tab:types}.

\begin{table}[htb!]
\begin{center}
\begin{tabular}{lccc} 
\hline
  & $u$-type & $d$-type & leptons \\
\hline
type~I & $\Phi_2$ & $\Phi_2$ & $\Phi_2$ \\
type~II & $\Phi_2$ & $\Phi_1$ & $\Phi_1$ \\
\hline
\end{tabular}
\caption{Fermion couplings in 
the 2HDM type~I and~II.}
\label{tab:types}
\end{center}
\end{table}

Here we work in the physical basis of the 2HDM, where most of the free
parameters 
in \refeq{eq:scalarpot} are expressed in terms of a set of ``physical''
parameters given by
\begin{equation}
c_{\be-\al} \; , \quad \tb \;, \quad v \; ,
\quad \Mh\;, \quad \MH \;, \quad \MA \;, \quad \MHp \;, \quad \msq \;, 
\label{eq:inputs}
\end{equation}
where we use the short-hand notation $s_x = \sin(x)$, $c_x = \cos(x)$. 
In our analysis we will identify the
lightest $\cp$-even Higgs 
boson, $h$, with the one observed at the LHC at $\sim 125 \gev$. 

The couplings of the Higgs bosons to SM particles are modified
w.r.t.\ the SM Higgs-coupling predictions because of the mixing in the
Higgs-boson sector.  The couplings of the neutral Higgs bosons to
fermions and to gauge bosons are given by,
\begin{eqnarray}
	\mathcal{L} &=&-\sum_{f=u,d,l}\frac{m_f}{v}\left[\xi_h^f\bar{f}fh + \xi_H^f\bar{f}fH +i \xi_A^f\bar{f}\gamma_5fA \right] \nonumber \\
&&+\sum_{h_i=h,H,A}	\left[   g m_W \xi_{h_i}^W  W_\mu W^\mu h_i + \frac{1}{2} g m_Z \xi_{h_i}^Z  Z_\mu Z^\mu h_i\right] .
\end{eqnarray}
Here $m_f$, $m_W$ and $m_Z$ are the fermion, $W$ boson and $Z$ boson masses,
respectively.   The modification factors in the couplings to fermions and gauge
bosons, $\xi_{h,H,A}^f$ and  $\xi_{h,H,A}^V$, 
for the 2HDM of type~I and~II are given in \refta{tab:coupling}.

\begin{table}
\begin{center}
\begin{tabular}{c|c|c}
 & type~I  & type~II\tabularnewline
\hline 
$\xi_{h}^{u}$  & $s_{\be-\al}+c_{\be-\al}\cot\be$  & $s_{\be-\al}+c_{\be-\al}\cot\be$\tabularnewline
$\xi_{h}^{d,l}$  & $s_{\be-\al}+c_{\be-\al}\cot\be$  & $s_{\be-\al}-c_{\be-\al}\tan\be$\tabularnewline
$\xi_h^V$ & $s_{\be-\al}$ & $s_{\be-\al}$\tabularnewline\hline
$\xi_{H}^{u}$  & $c_{\be-\al}-s_{\be-\al}\cot\be$  & $c_{\be-\al}-s_{\be-\al}\cot\be$\tabularnewline
$\xi_{H}^{d,l}$  & $c_{\be-\al}-s_{\be-\al}\cot\be$  & $c_{\be-\al}+s_{\be-\al}\tan\be$\tabularnewline
$\xi_H^V$ & $c_{\be-\al}$ & $c_{\be-\al}$\tabularnewline\hline
$\xi_{A}^{u}$  & $-\cot\be$  & $-\cot\be$\tabularnewline
$\xi_{A}^{d,l}$  & $\cot\be$  & $-\tan\be$\tabularnewline
$\xi_A^V$ & 0 & 0 \tabularnewline
\end{tabular}
\caption{Factors appearing in the couplings of the neutral Higgs 
bosons to fermions,  $\xi_{h,H,A}^f$,  and to gauge bosons,
$\xi_{h,H,A}^V$,  in the 2HDM of type~I and type~II.}
\label{tab:coupling}
\end{center}
\end{table}

The generic triple Higgs coupling
$\la_{h h_i h_j}$ involving at least one SM-like Higgs boson $h$ is defined
such that the Feynman rules are given by 
\begin{equation}
	\begin{gathered}
		\includegraphics{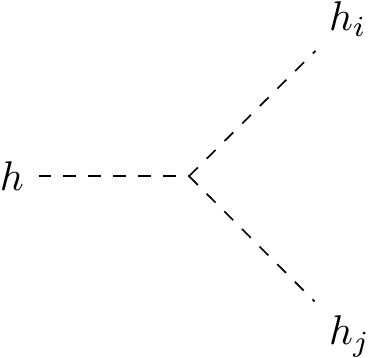}
	\end{gathered}
	= - ivn!\la_{h h_i h_j}
\label{eq:lambda}
\end{equation}
where $n$ is the number of identical particles in the vertex.
Relevant for our analysis here are $\lahhh$ and $\lahhH$.
We adopt this convention in \refeq{eq:lambda} so that the light Higgs
triple coupling $\lahhh$ has the same normalisation as $\laSM$ 
in the SM, which is given by $-6iv\laSM$ with $\laSM=\Mh^2/2v^2\simeq 0.13$.
We furthermore define $\kala \equiv \lahhh/\laSM$.

The explicit expressions of the two triple Higgs couplings are given by
\begin{align}
  \lahhh 
  &=  -\frac{1}{2} \Big\{ \la_1c_{\be } s_{\al }^3-\la _2 c_{\al }^3 s_{\be }
        +\left(\la_3+\la _4+\la_5\right) \left(c_{\al }^2 c_{\be }
        s_{\al }-c_{\al } s_{\al }^2 s_{\be}  \right) \Big\} \nonumber \\
\label{eq:hhh_phys}
        &= \frac{1}{2v^2} \bigg\{ \Mh^2 s_{\be -\al }^3 
        +\left(3 \Mh^2-2 \bar{m}^2\right) c_{\be -\al }^2 s_{\be -\al }
        +2 \cot 2 \be \left( \Mh^2-\bar{m}^2\right) c_{\be -\al
        }^3\bigg\}, \\[.5em]
\notag
 \lahhH 
 &= \frac{1}{2} \Big\{  3 \la _1  c_{\al } c_{\be } s_{\al }^2
 + 3 \la _2 c_{\al}^2 s_{\al } s_{\be } \\
&\quad\qquad +\left(\la_3+\la _4+\la_5\right)
 \left(c_{\al }^3 c_{\be }-2 c_{\al }^2 s_{\al }
   s_{\be }-2 c_{\al } c_{\be } s_{\al }^2+s_{\al }^3
   s_{\be}\right)\Big\} \nonumber \\
\label{eq:hhH_phys} \notag
        &= -\frac{\CBA}{2v^2} \bigg\{  \left(2\Mh^2+\MH^2-4\mbar\right)\SBA^2
        +2\cot{2\be}\left(2\Mh^2+\MH^2-3\mbar\right) \SBA\CBA \\
        &\quad\qquad -\left(2\Mh^2+\MH^2-2\mbar \right) \CBA^2
         \bigg\},  
\end{align}
where $\bar{m}^2$, derived from $\msq$, is given by: 
\begin{eqnarray} 
 \bar{m}^2&=&\frac{\msq}{\sin\be\cos\be} \,.
\label{eq:mbar}
\end{eqnarray} 

The triple Higgs couplings depend on $\CBA$. In particular, in the
``alignment limit'', $\CBA \to 0$, where 
the light $\cp$-even Higgs couplings to the SM particles recover SM
values, and the
triple Higgs couplings approach the values $\lahhh = \laSM$ and
$\lahhH = 0$, respectively.


\subsection{Theoretical and experimental constraints}
\label{sec:constraints}

In this subsection we briefly summarize the various theoretical and
experimental constraints considered in our analysis (more details can be
found in \citeres{Arco:2020ucn,Arco:2022xum}). Note, that we did not
check for constraints arising from di-Higgs measurements at the
LHC. The analysis performed in \cite{Abouabid:2021yvw} showed that
non-resonant and resonant di-Higgs searches start to cut in the
parameter spaces of extended Higgs sector models. However, the
parameter spaces of the CP-conserving 2HDM investigated here are
not affected yet. 

\begin{itemize}

\item {\bf Theoretical constraints}\\
The important theoretical constraints come from
tree-level perturbartive unitarity and stability of the electroweak vacuum.
They are ensured by an explicit test of the underlying Lagrangian
parameters (details of our approach can be found in \citere{Arco:2020ucn}).
The parameter space allowed by these two constraints can be enlarged,
if we allow for a mass term breaking the imposed $\mathbb{Z}_2$
symmetry softly, i.e.\ we choose a non-zero $m_{12}^2$. In some of
the sample scenarios that we will investigate later, we chose $m_{12}^2$ as 
\begin{equation}
  \msq = \frac{\MH^2\cos^2\al}{\tb}~.
  \label{eq:m12special}
\end{equation}

\item {\bf Constraints from electroweak precision data}\\
For SM extensions based solely on extensions of the Higgs sector 
the constraints from the electroweak precision observables (EWPO)
can be expressed in terms of the oblique parameters $S$, $T$ and
$U$~\cite{Peskin:1990zt,Peskin:1991sw}. 
Most constraining in the 2HDM is the $T$~parameter, requiring
either $\MHp \approx \MA$ or $\MHp \approx \MH$.
In \citere{Arco:2020ucn} three scenarios were defined to meet this constraint:
(A)\;$\MHp = \MA$, (B)\;$\MHp = \MH$, and (C)\;$\MHp = \MA = \MH$.

Here it should be kept in mind that the EWPO used to set these
constraints do not take into account the recent measurement of $\MW$ at
CDF~\cite{CDF:2022hxs}, which deviates from the SM prediction by
$\sim 7\,\sig$.  
After this result for $\MW$ was published, many articles appeared
to describe the CDF value in BSM models, including analyses in the
2HDM, see \citeres{Song:2022xts,Bahl:2022xzi,Babu:2022pdn} for the
first papers. It was shown that the large $\MW$ value can be accomodated
by introducing a certain amount of splitting between the masses of the
heavy 2HDM Higgs bosons. This also holds (albeit with a smaller amount
of splitting) if a possible new world average, see \citere{PDG2022}, is
taken into account~\cite{Arco:2022jrt}. However, we will not include
this possibility into our analysis.

\item {\bf Constraints from direct Higgs-boson searches at colliders}\\
The exclusion limits at the $95\%$ confidence level
of all relevant BSM Higgs boson searches (including Run~2 data from the
LHC) are included in the public code
\HB\,\texttt{v.5.9}~\cite{Bechtle:2008jh,Bechtle:2011sb,Bechtle:2013wla,Bechtle:2015pma,Bechtle:2020pkv}.%
\footnote{See \citere{Bahl:2022igd} for the latest version.}%
~For a parameter point in a particular model, \HB\ determines on the
basis of expected limits which is the most
sensitive channel to test each BSM Higgs boson.
Then, based on this most sensitive channel, \HB\ determines whether the
point is allowed or not at the $95\%$~CL. 
As input \HB\ requires some specific predictions from the model,
like branching ratios or Higgs couplings, that were computed with the
code \texttt{2HDMC-1.8.0}~\cite{Eriksson:2009ws}.%
\footnote{Alternatively, the code
  \texttt{HDECAY}~\cite{Djouadi:1997yw,Djouadi:2018xqq} can be used. For
  a comparison of the two codes, see \cite{Harlander:2013qxa}.} 

\item {\bf Constraints from the properties of the \boldmath{$\sim 125 \gev$}
Higgs boson}\\
Any model beyond the SM has to accommodate a Higgs boson 
with mass and signal strengths as they were measured at the LHC.
In the parameter points used the compatibility of the $\cp$-even scalar
$h$ with a mass of $125.09\gev$, $h_{125}$, with the measurements of signal
strengths at the LHC is tested with the code
\texttt{HiggsSignals
  v.2.6}~\cite{Bechtle:2013xfa,Bechtle:2014ewa,Bechtle:2020uwn}. 
The code provides a statistical $\chi^2$ analysis of the $h_{125}$
predictions of a certain model compared to the measurement of
Higgs-boson signal rates and masses from the LHC.
As for the BSM Higgs boson searches, the predictions of the 2HDM
have been obtained with {\tt{2HDMC}}~\cite{Eriksson:2009ws}.
For a 2HDM parameter point to be allowed it was
required~\cite{Arco:2022xum} that  
the corresponding $\chi^2$ is within $2\,\sig$ ($\De\chi^2 = 6.18$)
from the SM fit: $\chi_\mathrm{SM}^2=84.73$ with 107 observables.%
\footnote{These values have changed with the latest
  \HS\ version~\cite{Bahl:2022igd}, but we do not expect a qualitative
  impact of this change on our results.} 

\item {\bf Constraints from flavor physics}\\
Constraints from flavor physics can be significant in the 2HDM, 
in particular because of the presence of the charged Higgs boson. 
Flavor observables like rare $B$~decays,
mixing parameters of $B$~mesons, 
and LEP constraints on $Z$ decay partial widths
etc.\ are sensitive to charged Higgs boson
contributions~\cite{Enomoto:2015wbn,Arbey:2017gmh}. 
To test the parameter space, taking into account the most constraining decays
$B \to X_s \ga$ and $B_s \to \mu^+ \mu^-$, 
the code \texttt{SuperIso}~\cite{Mahmoudi:2008tp,Mahmoudi:2009zz} was
used, again with the model input given by {\tt{2HDMC}} (see
\citere{Arco:2020ucn} for more details).

\end{itemize}


\subsection{Benchmark planes}
\label{sec:planes}

\noindent
Based on the analysis in \citere{Arco:2020ucn} we define four benchmark
planes that exhibit an interesting phenomenology w.r.t.\ the di-Higgs
production cross sections in the gluon fusion channel, $gg \to hh$.
The four planes are all in the 2HDM Yukawa type~I,
as this type allows for larger deviations from the alignment limit
taking into account the experimental constraints. 
The parameters are chosen as:

\begin{enumerate}

\item
  $\MHp = \MH = \MA = 1000 \gev$,
  $\msq$ fixed via \refeq{eq:m12special},\\
  free parameters: $\CBA$, $\tb$

\item
  $\MHp = \MH = \MA = 650 \gev$,
  $\tb = 7.5$,\\
  free parameters: $\CBA$, $\msq$

\item
  $\tb = 10$, 
  $\msq$ fixed via \refeq{eq:m12special},\\
  free parameters: $\CBA$, $\MHp = \MH = \MA$
  
\item
  $\tb = 10$,
  $\CBA= 0.2$,
  $\msq$ fixed via \refeq{eq:m12special}\\
  free parameters: $\MH$, $\MA =\MHp$
  
\end{enumerate}
\noindent



\section{Cross section results}
\label{sec:xs}

In this section we start our numerical analysis with the evaluation of
the di-Higgs production cross sections in the benchmark planes defined
in \refse{sec:planes}. We first discuss details of the calculation and
then present the results, where we analyze the impact of a possible heavy
Higgs, $H$, in the $s$-channel. We perform this analysis in all
benchmark planes listed above to give a broad overview about the
possible phenomenology of di-Higgs production in the 2HDM. In the
following sections we will discuss selected planes and points to further
examine the effects of the various triple Higgs couplings and the
properties of the heavy $\cp$-even Higgs boson.


\subsection{Calculation of \boldmath{$gg \to hh$}}
\label{sec:gghh}

\begin{figure}[ht!]
\vspace{-1em}
  \begin{center}
	\begin{subfigure}[b]{.32\textwidth}
            \includegraphics[width=1\linewidth]{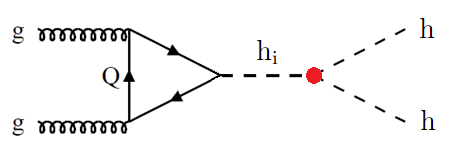}
		\caption{Triangle diagram.}
		\label{fig:BSMtriangle}
	\end{subfigure} 
	\begin{subfigure}[b]{.32\textwidth}
            \includegraphics[width=1\linewidth]{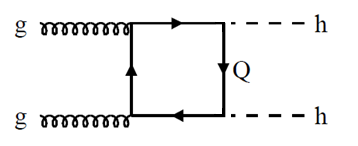}
		\caption{Box diagram.}
		\label{fig:BSMbox}
	\end{subfigure}
  \end{center}
\caption{Leading-order diagrams to SM-like Higgs pair production in
  gluon fusion processes at hadron colliders. The red dot shows the triple
  Higgs coupling, and $h_i=h,H$.}  
\label{fig:gghhdiagrams}
\end{figure}

The main di-Higgs production process at the LHC is given by gluon
fusion. The diagrams contributing at leading order are shown in 
\reffi{fig:gghhdiagrams}. They both involve a heavy quark loop (top or
bottom), where
for small $\tan\beta$ the bottom-quark loop only plays a minor role. In the SM the
triangle diagram, \reffi{fig:BSMtriangle}, gives access to the trilinear
Higgs coupling, $\lahhh$, with the SM Higgs exchange in the $s$-channel.
The box diagram, \reffi{fig:BSMbox}, interferes destructively with the
triangle diagram, resulting in a small cross section.
In BSM theories additional diagrams can contribute. In particular 
in the case of the 2HDM the second $\cp$-even Higgs can be exchanged in
the $s$-channel, involving $\lahhH$. This diagram will usually be referred to
as the ``resonant diagram''%
\footnote{Owing to the fact that for $m_H > 2 m_h$ the $H$ can be
  resonantly produced which can largely enhance 
the di-Higgs cross section w.r.t.~to the SM value. However, also scenarios with
$m_H < 2 m_h$ can be realized in the 2HDM where no such enhancement
is observed. Still, for simplicity of the notation, we will call it
``resonant diagram''.}%
, whereas the SM-like contributions will be referred to as
``the continuum''. Note that the Yukawa coupling and the trilinear Higgs
self-coupling 
$\lahhh$ of the SM-like Higgs boson $h$ can deviate from the SM values
so that the observed destructive interference between the triangle and
box diagram in the SM may not be effective. 

For our numerical evaluation we use the code \texttt{HPAIR}
\cite{Plehn:1996wb,Dawson:1998py,Grober:2017gut,Abouabid:2021yvw},
adapted to the 2HDM. The original code evaluates the cross section of the
production of two neutral Higgs bosons 
through gluon fusion at the LHC for the SM and the MSSM. The calculation is done at
leading order (LO, see \reffi{fig:gghhdiagrams}), and includes next-to-leading
order (NLO) QCD corrections in the heavy top-mass limit. In this
limit, it is assumed that the contribution of 
the bottom quark is negligible (it would introduce modifications of
less than 1\% in the SM) and then the top mass is taken to
infinity. This assumption becomes less accurate at high values of
$\tan\be$ because the bottom quark loop contribution gets larger. 

At LO the calculation includes top- and bottom-quark loops 
with full mass dependence. It is equivalent to
the calculation done in the Minimal Supersymmetric Standard Model (MSSM)
since its Higgs sector is equivalent to the 2HDM Type II. The only
changes that are
implemented in the case of the 2HDM are the modification of the Yukawa
couplings of the MSSM according to \refta{tab:coupling}, for the
corresponding type and the change of the triple Higgs couplings.%
\footnote{It should be noted that {\tt HPAIR} for the MSSM
does not include any squark loop contributions.}%
~As the QCD corrections
in the heavy top-quark limit only involve couplings between coloured
particles, they can straightforwardly be taken over from the MSSM to
the 2HDM. For further details, we refer to
\citeres{Plehn:1996wb,Dawson:1998py,Abouabid:2021yvw,Grober:2017gut}.


\subsection{Analysis of the cross section predictions}
\label{sec:xs-anal}

\begin{figure}[ht!]
\vspace{-1em}
  \begin{center}
\includegraphics[width=0.41\textwidth]{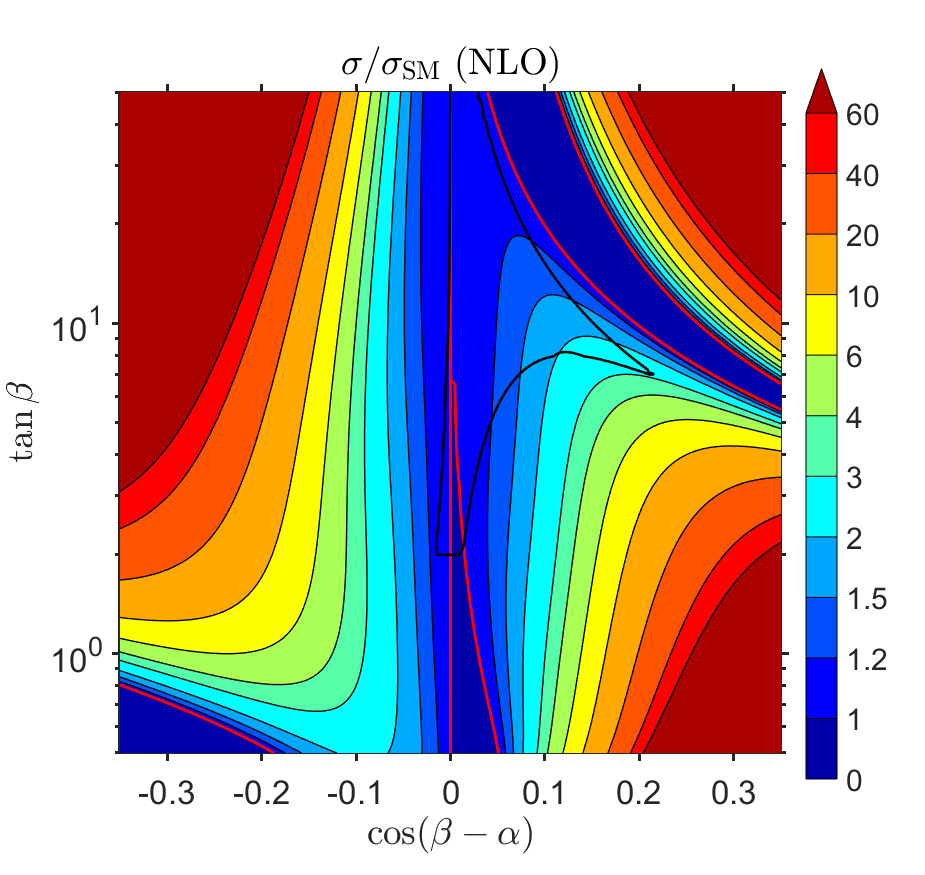}
\includegraphics[width=0.41\textwidth]{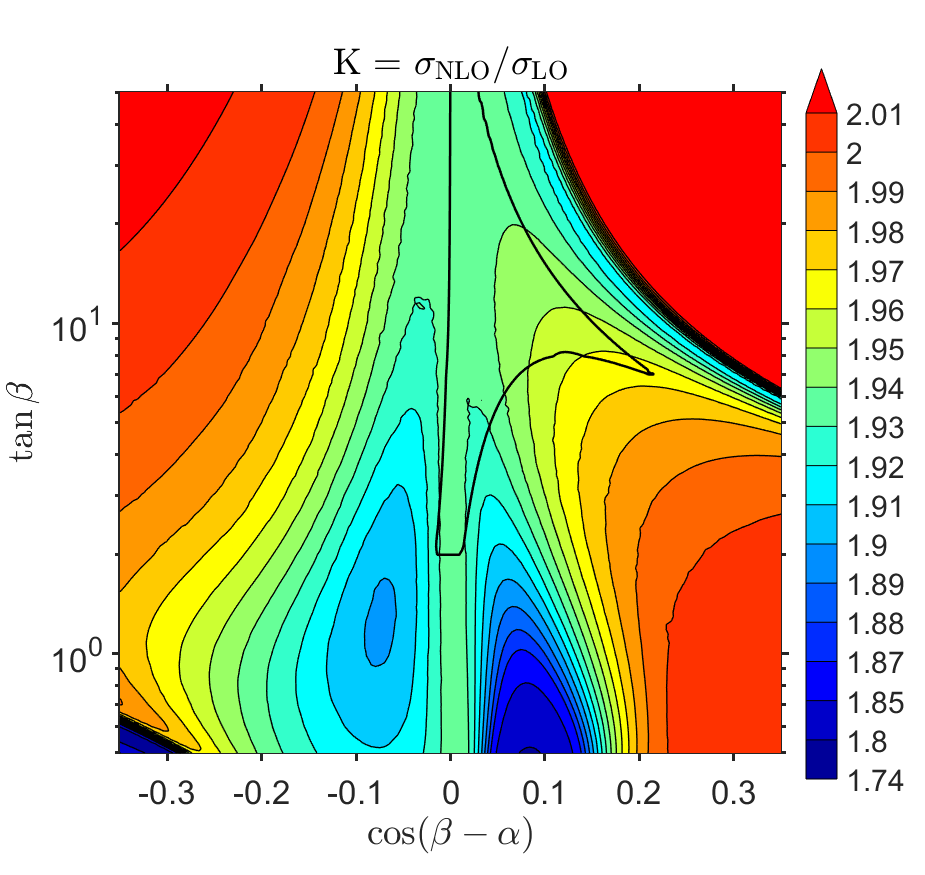}
\includegraphics[width=0.41\textwidth]{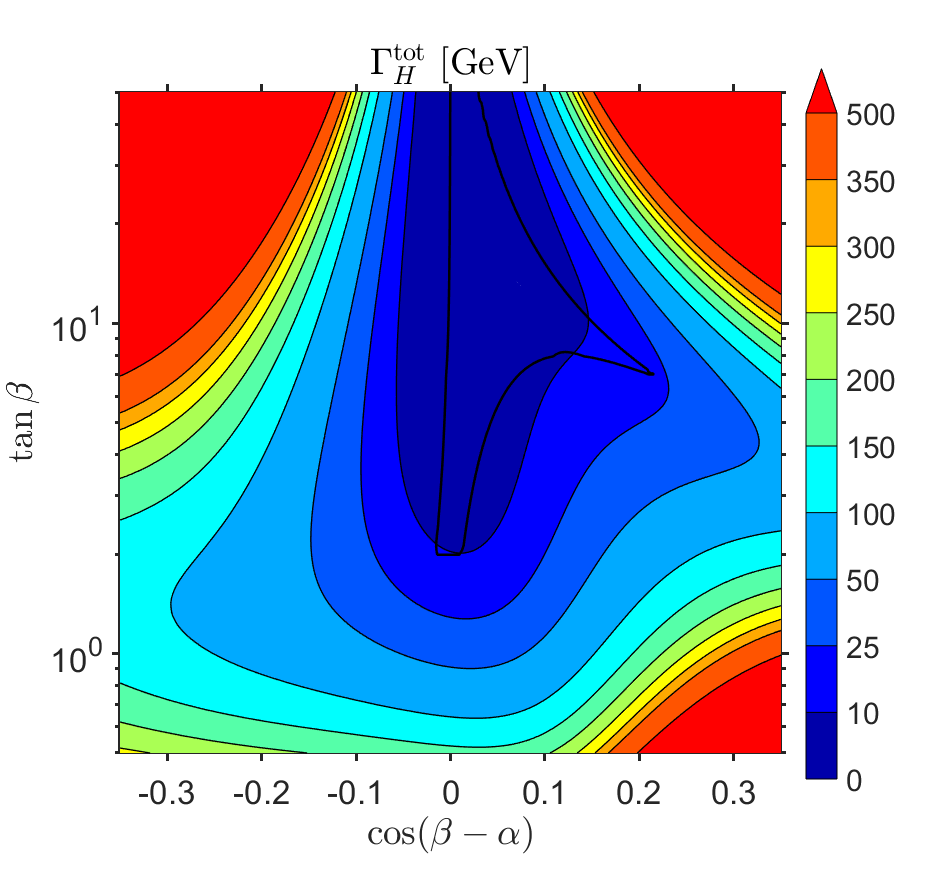}
\includegraphics[width=0.41\textwidth]{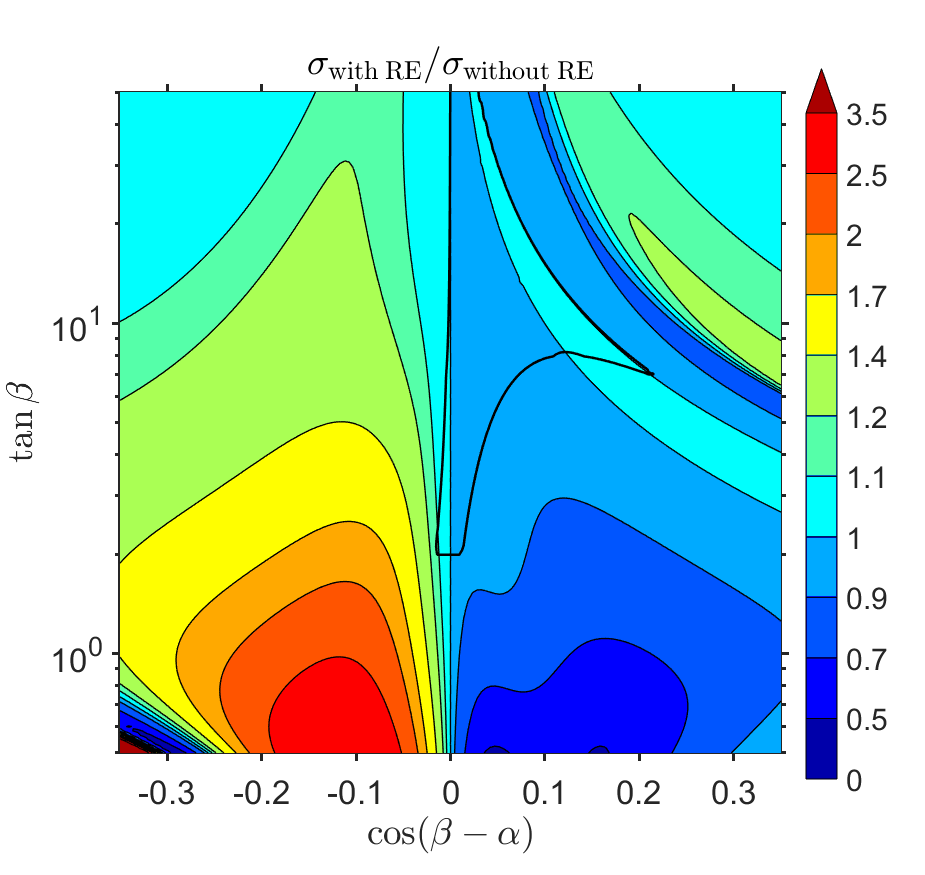}
	\end{center}
  \caption{\textbf{Plane 1}. 2HDM type I,
    \ $\tan\beta$ versus $\cos(\beta-\alpha)$.\
  \textit{Upper left}: Cross section 
  prediction for di-Higgs production in 2HDM normalized to the SM
  value, both evalued at NLO QCD in the heavy
    top-quark limit. Allowed area inside the black contour. Red lines indicate
  the values at with the ratio is 1 (i.e. $\sig = \sig_{\rm SM}$). The color
  coding indicates the size of this ratio. \textit{Upper right}:
  $K$-factor, defined as the ratio of the NLO and LO cross
  sections. \textit{Lower left}: Total decay width of the heavy Higgs
  $H$. \textit{Lower right}: Ratio of the cross section with and without
  resonant enhancement, both evaluated at LO.}
\vspace{-1em}
\label{fig:xsec1}
\end{figure}

\begin{figure}[ht!]
  \begin{center}
\includegraphics[width=0.41\textwidth]{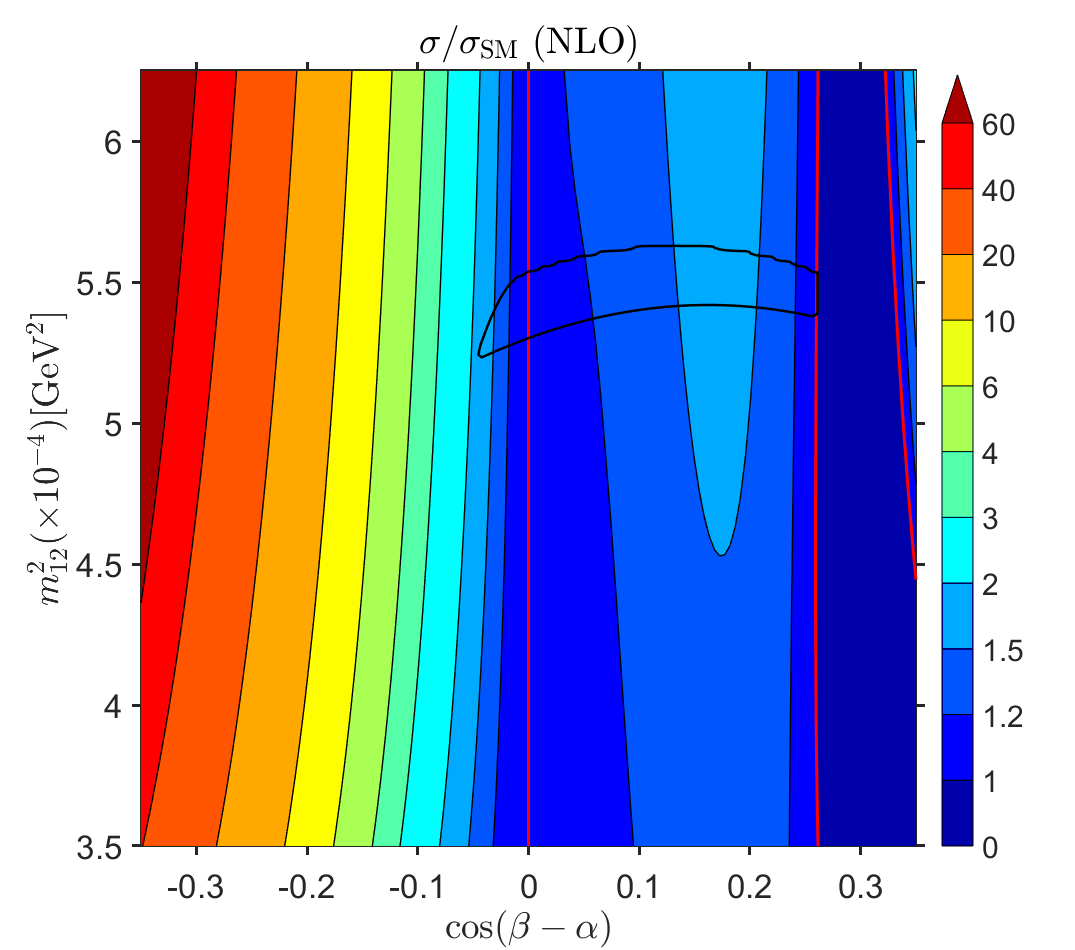}
\includegraphics[width=0.41\textwidth]{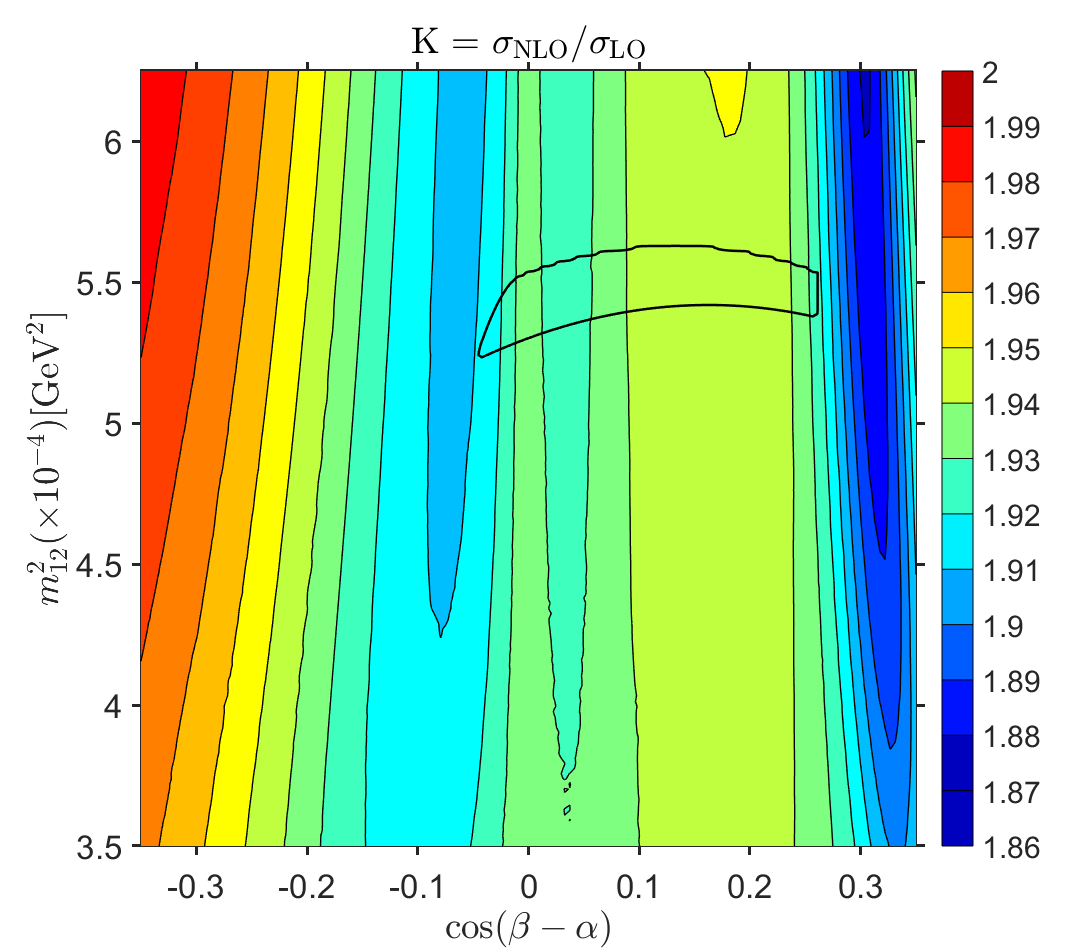}
\includegraphics[width=0.41\textwidth]{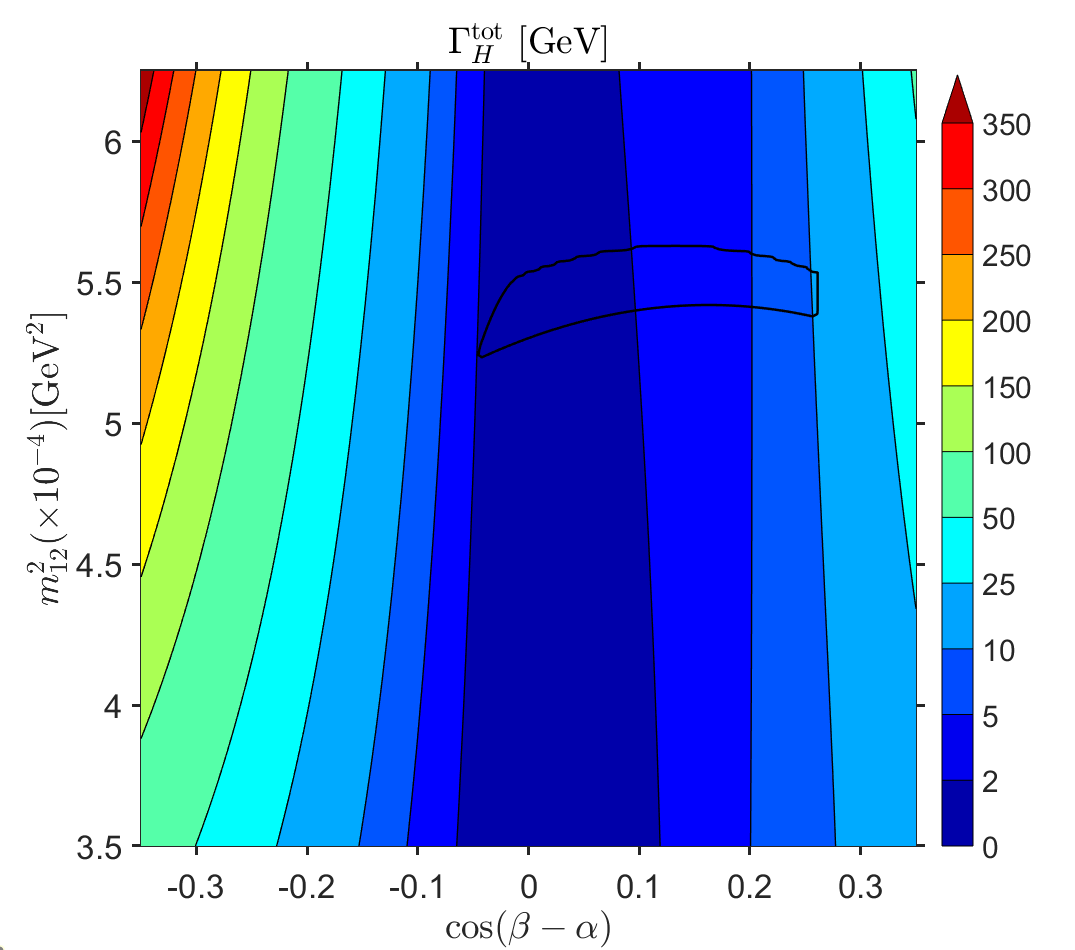}
\includegraphics[width=0.41\textwidth]{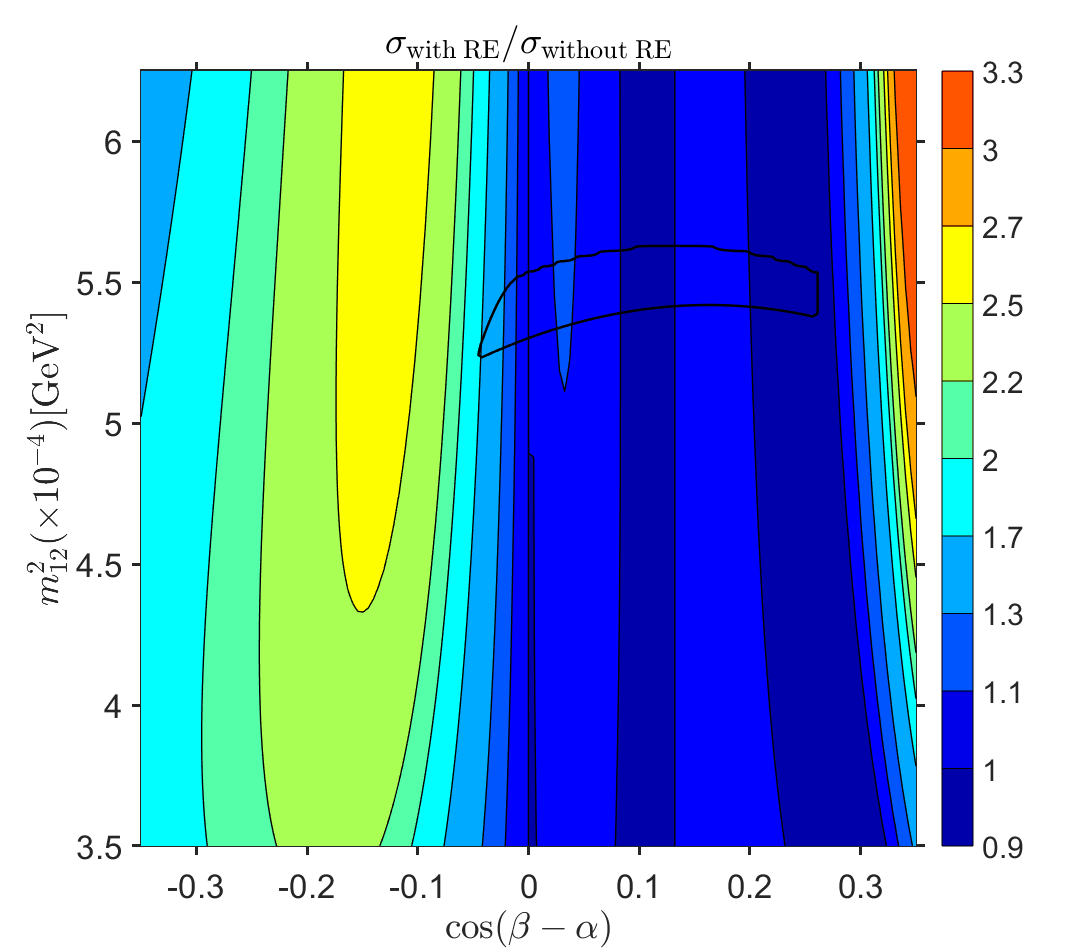}
	\end{center}
  \caption{\textbf{Plane 2}. 2HDM, type I, $m_{12}^2$ versus
    $\cos(\beta-\alpha)$. Otherwise plots as in 
    \protect\reffi{fig:xsec1}.}
\vspace{-1em}
\label{fig:xsec2}
\end{figure}

\begin{figure}[ht!]
\vspace{-1em}
  \begin{center}
\includegraphics[width=0.41\textwidth]{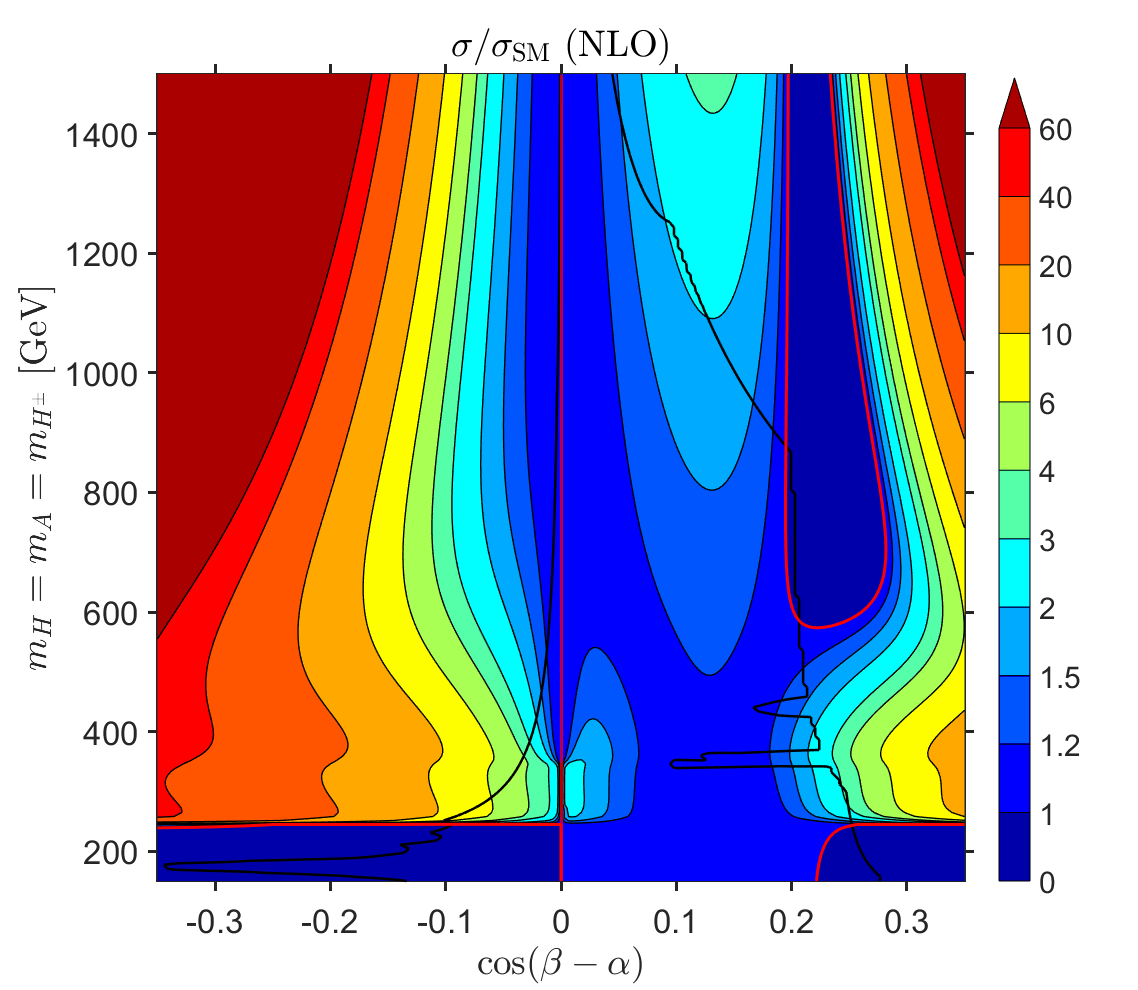}
\includegraphics[width=0.41\textwidth]{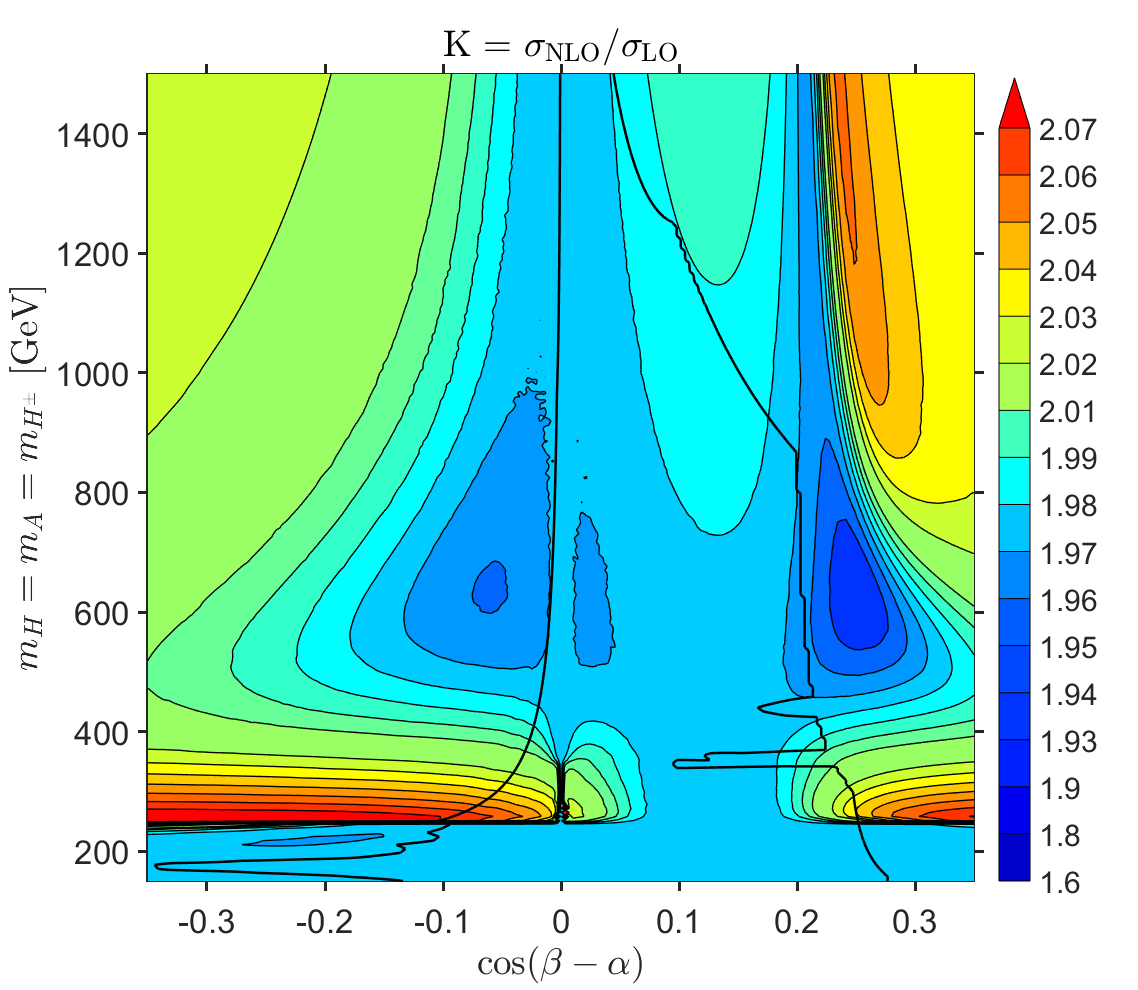}
\includegraphics[width=0.41\textwidth]{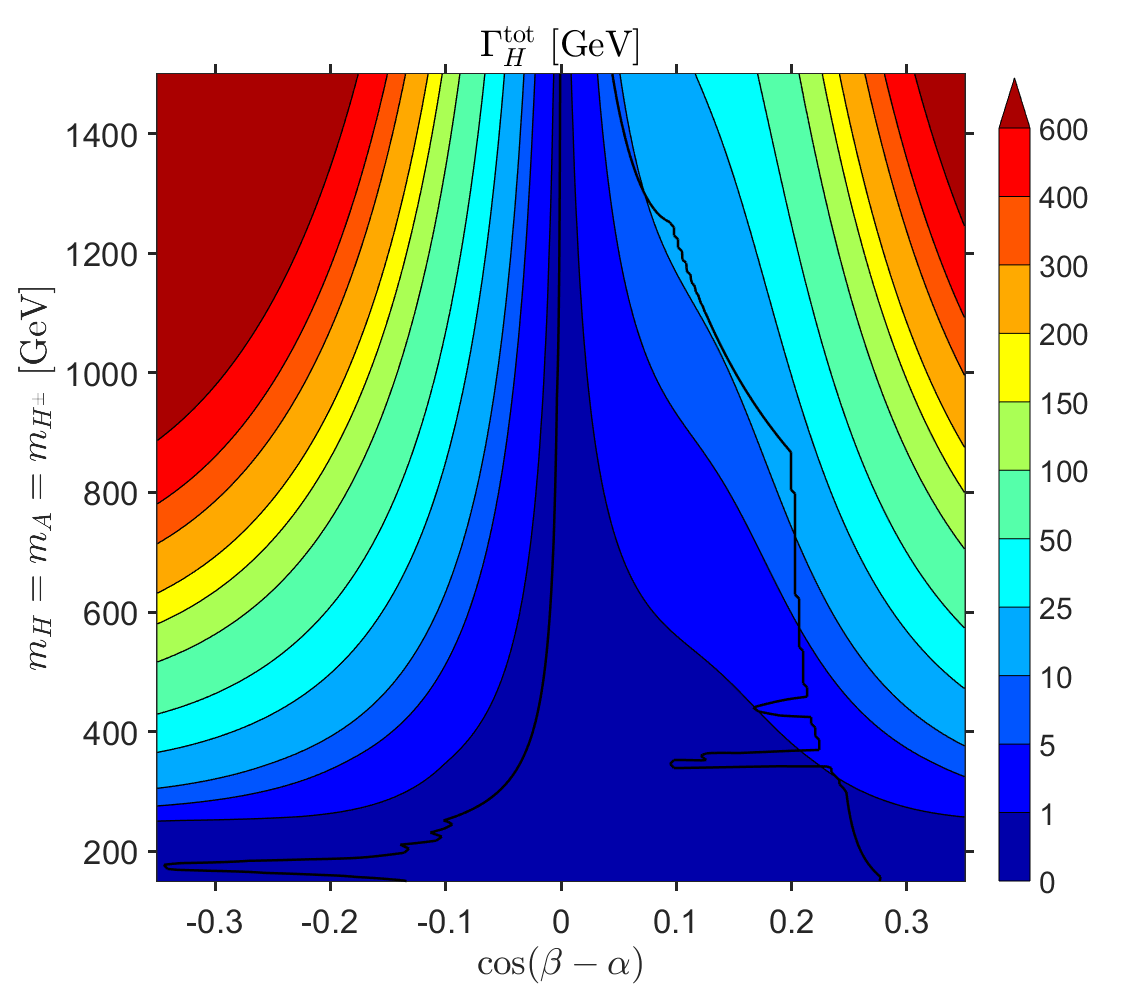}
\includegraphics[width=0.41\textwidth]{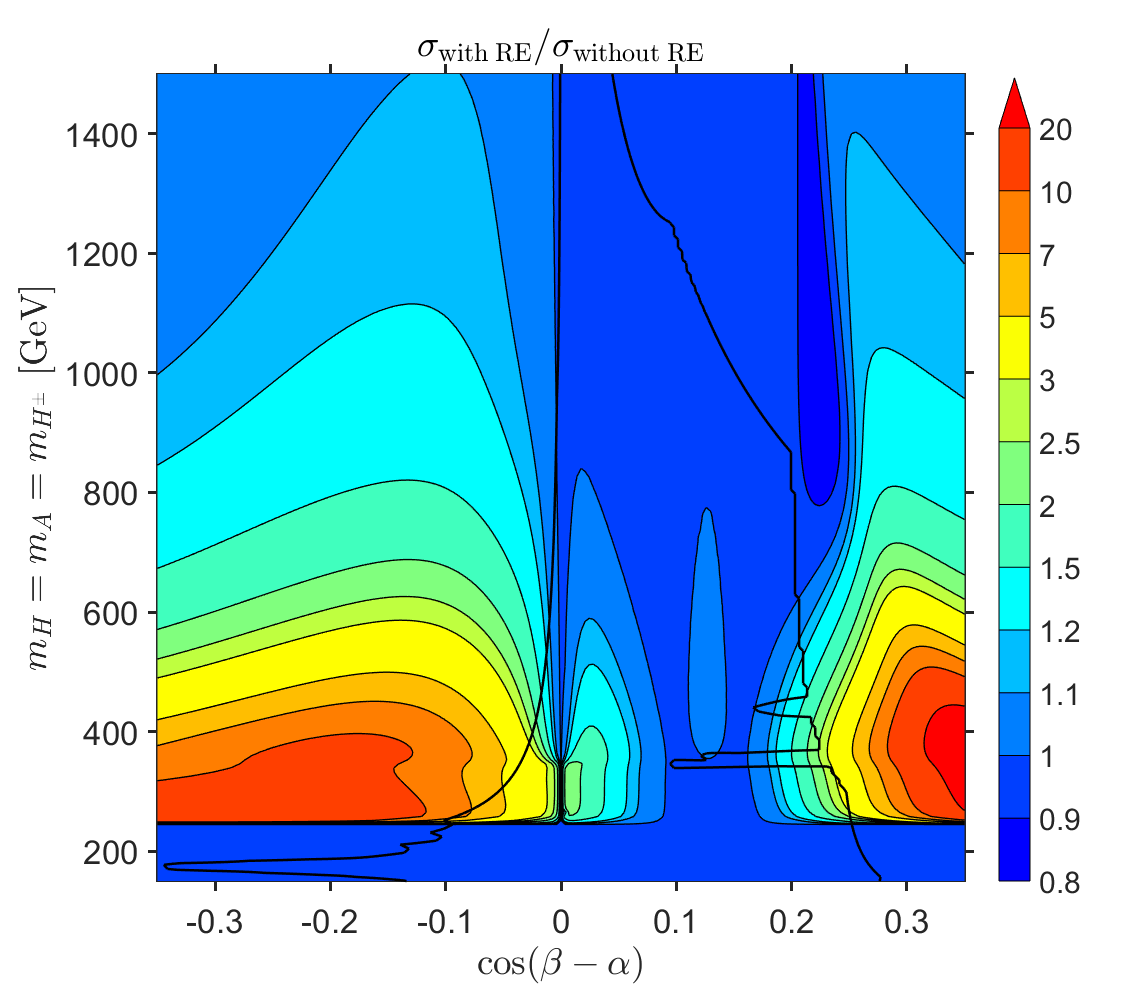}
	\end{center}
  \caption{\textbf{Plane 3}. 2HDM type I, $m_H=m_A=m_{H^\pm}$ versus
    $\cos(\beta-\alpha)$. Otherwise plots as in
    \protect\reffi{fig:xsec1}.}
\vspace{-1em}
\label{fig:xsec3}
\end{figure}

\begin{figure}[ht!]
\vspace{-1em}
  \begin{center}
\includegraphics[width=0.41\textwidth]{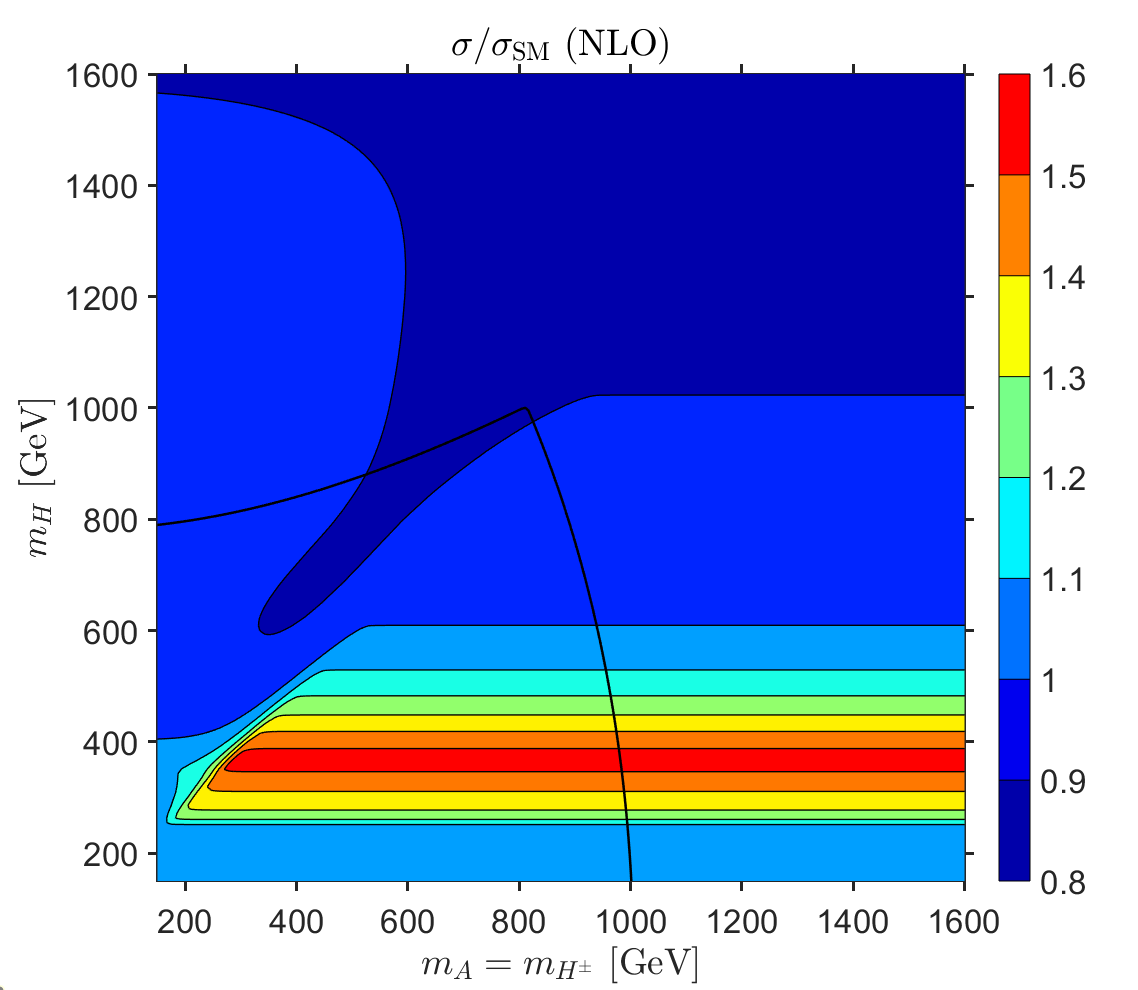}
\includegraphics[width=0.41\textwidth]{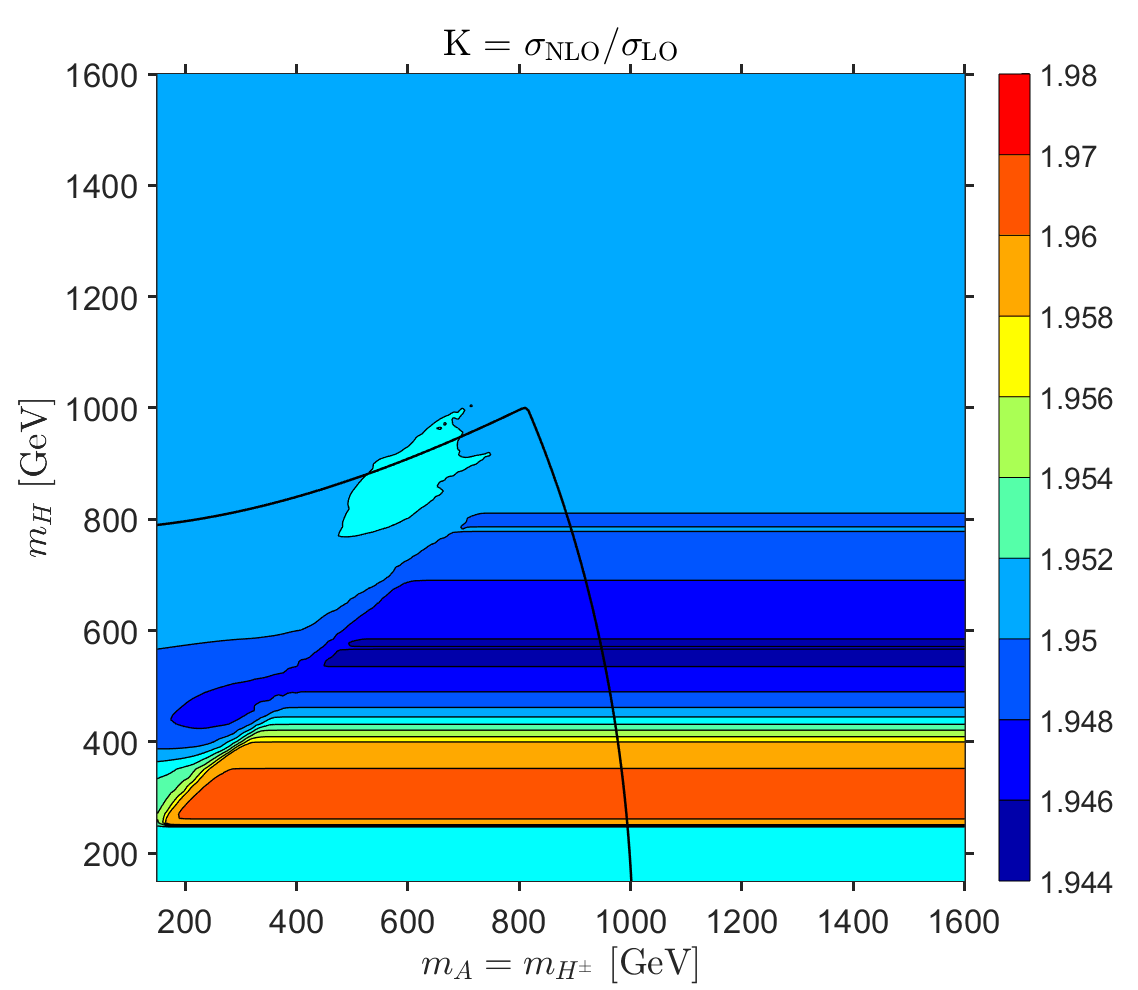}
\includegraphics[width=0.41\textwidth]{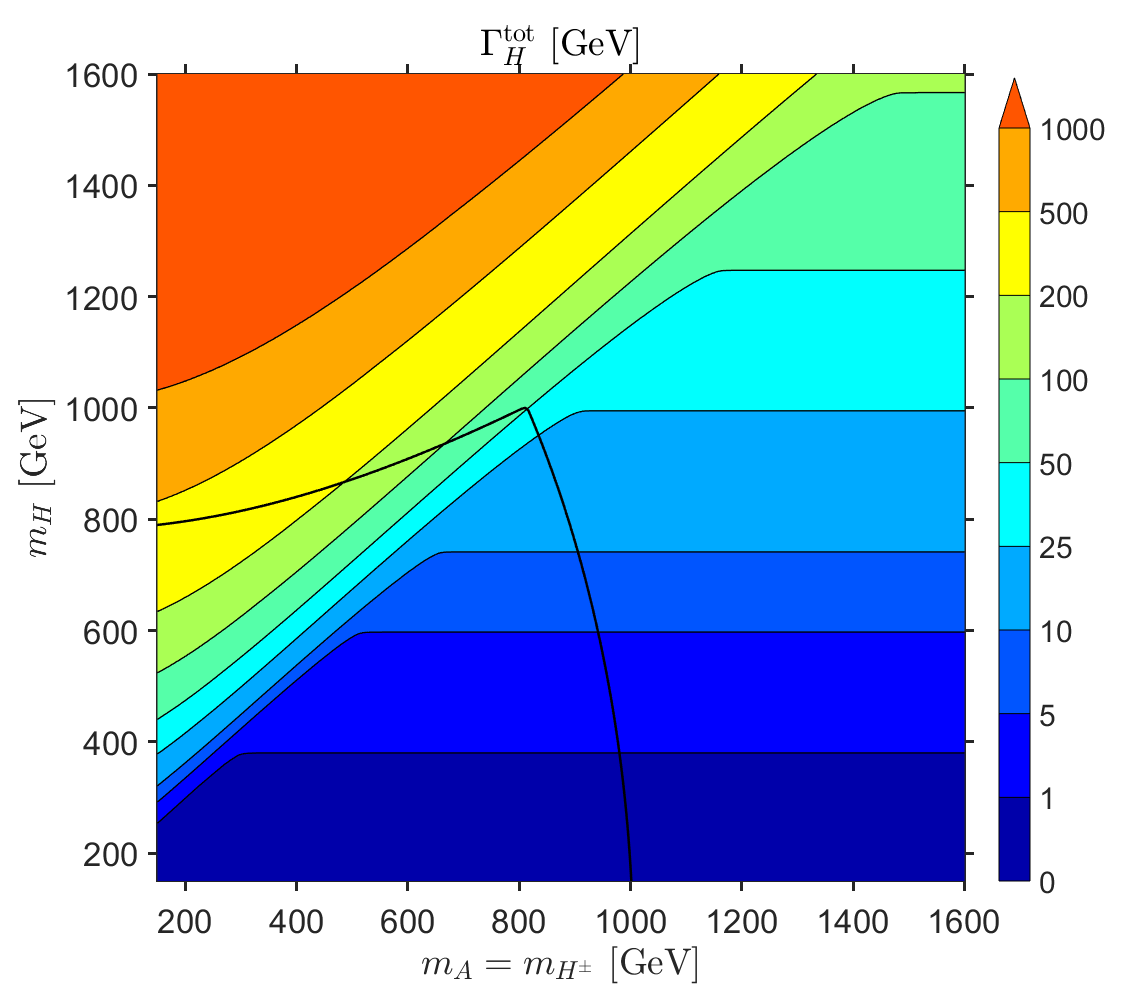}
\includegraphics[width=0.41\textwidth]{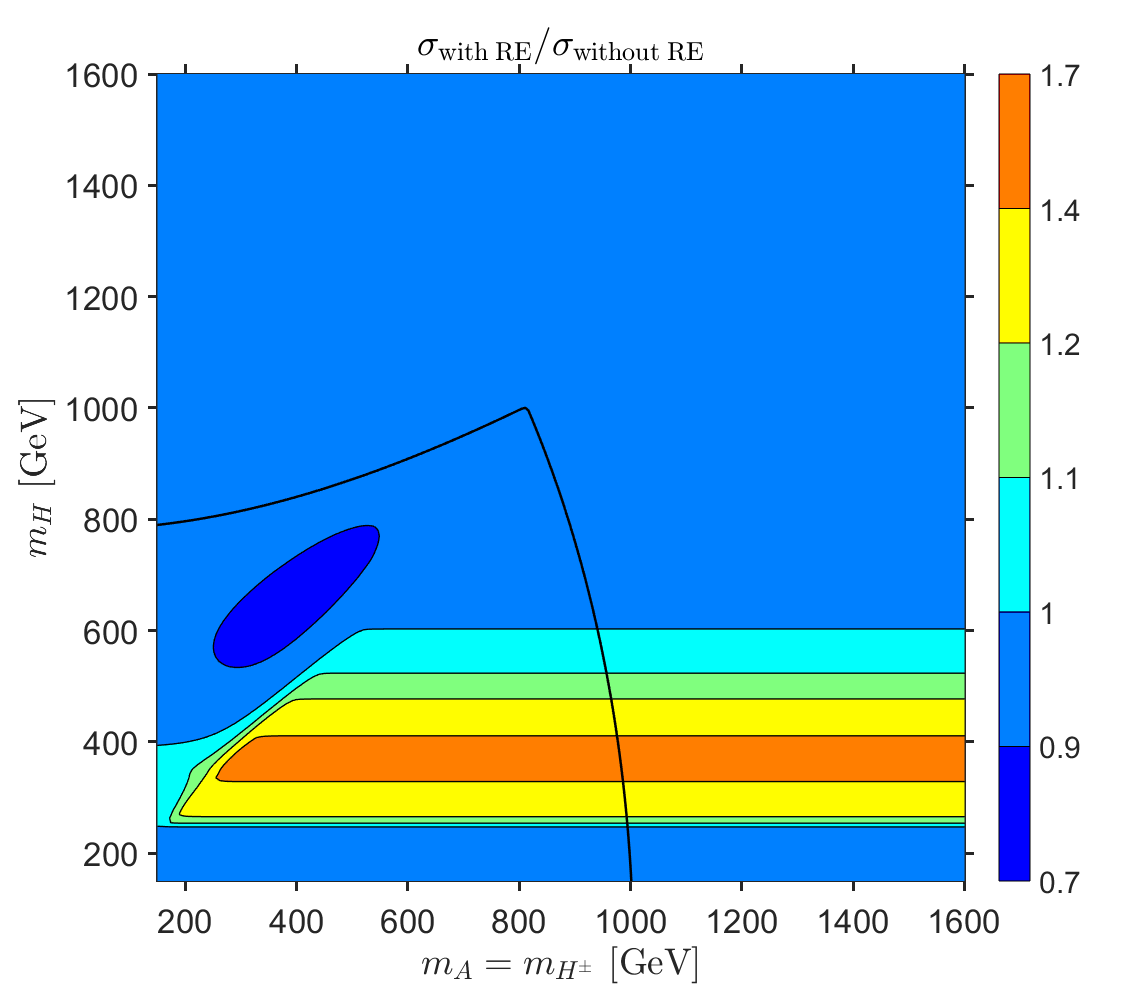}
	\end{center}
  \caption{\textbf{Plane 4}. 2HDM type I, $m_H$ versus $m_A =
    m_{H^\pm}$. Otherwise plots as in 
    \protect\reffi{fig:xsec1}.}
\vspace{-1em}
\label{fig:xsec8}
\end{figure}

In \reffis{fig:xsec1} - \ref{fig:xsec8} we show the results for the
di-Higgs production cross section in the 2HDM normalized to the SM value
calculated at NLO for the gluon fusion process. The SM prediction was
obtained with \texttt{HPAIR} assuming the alignment limit and coincides
with the values given in \citere{Abouabid:2021yvw},\footnote{As the
  NLO corrections are computed in the heavy top-quark limit it differs
from the NLO SM-value including the full mass dependence
\cite{borowka1,borowka2,baglio1,baglio2,baglio3}. For a
recent review of higher-order corrections to SM di-Higgs production,
see \cite{DiMicco:2019ngk}. Further, more recent results on
higher-order corrections to Higgs pair production in gluon fusion can
be found in \cite{Chen:2019lzz, Chen:2019fhs,Heinrich:2020ckp,deFlorian:2021azd,Heinrich:2022idm,Muhlleitner:2022ijf,Davies:2022ram,Ajjath:2022kpv}.}

\begin{equation}
    \sig^{\rm LO}_{\rm SM}\; (pp \to hh) = 19.76\; {\rm fb}\,, \quad
    \sig^{\rm NLO}_{\rm SM}\; (pp \to hh) = 38.24\; {\rm fb}\,.
    \label{eq:smvalues}
\end{equation}

The results for all the benchmark planes are shown as follows. In the upper
left plot of each figure, we present the NLO 2HDM cross sections
normalized to the NLO SM value, as
indicated by the color coding. The red line shows where the ratio is
one, and the inner part of the solid black line is allowed by all
theoretical and experimental constraints (as evaluated in
\citeres{Arco:2020ucn,Arco:2022xum}), see \refse{sec:constraints}. 
The upper right plot shows the
$K$-factor, $K = \sig_{\rm NLO}/\sig_{\rm LO}$ of the 2HDM $hh$ cross
sections. It should be noted that for the determination of the
$K$-factor we 
consistently evaluated the LO cross section with LO pdfs and
the strong coupling $\alpha_s$ at LO and the NLO cross section is
evaluatied with NLO pdfs 
and NLO $\al_s$~\citeres{Buckley:2014ana,Dulat:2015mca}.
The lower left plot 
indicates the total width of the heavy $\cp$-even Higgs boson, which
contributes via the $s$-channel diagram (\reffi{fig:BSMtriangle} with
$h_i = H$). This quantity will be relevant 
for the discussion of the dependence of the cross section on the
underlying parameter space. Finally, the lower right plot shows the
ratio of the full cross section, devided by the cross
section obtained omitting the diagram with $H$ in the $s$-channel,
both evaluated at leading order. Large
deviations from unity indicate an important contribution from the resonant
$H$~diagram. 

\medskip
The analysis for the benchmark \textbf{plane 1} is presented in
\reffi{fig:xsec1}, where $\tan\beta$ is plotted against
$\cos(\beta-\alpha)$ for the four quantities described in the previous
paragraph. We observe that the maximum deviation from the
SM prediction within the allowed area occurs precisely at the ``tip'' that
is furthest away from the alignment limit $\cos (\beta - \alpha)=0$
(see upper left plot). Here the enhancement factor 
in the cross section is $\sim 3$. This corresponds to the minimum size
of the triple Higgs coupling \lahhh\ that was obtained in the allowed region of
this benchmark plane ($\kala \sim -0.4$). If a
deviation between the SM prediction and experiment is observed, this
could point to a deviation of the 
\kala\ coupling. The $K$-factor in this region results in a factor close
to 2, more 
specifically in the allowed region of $\sim 1.92$ to
$\sim 1.97$ (upper right plot). One can see that the
decay width of $H$ in this region can amount to  
$\sim 25 \gev$ (lower left plot). A large total width $\Gamma^{\text{tot}}_H$ has
a suppressing effect on 
the $H$ contribution to the cross section due to its appearance in the
denominator of the $s$-channel propagator, as will be 
discussed below. We see that the $H$~contribution has no
enhancing effect on $hh$ production within the allowed area given by
the black lines (lower right plot). Including the resonance either
slightly suppresses the cross section or leaves it unchanged (we find
the ratio 1 exactly at the ``tip'', where the $hh$ cross section is
maximal within the allowed region). In
conclusion, in this plane the maximum enhancement of the cross section is
precisely due to the deviation of the triple Higgs coupling from the
expected SM value. 

\medskip
In the case of the benchmark \textbf{plane 2}, shown in
\reffi{fig:xsec2}, where we now plot 
$m_{12}^2$ versus $\cos(\beta-\alpha)$, it can be observed that the
cross section does not 
have a significant  enhancement in the allowed region, as it does not
even reach a factor of $\sim 2$ times the SM value. Also for this
benchmark plane we observe that the largest value of the 
cross section falls in the region of the minimum value of \kala,
i.e.\ where
the destructive interference between box diagram 
and $h$~exchange is minimal. The $K$ factor
in the allowed region is roughly $\sim 1.91$ to $\sim 1.95$, again close
to~2. In the evaluation of the effect of the heavy Higgs $H$, we observe that
neither the decay widths nor the resonant enhancement are significant in
the allowed region. The $H$ contribution almost has no effect on the
$hh$ production cross section. 

\medskip
In the benchmark \textbf{plane 3}, shown in \reffi{fig:xsec3}, where
we plot $m_H=m_A=m_{H^\pm}$ against $\cos(\beta-\alpha)$, 
we allow for a variation of the
heavier Higgs masses by fixing the value of $\tb$ and the mass
$m_{12}^2$ according to the \refeq{eq:m12special}. 
Concerning the normalized NLO cross section there is no
enhancement below $250 \gev$ since the heavy Higgs is not produced
on-shell. Above this threshold there is resonant enhancement due
to the 
contribution of the heavy Higgs in the $s$-channel. In particular, the
cross section is enhanced by up to a factor of $\sim 8$ in the ``tail'' at
$\CBA$ $\sim -0.1$. 
In this region one finds \kala\ close to~1, so that the enhancement
of the cross section w.r.t.~the SM is given by the diagram with $H$ in
the $s$-channel that resonantly decays into $hh$. 
This can also be seen in the lower right plot of \reffi{fig:xsec3},
where a ratio of resonant over non-resonant cross section of up to $\sim 8$
is found. 
The $K$-factor in this region is slightly above 2 (up to $\sim 2.06$). 

Since in this plane we allow for a variation of the mass of the heavy
Higgs in the propagator one can observe the enhancement of the cross section
around $\MH \sim 350-400 \gev$ that is expected from single Higgs
production above the di-top threshold, $\MH \sim 2 \mt$.
This enhancement is clearly visible in the lower right plot
of \reffi{fig:xsec3} (particularly for negative $\CBA$ within the
allowed region).  

The final plane we investigate in detail is \textbf{plane 4} as shown in
\reffi{fig:xsec8}. Here the two free parameters, that we plot against
each other, are $\MH$ and $\MA = \MHp$. Correspondingly, one expects a
large enhancement of the 2HDM 
cross section for $\MH \sim 400 \gev$ where $m_H > 2 m_h$ and above
the di-top threshold. Exactly this can be observed in
the upper left plot (total cross section) and the lower right plot
(relative enhancement from the resonant diagram). The cross section can
be up to 60\% larger than the SM di-Higgs production cross section, with
a $K$~factor again close to~2. The total width of the heavy $\cp$-even
Higgs ranges from very small values up to larger than $200 \gev$ within
the allowed region (found for low $\MA = \MHp$ and large
$\MH$). For large total widths the resonance enhancement is not
effective. The largest enhancements of the cross section are found
for relatively small values of $\Ga_H^{\rm tot} \lsim 10 \gev$.



\subsection{Dependence on triple Higgs couplings}
\label{sec:sensitivity}

In this subsection we analyze the cross section with respect to the
triple Higgs couplings involved in the di-Higgs production process.
In particular, we will show in which part of the parameter space the
total di-Higgs production cross section has a relevant
dependence on \lahhh\ and/or \lahhH.

Here we focus on a statistical treatment of the errors of the total
cross section measurement, which is assumed to be Gaussian, neglecting
systematic uncertainties. 
It was found in~\citeres{Cepeda:2019klc,DiMicco:2019ngk} that the statistical
uncertainty of the total di-Higgs cross section measurement, assuming SM
values, will reach a level of $4.5\,\sig$ at the end of the HL-LHC,
combining ATLAS and CMS. (Taking into account systematic effects could
lower this value to $\sim 4\,\sig$.)
Consequently, we will approximate the corresponding error
in the measurement as $\de \mathbf{xs} = \mathbf{xs} / 4.5$.%
\footnote{It should be noted that we usually denote the cross section as $\sig$,
  but in this discussion we change our notation to $\mathbf{xs}$ since it
  can be misunderstood as the standard deviation in statistics which
  is also denoted as~$\sig$.}%
~The significance of the deviation of the (to be measured) 2HDM cross
section w.r.t.\ the SM value can then be expressed as 
\begin{equation}
  \De \sig_{\rm SM} \equiv
  \frac{\mathbf{xs}_{\rm 2HDM} - \mathbf{xs}_{\rm SM}}{\de \mathbf{xs}}.
    \label{eq:expsensit}
\end{equation}
It should be noted that this approximation becomes worse for larger
deviations of $\mathbf{xs}_{\rm 2HDM}$ from $\mathbf{xs}_{\rm SM}$,
since the precision of the measurments, $\de\mathbf{xs}$, has been
evaluated assuming the SM value. For higher cross sections a more
precise measurement can be expected. 

\begin{figure}[ht!]
\vspace{-1em}
  \begin{center}
\includegraphics[width=0.32\textwidth]{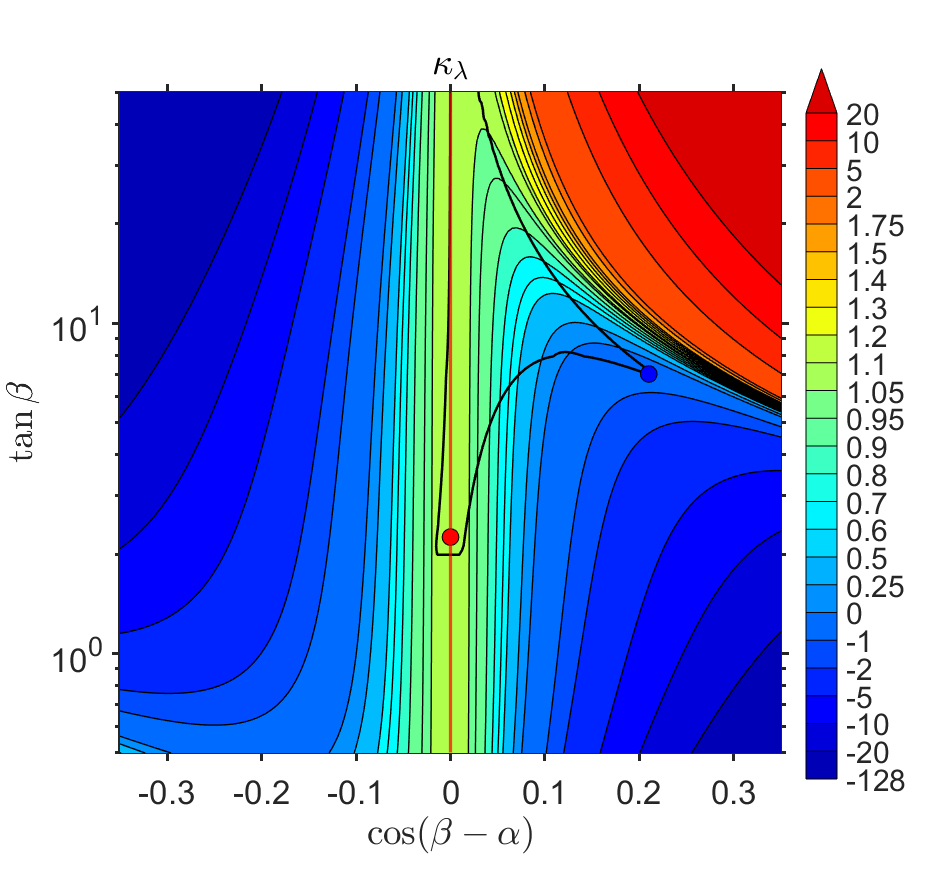}
\includegraphics[width=0.32\textwidth]{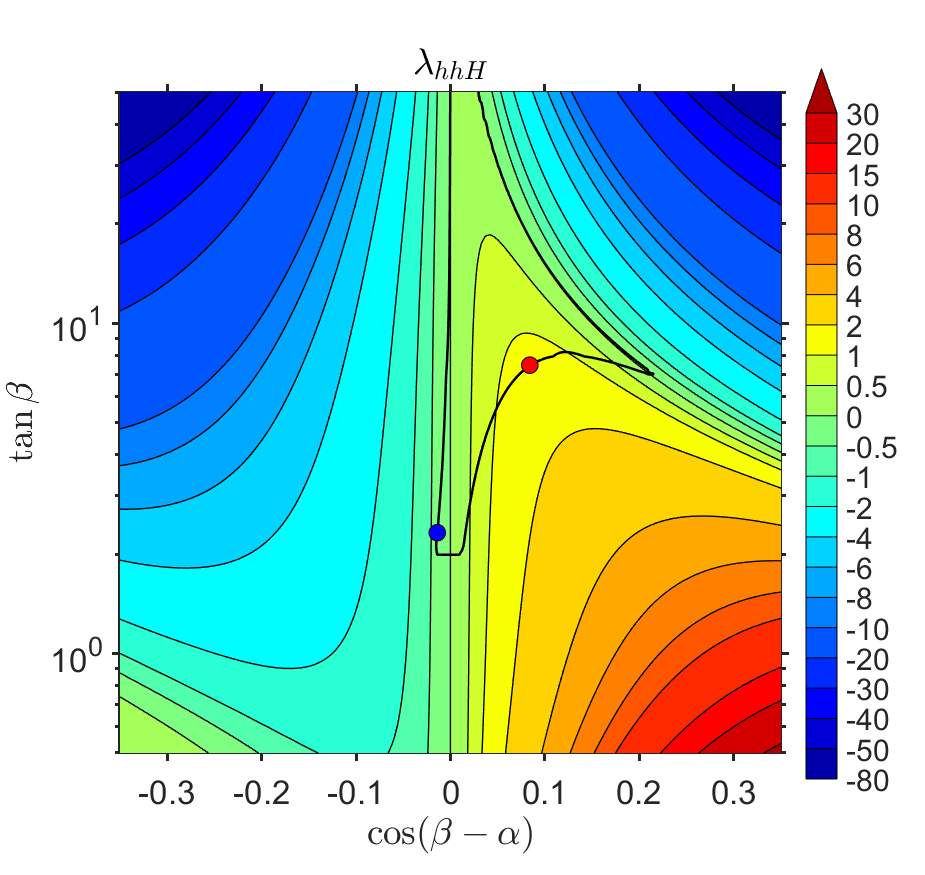}
\includegraphics[width=0.32\textwidth]{figs/sigma_1nlo.png}
\includegraphics[width=0.32\textwidth]{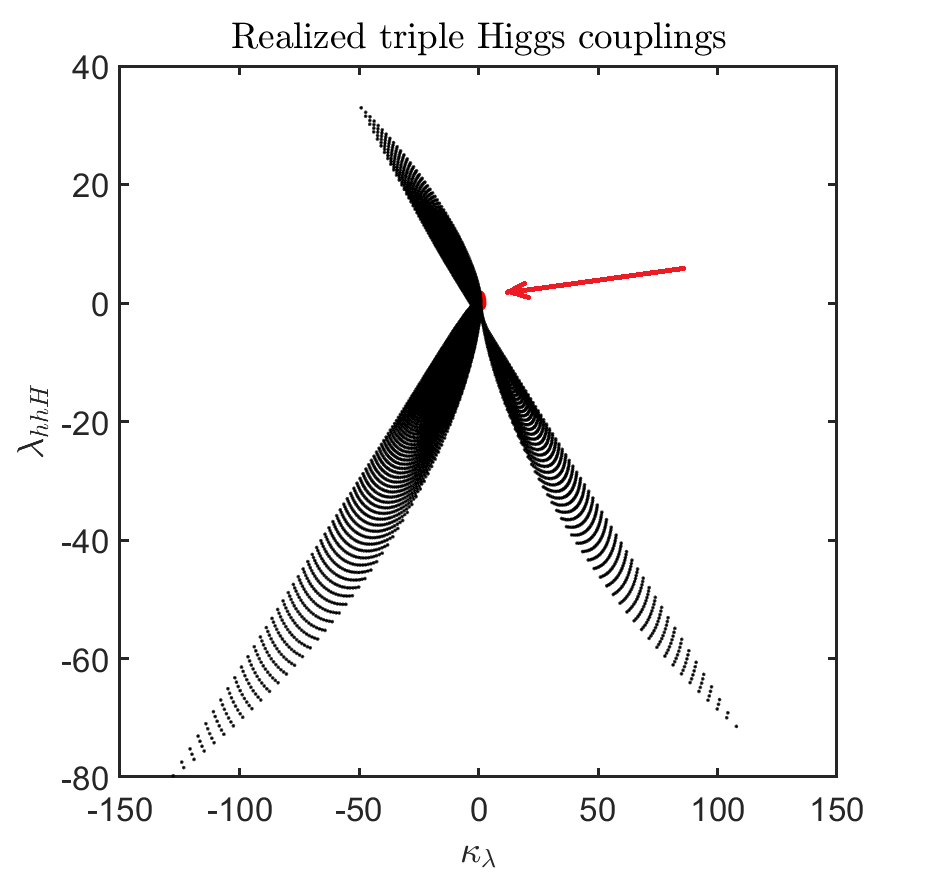}
\includegraphics[width=0.32\textwidth]{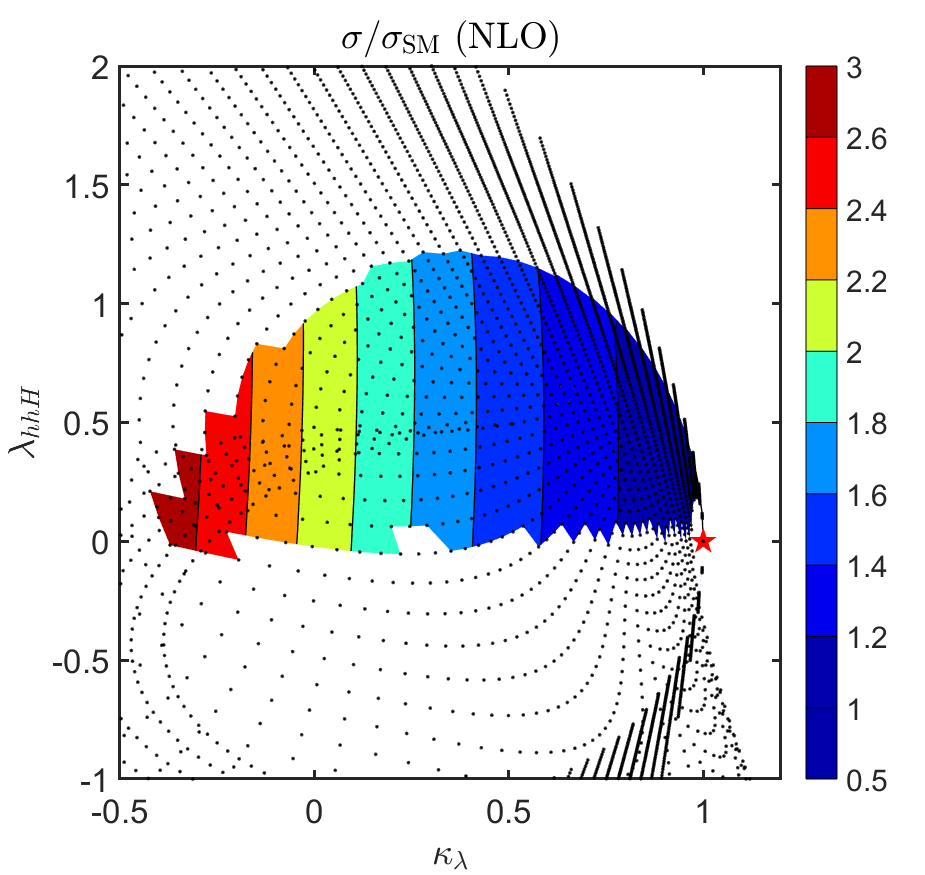}
\includegraphics[width=0.32\textwidth]{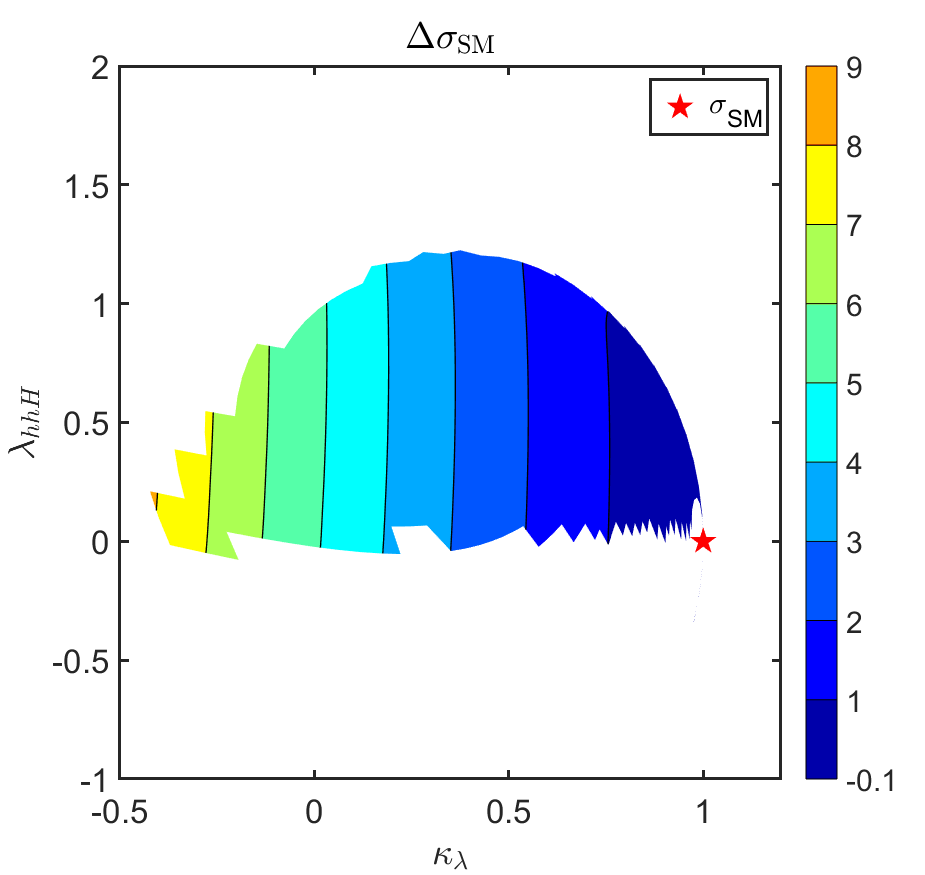}
  \end{center}
\caption{\textbf{Plane 1}. \textit{Upper line}: Triple Higgs coupling
  predictions in the 2HDM and value of the normalized cross section (w.r.t.\
  the SM value) evaluated at NLO QCD in the same parameter
  space. Red (blue) dots show the maximum (minimum) value of the
  trilinears that is realized in the allowed region.
  \textit{Lower left}: Points realized in the couplings
  plane, the red area around $\kala = 1$ and $\lahhH = 0$ represents the
  points that fall into the allowed region, indicated by a red arrow.
  \textit{Lower middle}: Zoom into the previous
  plot, color code indicates the normalized cross section within the
  allowed region for different values of triple Higgs couplings. Black
  dots indicate the existent scan parameter points in the $\kala$-$\lahhH$
  plane within and outside the allowed region that was obtained from the
  figures in the upper row. The red star indicates the SM limit ($\kala
  = 1$ and $\lahhH = 0$). \textit{Lower right}: Expected sensitivity to
  the deviation of the cross section from the SM value. Red star
  indicates the SM.}  
\vspace{-1em}
\label{fig:couplings1}
\end{figure}

\begin{figure}[ht!]
\vspace{-1em}
  \begin{center}
\includegraphics[width=0.32\textwidth]{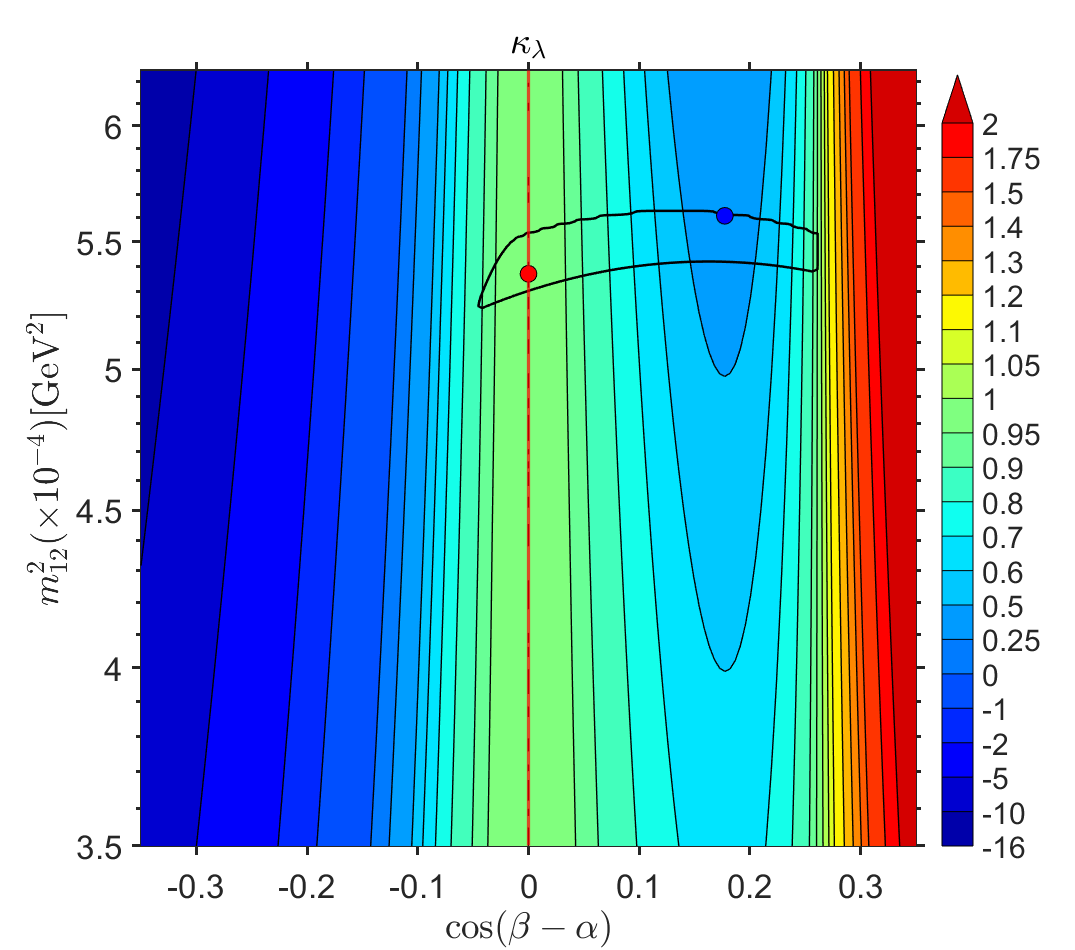}
\includegraphics[width=0.32\textwidth]{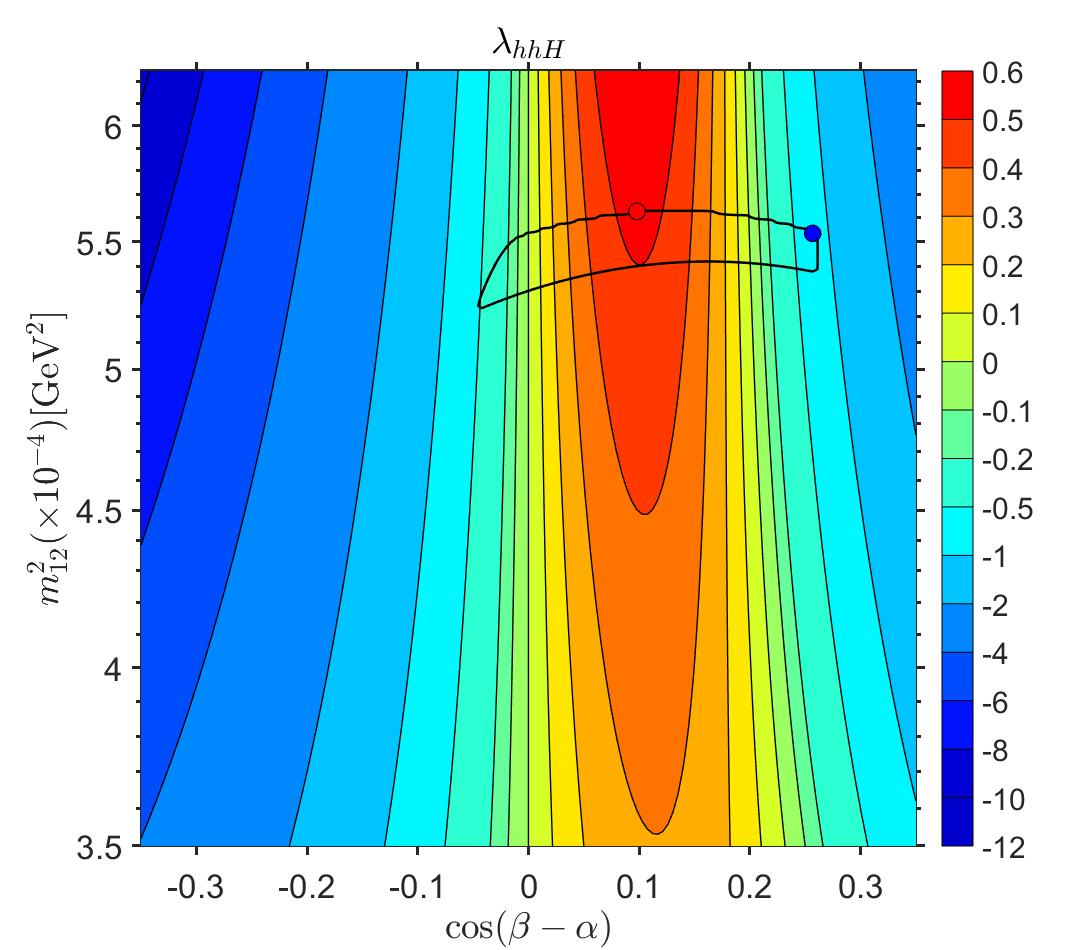}
\includegraphics[width=0.32\textwidth]{figs/sigma_2nlo.png}
\includegraphics[width=0.32\textwidth]{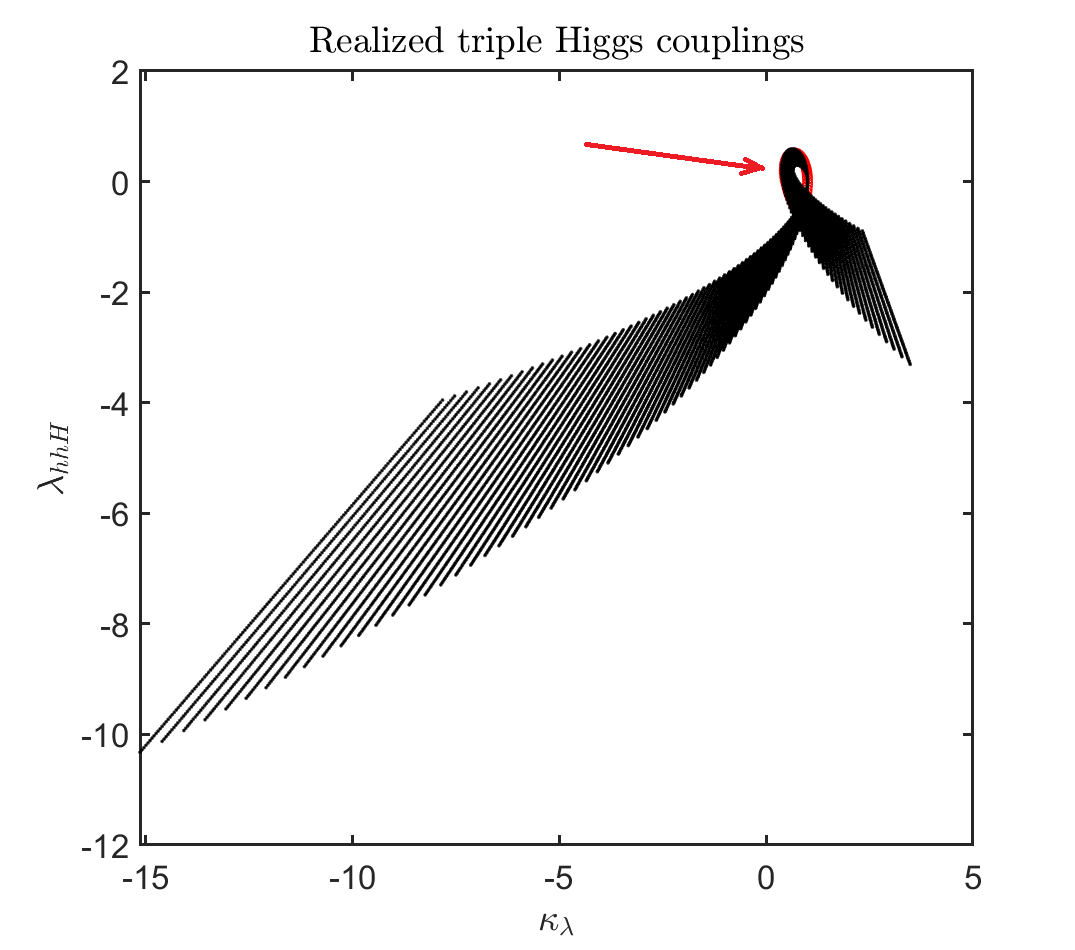}
\includegraphics[width=0.32\textwidth]{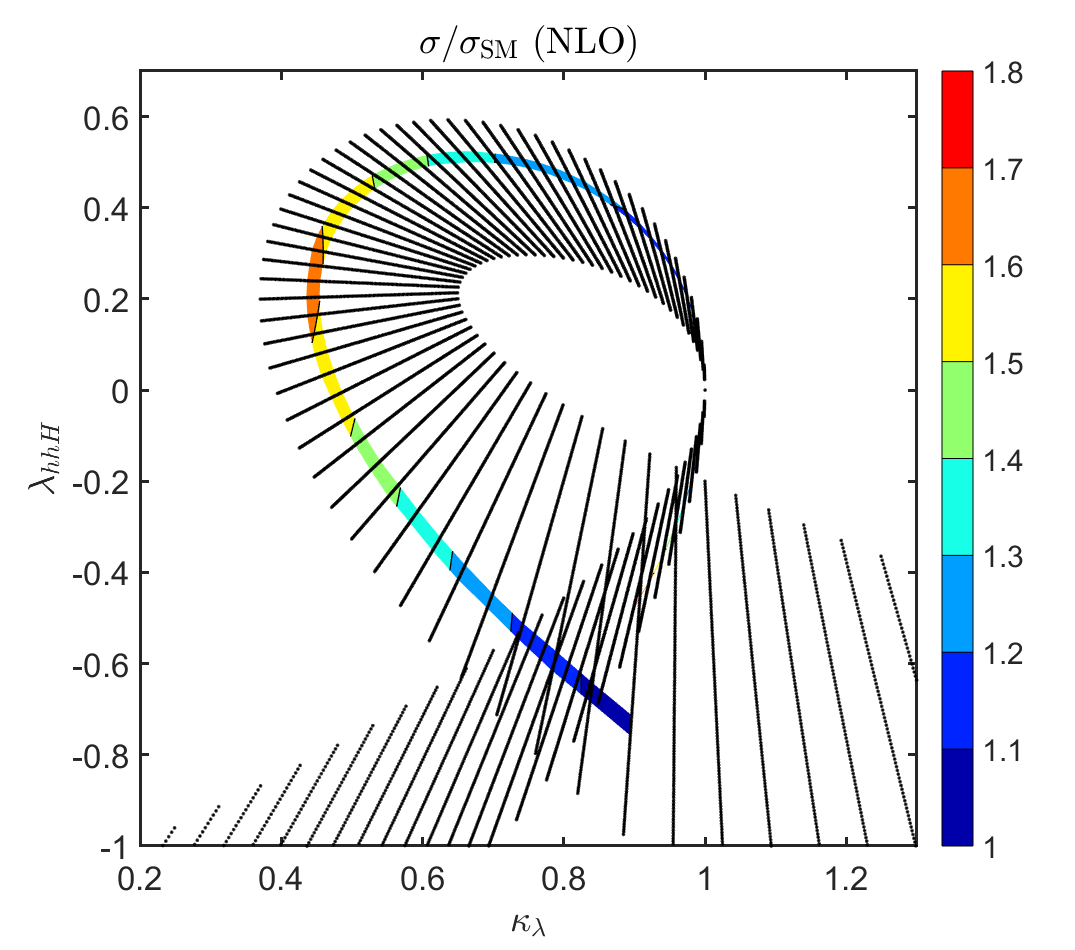}
\includegraphics[width=0.32\textwidth]{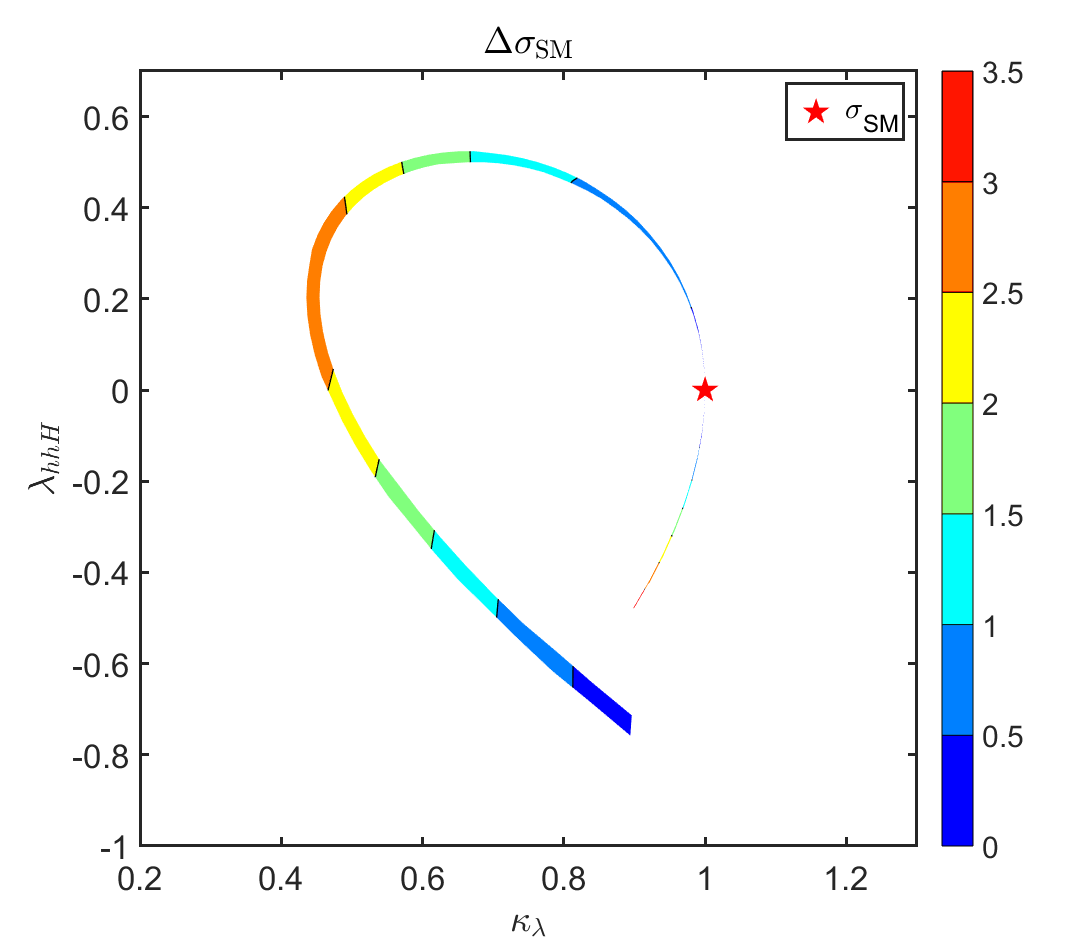}
  \end{center}
  \caption{\textbf{Plane 2}. 2HDM, type I, $m_{12}^2$ versus
    $\cos(\beta-\alpha)$. Otherwise plots as in \protect\reffi{fig:couplings1}.}
\label{fig:couplings2}
\end{figure}

\begin{figure}[ht!]
  \begin{center}
\includegraphics[width=0.32\textwidth]{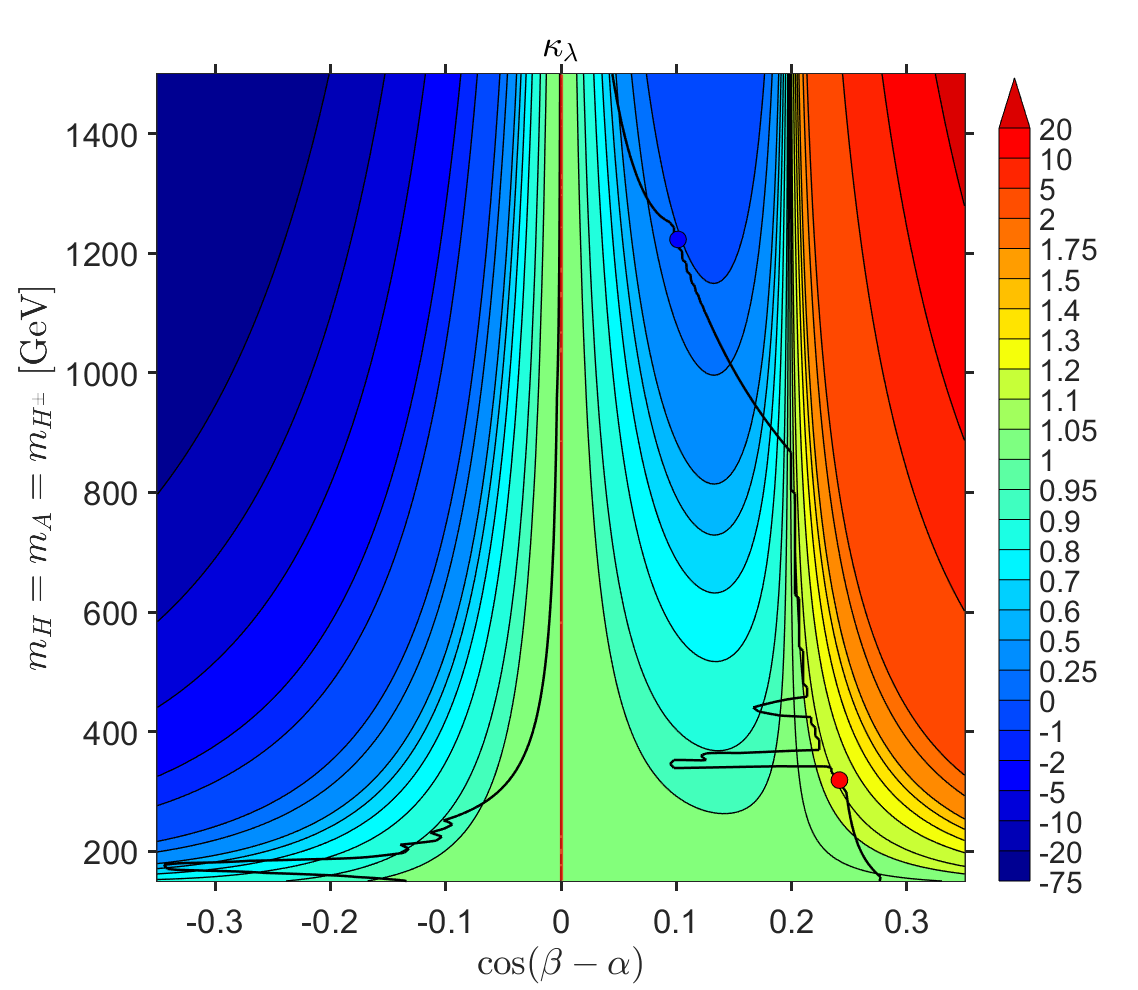}
\includegraphics[width=0.32\textwidth]{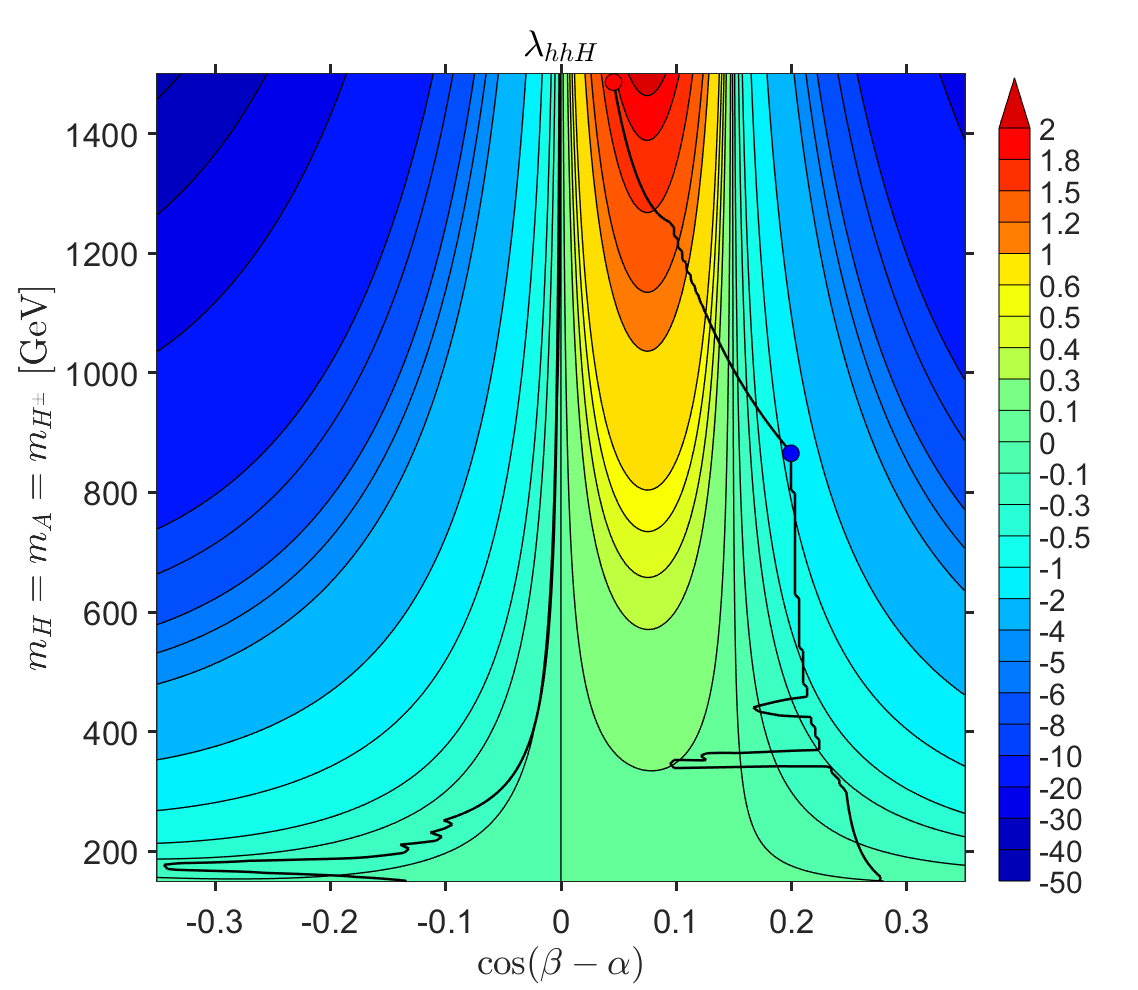}
\includegraphics[width=0.32\textwidth]{figs/sigma_3nlo.png}
\includegraphics[width=0.32\textwidth]{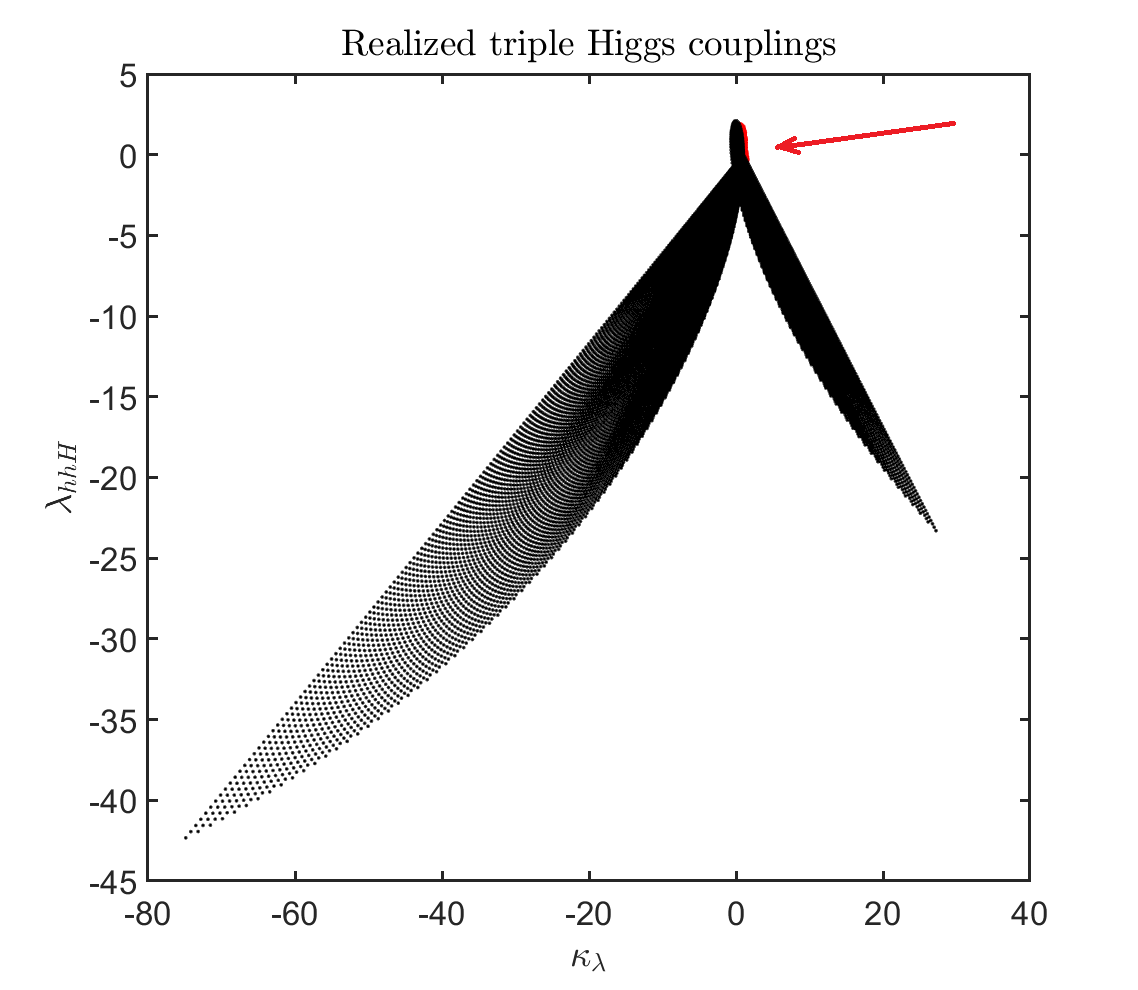}
\includegraphics[width=0.32\textwidth]{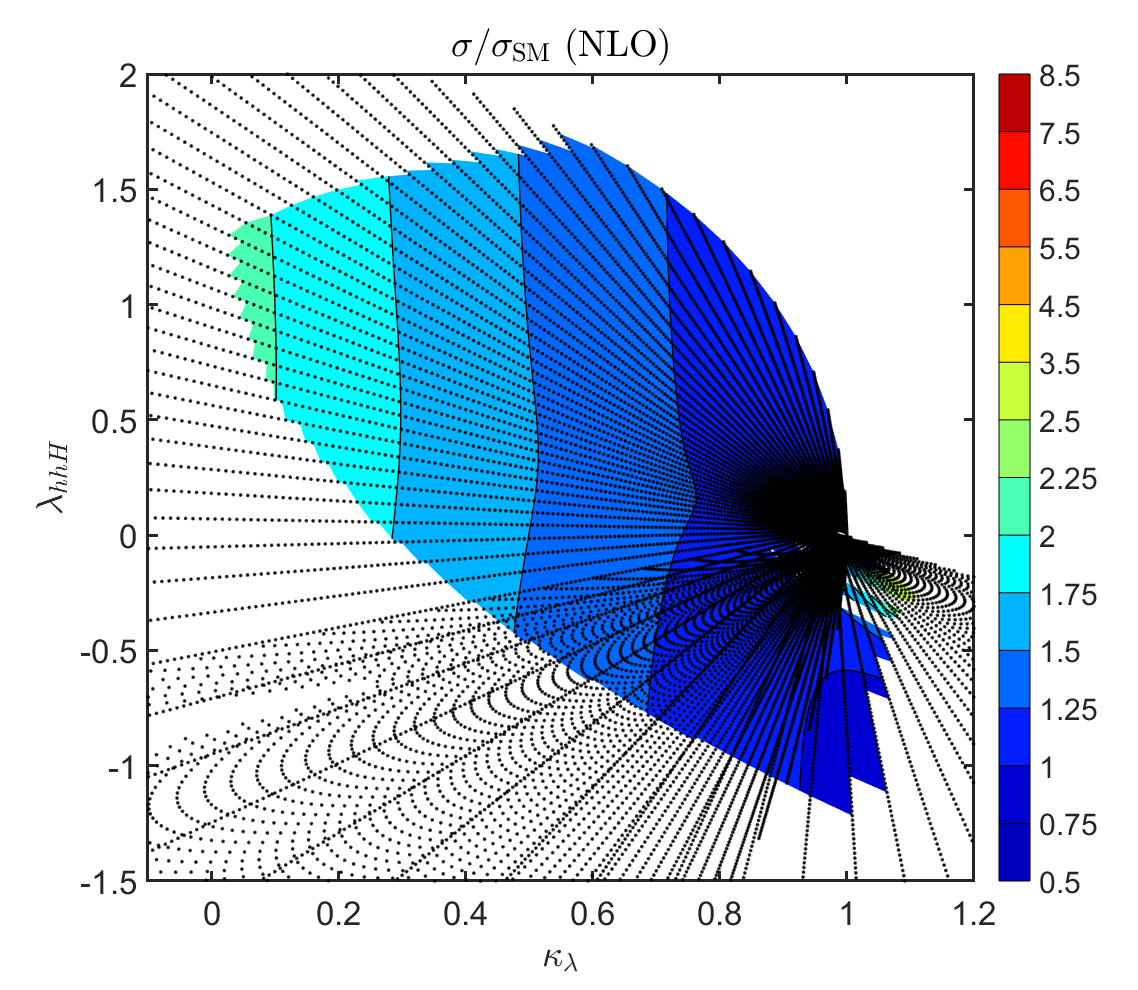}
\includegraphics[width=0.32\textwidth]{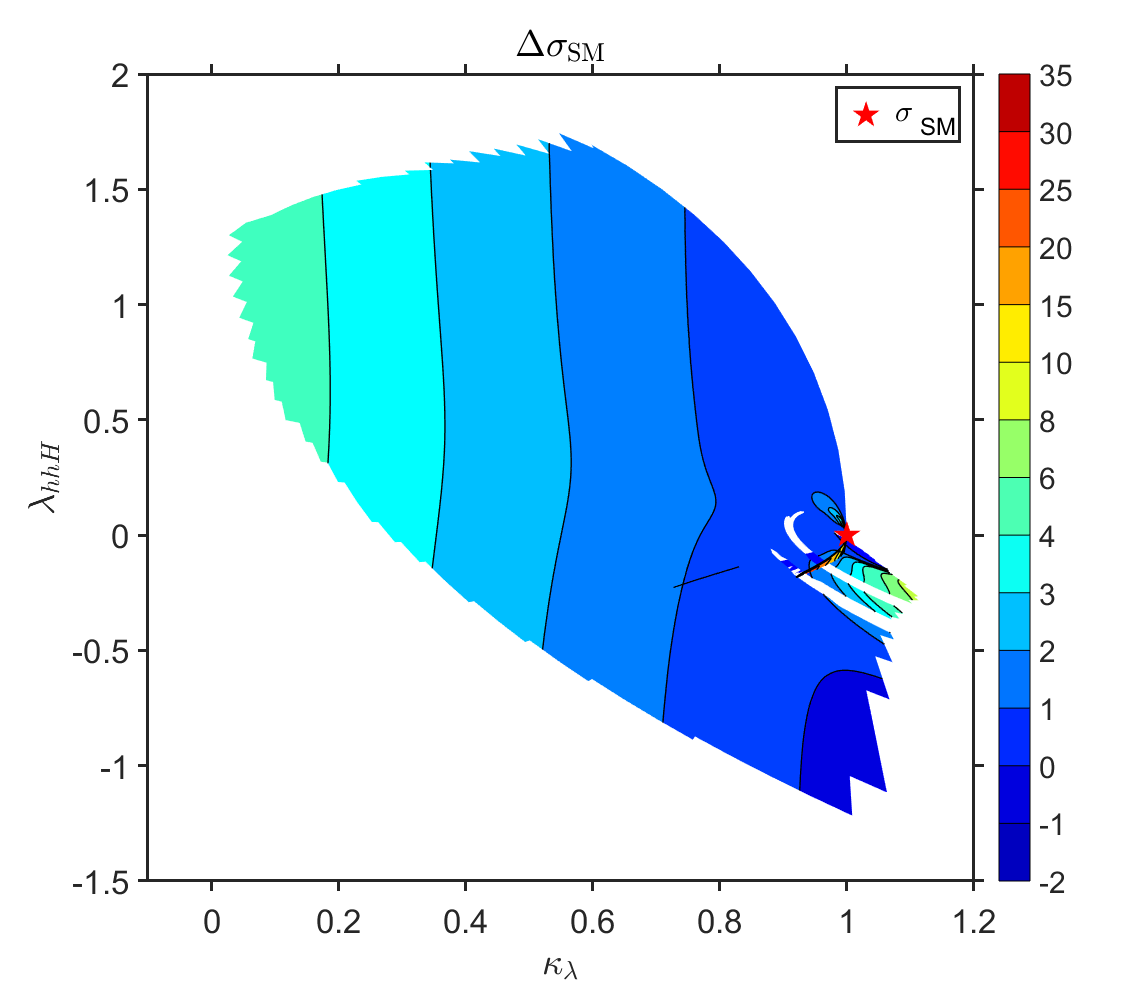}
  \end{center}
  \caption{\textbf{Plane 3}.  2HDM type I, $m_H=m_A=m_{H^\pm}$ versus
    $\cos(\beta-\alpha)$. Otherwise plots as in
    \protect\reffi{fig:couplings1}.}
\label{fig:couplings3}
\end{figure}

\begin{figure}[ht!]
\vspace{-1em}
  \begin{center}
\includegraphics[width=0.41\textwidth]{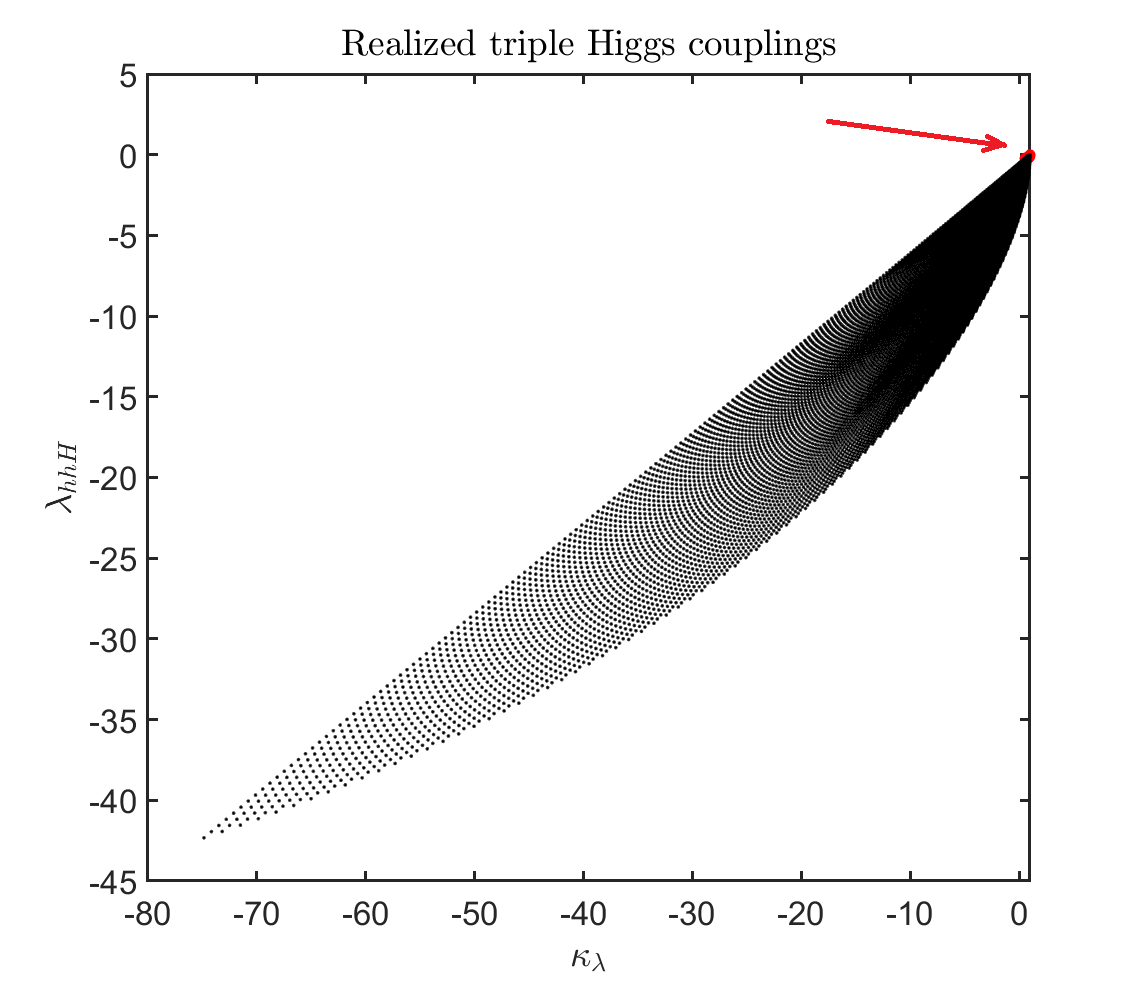}
\includegraphics[width=0.41\textwidth]{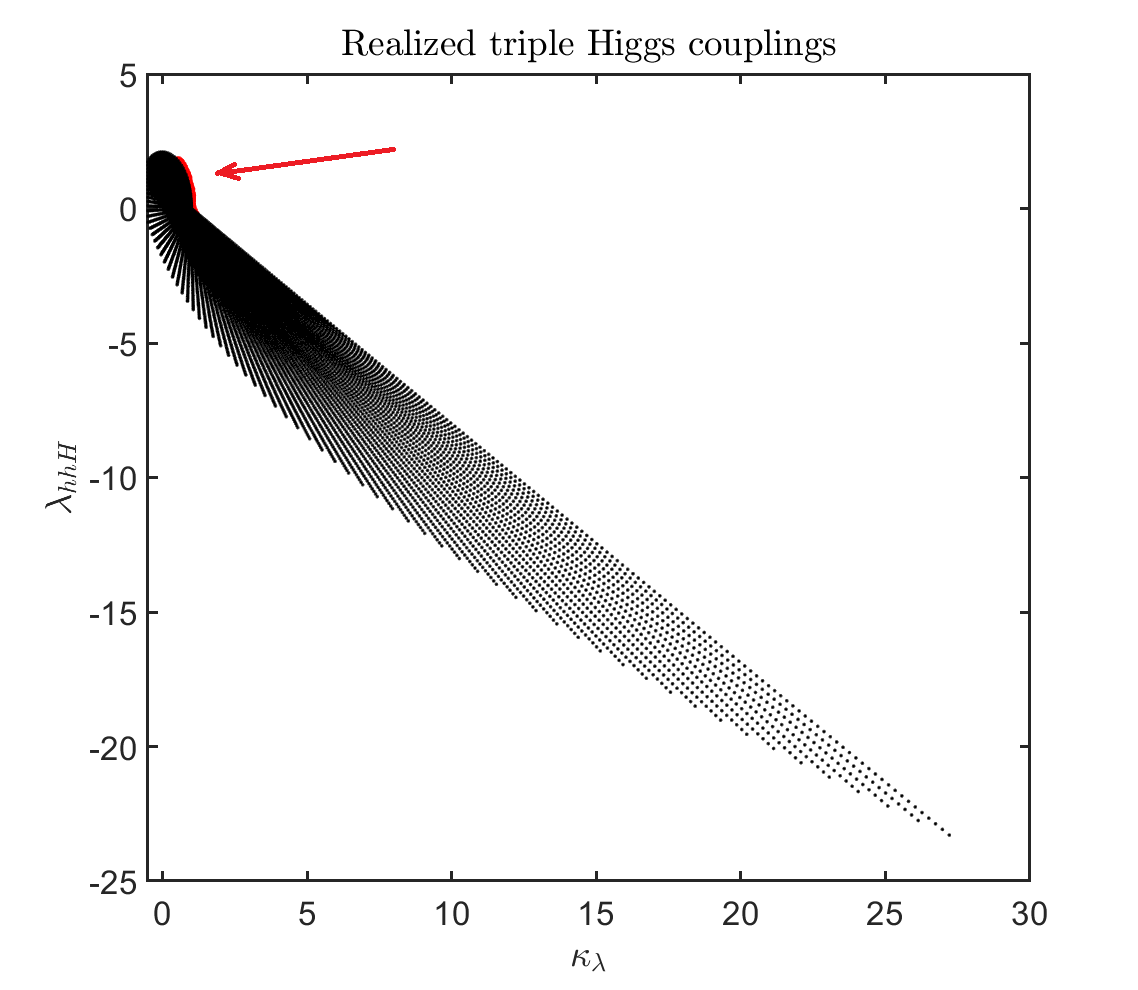}
\includegraphics[width=0.41\textwidth]{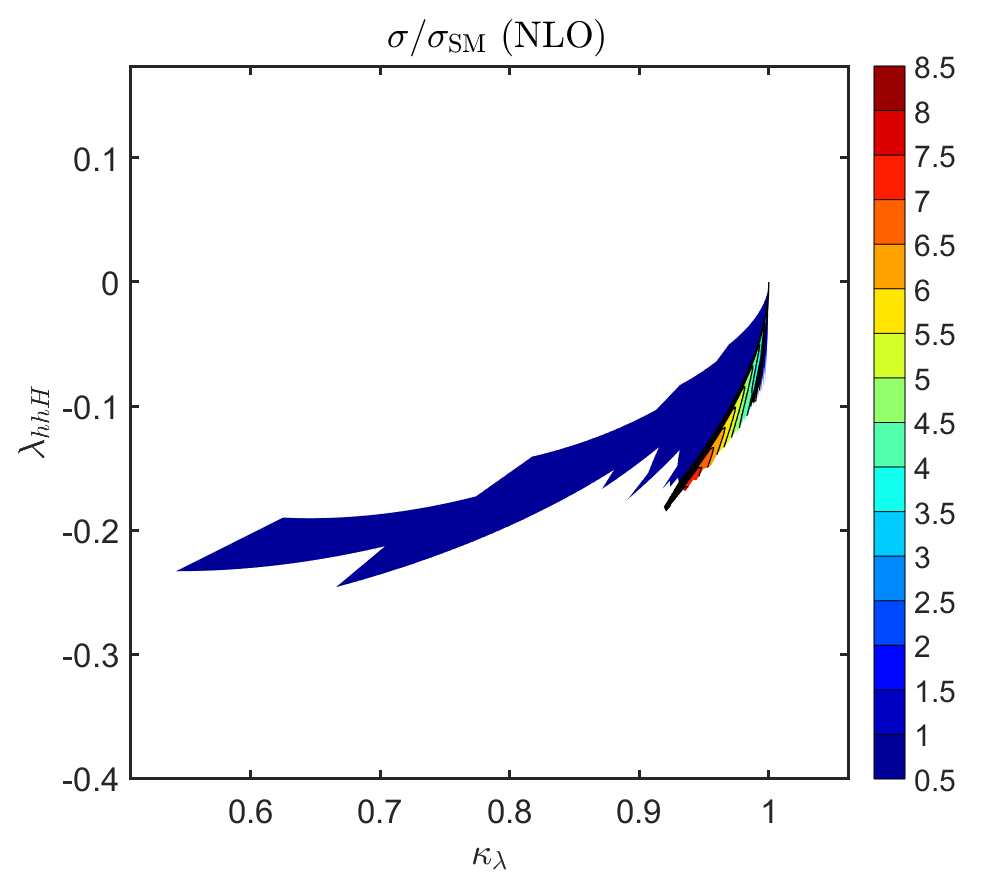}
\includegraphics[width=0.41\textwidth]{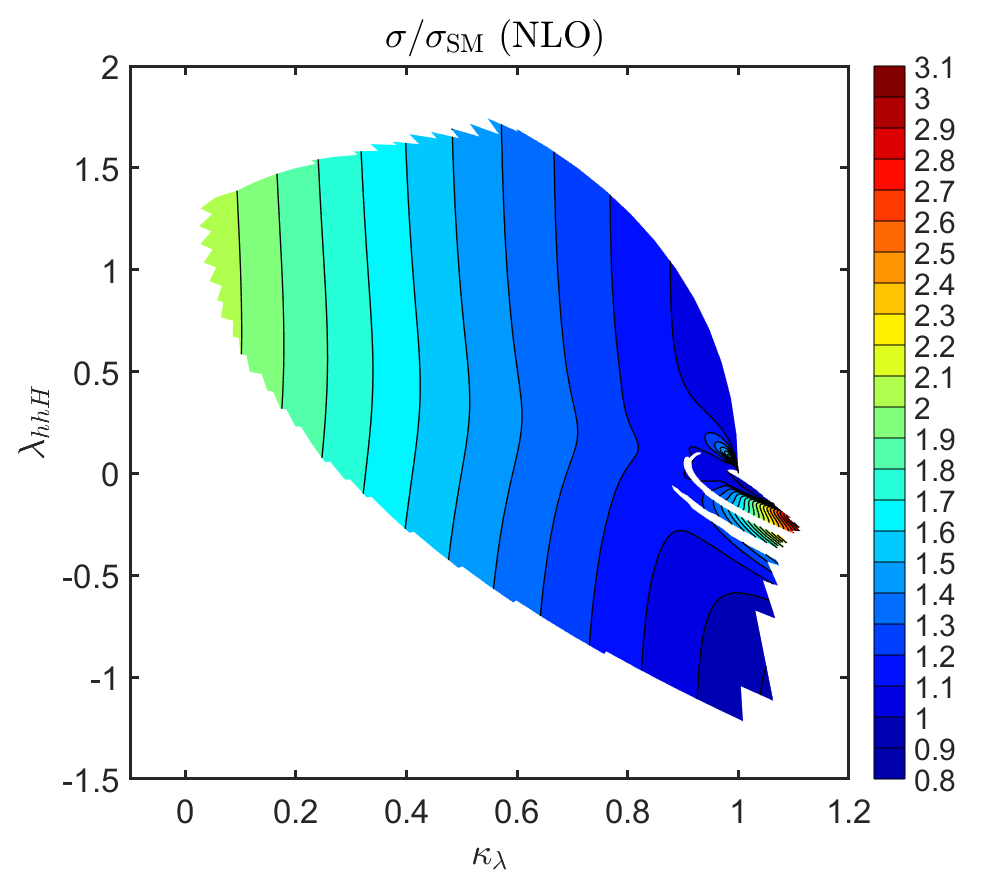}
\includegraphics[width=0.41\textwidth]{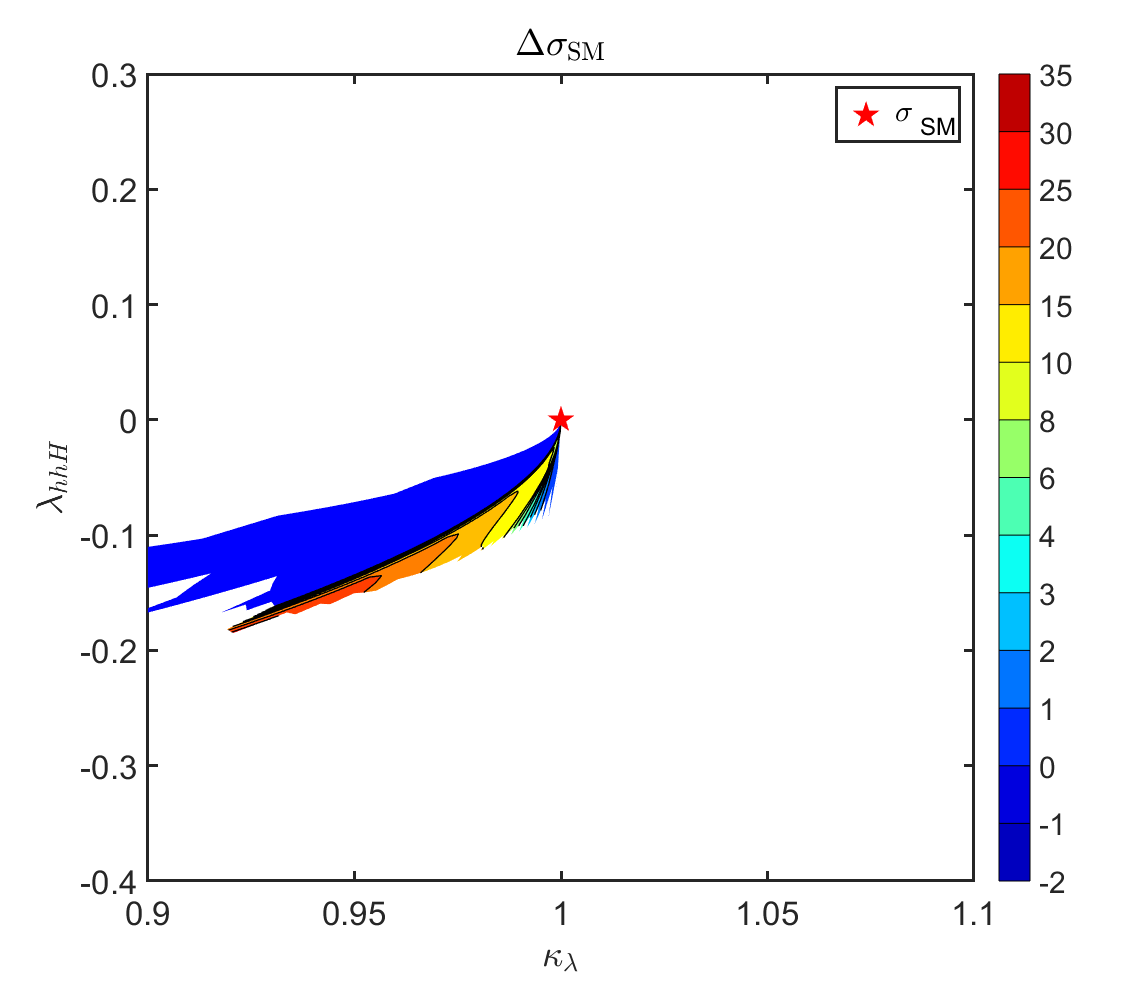}
\includegraphics[width=0.41\textwidth]{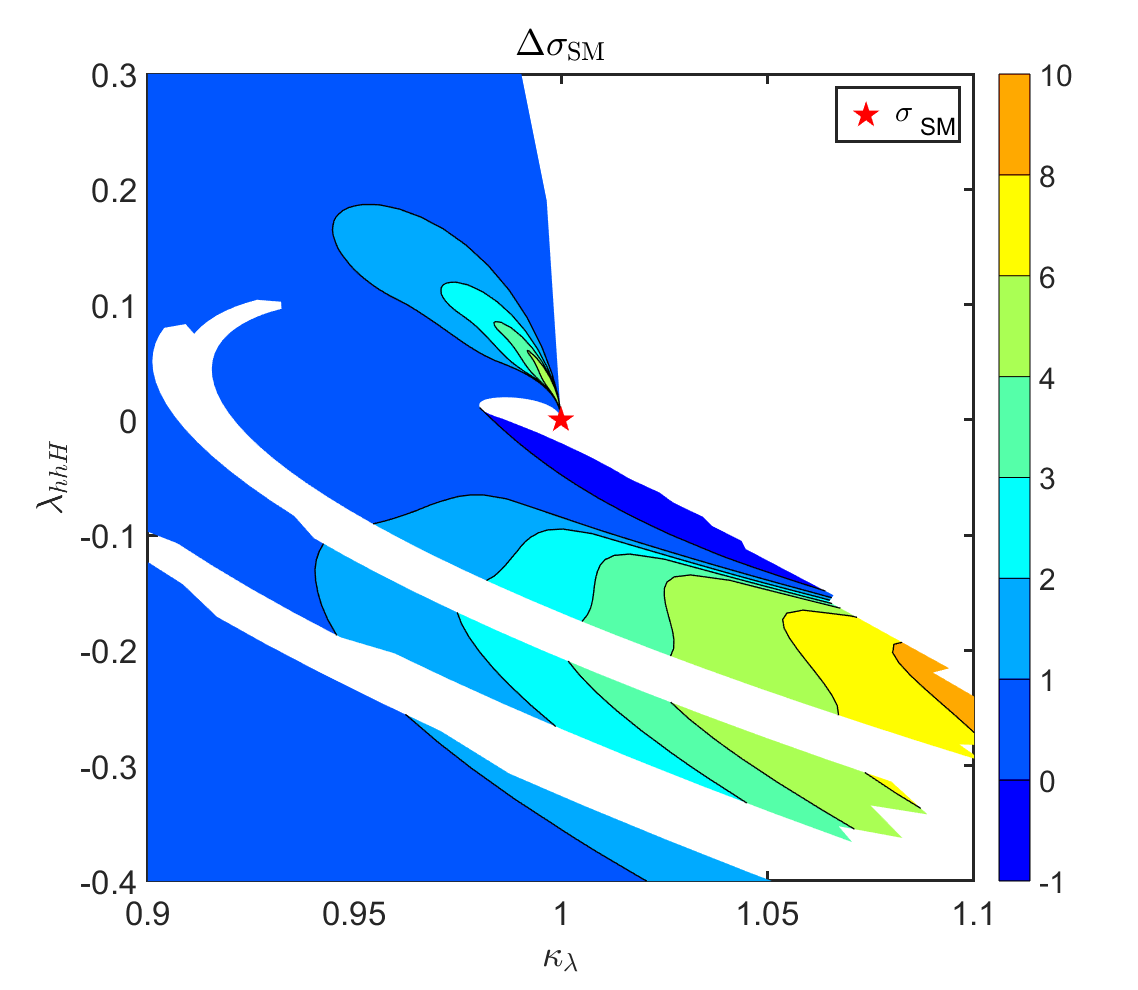}
  \end{center}
\caption{\textbf{Plane 3}. Split for negative (left) and positive
  (right) values of $\CBA$. \textit{Upper line}: Allowed
  region. \textit{Middle line.} Total di-Higgs production cross section
  at NLO QCD w.r.t the SM. \textit{Lower line.} Expected sensitivity to the
  cross section deviation (zoomed into the interesting region).}  
\vspace{-1em}
\label{fig:couplings3split}
\end{figure}

\begin{figure}[ht!]
\vspace{-1em}
  \begin{center}
\includegraphics[width=0.32\textwidth]{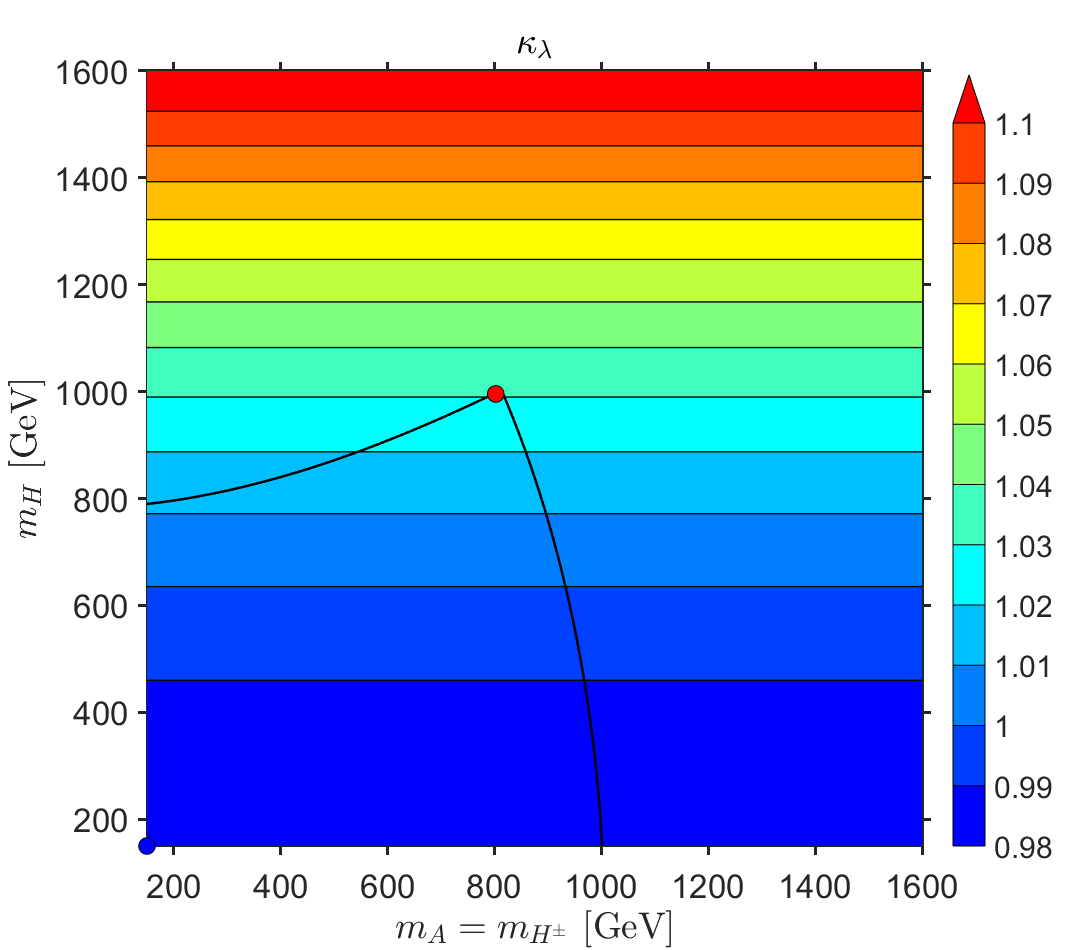}
\includegraphics[width=0.32\textwidth]{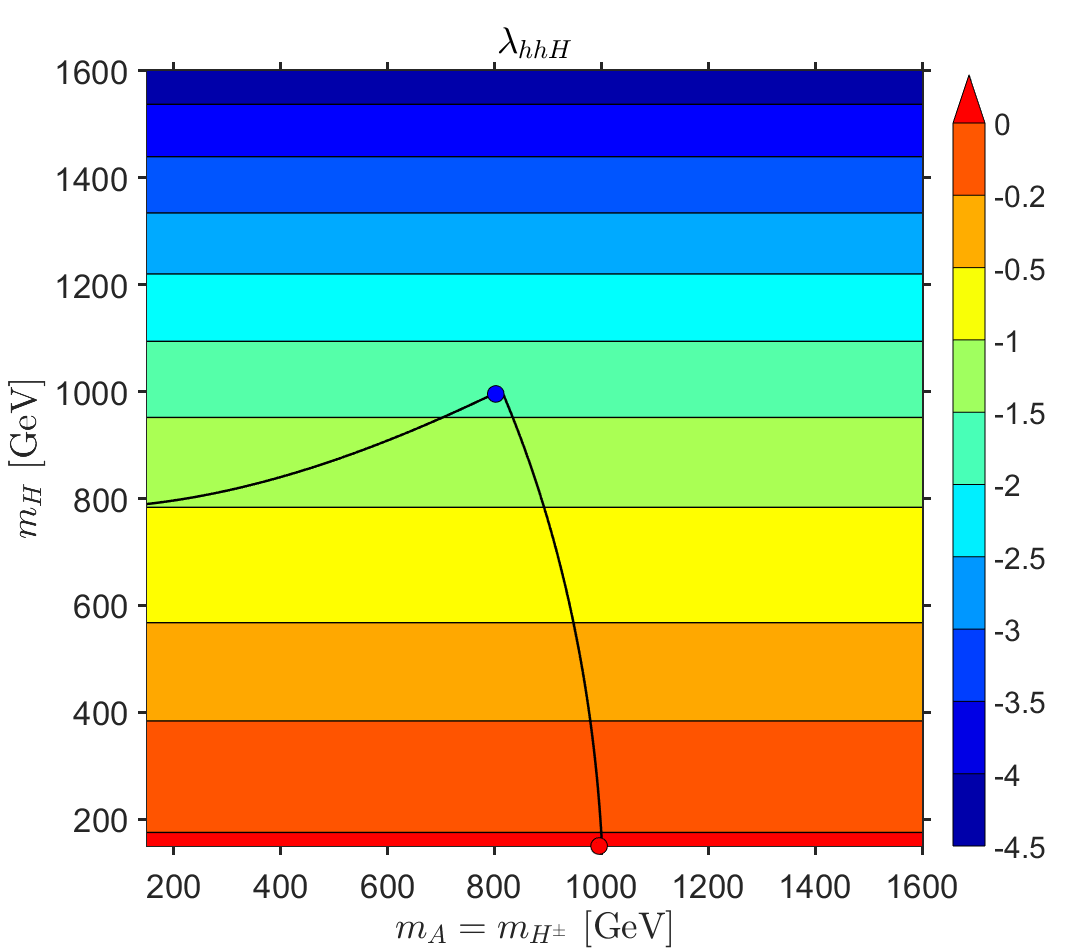}
\includegraphics[width=0.32\textwidth]{figs/sigma_8nlo.png}
\includegraphics[width=0.32\textwidth]{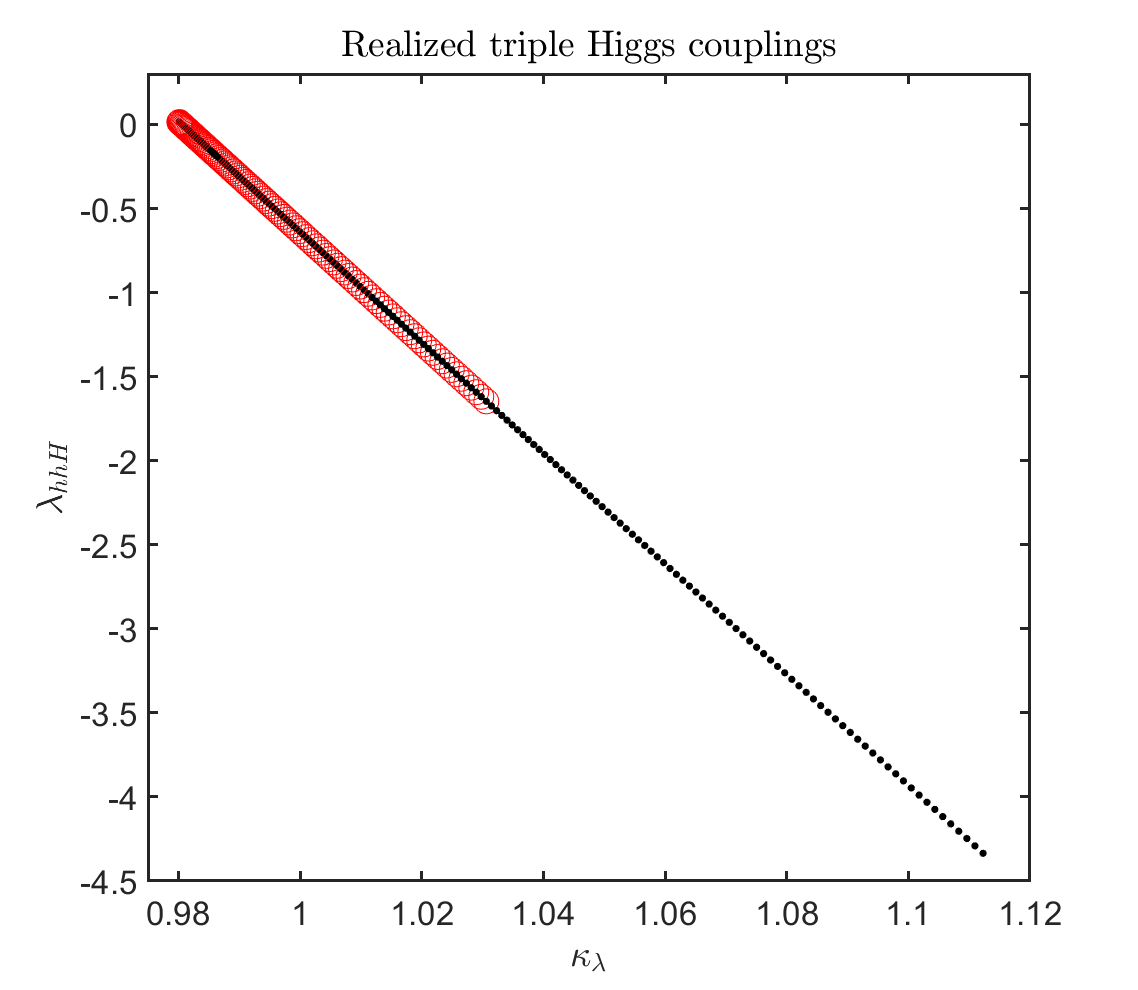}
\includegraphics[width=0.32\textwidth]{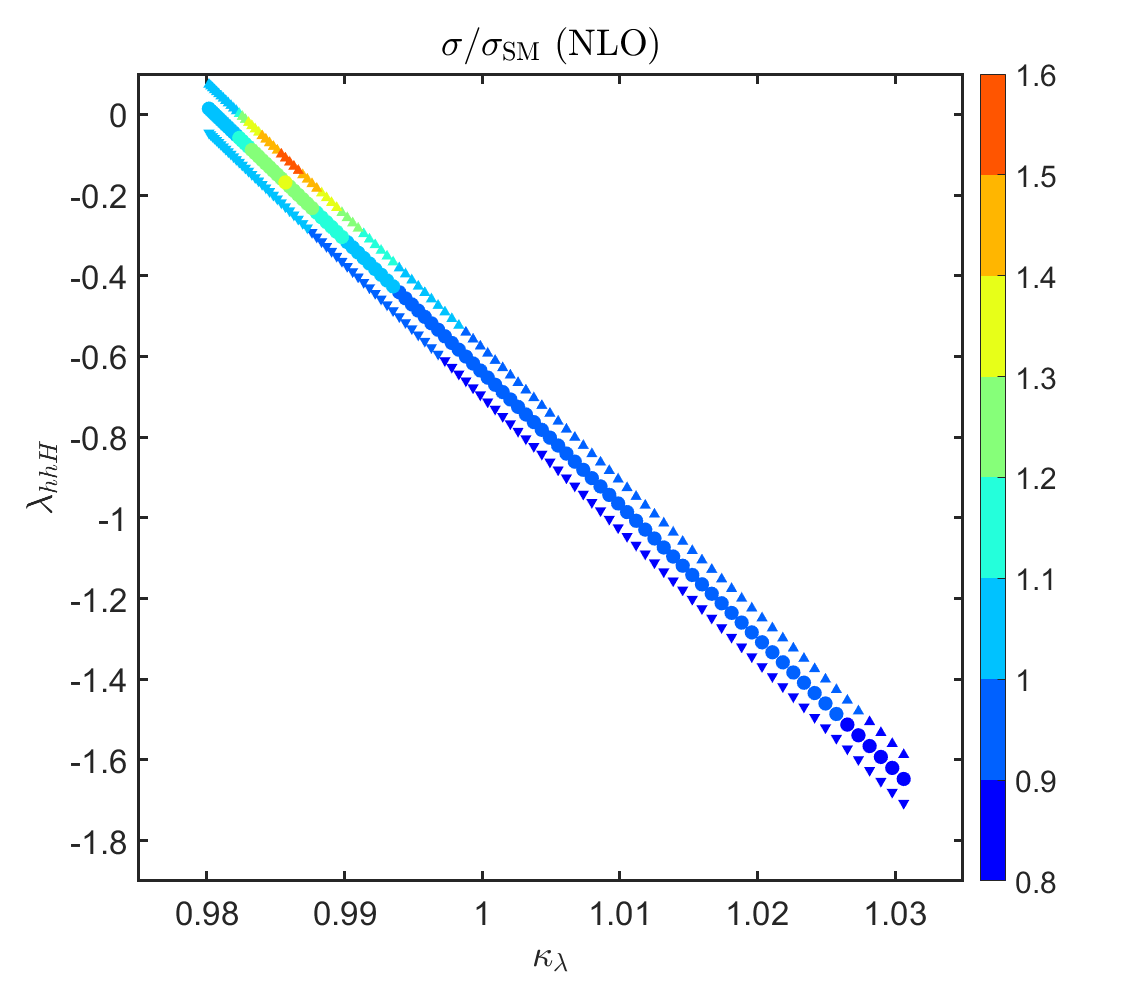}
\includegraphics[width=0.32\textwidth]{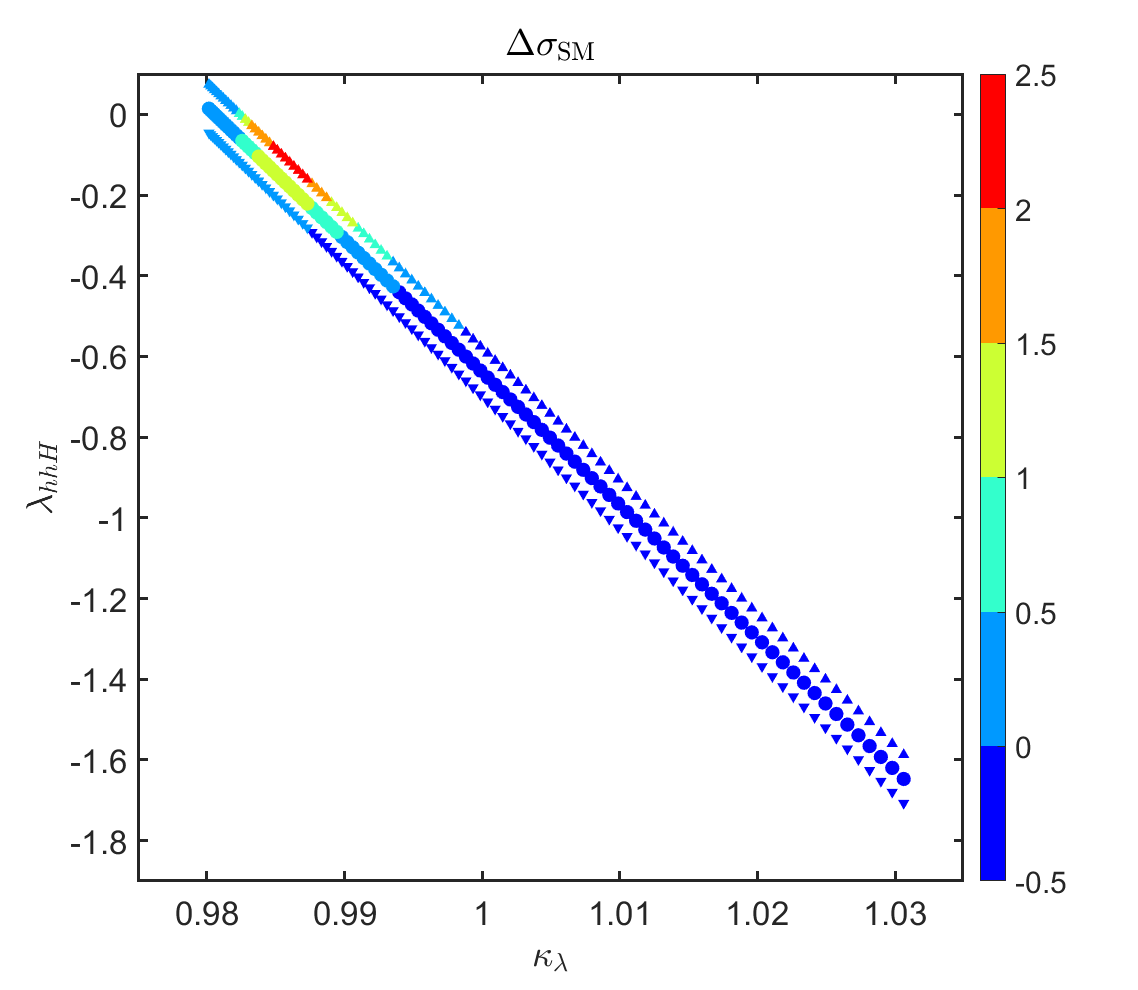}
  \end{center}
  \caption{\textbf{plane 4}. 2HDM type I, $m_H$ versus $m_A =
    m_{H^\pm}$. Otherwise plots as in \reffi{fig:couplings1}.}
\vspace{-1em}
\label{fig:couplings8}
\end{figure}

We present our results within the four benchmark planes discussed above
in \reffis{fig:couplings1} -\ref{fig:couplings8}.
For each plane in the upper left and middle plot we will show the
predictions of \kala\ and \lahhH, as obtained in
\citeres{Arco:2020ucn,Arco:2022xum}. The upper right plot, for better
comparison, repeats the results of $\sig_{\rm 2HDM}/\sig_{\rm SM}$ at
NLO QCD as presented in the upper left plots in \reffis{fig:xsec1} -
\ref{fig:xsec8}, where we here also show the maximum (minimum) value of the
coupling that is realized within the allowed region as red (blue) dots.
The three upper plots are always given in the plane 
of the two free parameters involved in the respective benchmark
choice. 
The lower left plot shows which combinations of \kala\
and \lahhH\ can be reached in each plane, where the points inside the
area allowed by theoretical and experimental constraints are marked in
red (and indicated with a red arrow). The lower middle plot, focusing on the allowed region in the
\kala-\lahhH\ plane, presents the values of $\sig_{\rm 2HDM}/\sig_{\rm SM}$
at NLO QCD in this plane (with the SM point $\kala = 1$, $\lahhH = 0$ marked by
a red star). This indicates the dependence of the total 2HDM di-Higgs
production cross section on the two involved triple Higgs couplings. 
The black points represent the values
of the THCs that are reached in this plane.
The lower right plot shows the same area in the $\kala$-\lahhH\ plane,
now indicating the expected number of $\De\sig_{\rm SM}$, see
\refeq{eq:expsensit}, that the (to be measured) 
2HDM result differs from the SM prediction, i.e.\ with which
significance such a deviation can be measured experimentally.

In benchmark \textbf{plane 1}, \reffi{fig:couplings1}, one can observe
from the comparison of the upper left and right plots that within the
allowed range, as discussed above, the smallest 
$\kala$ value gives rise to the 
largest value of $\sig_{\rm 2HDM}$.
As can be inferred from the lower middle plot, in this benchmark plane
the cross section depends 
strongly on \kala, but effectively not on \lahhH. This is due to the
fact, as discussed above, that the heavy $\cp$-even Higgs is too heavy
to give a sizable $s$-channel contribution. Overall, one can observe
that for the smallest allowed \kala\ values, $\kala \sim -0.4$, a cross
section enhancement of up to $\sim 3$ can be found. 

Finally, we see from the lower right plot that for the smallest \kala,
corresponding to the largest $\sig_{\rm 2HDM}$, a 
deviation of up to $\De\sig_{\rm SM} \sim 9$ can be expected. This
indicates that 
within this 2HDM benchmark plane a clear distinction between the 2HDM
and the SM via the di-Higgs production cross section can be possible.
Deviations of more than $2\sig$ can be
expected for $\kala \lsim 0.6$. 

In benchmark plane 2 a similar result as in plane 1 can be observed, as shown in \reffi{fig:couplings2}. The largest cross sections are found for the smallest
\kala\ values, and the predicted 2HDM di-Higgs cross section depends
only mildly on \lahhH. The latter can again be understood because of the
relatively large value of $\MH = 650 \gev$ in this benchmark
plane. The maximum significance of the 2HDM deviation w.r.t.~the SM
value is less than for plane~1 with a value of at most~$3.5\,\sig$,
reached for $\kala \sim 0.9$ and $\lahhH \sim -0.5$.

The situation is more involved in \textbf{plane 3}, which we show
in \reffi{fig:couplings3}.
As discussed in the previous subsection, very large enhancements can be
reached in this parameter plane, and larger allowed regions are found
for both signs of $\CBA$. The lower middle plot, showing the cross
section enhancement w.r.t.\ the SM seems to show a relatively small
enhancement of up to $\sim 3$. The larger effects that actually occur
(enhancements of up to $\sim 8$) are found in comparably small regions and
are thus not well visible in this figure (but will be shown clearly
below). The lower right plot shows the $\De\sig_{\rm SM}$, and for large
parts of the parameter space we find the same feature as in the two
benchmark planes above: the largest deviations are found for the
smallest \kala\ values, and independent of \lahhH, reaching
about $5\,\sig$. However, some ``overlaid'' structure is visible around
$\kala \sim 1$. Here the same combinations of \kala\ and \lahhH\ are
reached for different points in the parameter space, in particular for
both signs of $\CBA$. To analyze this scenario in more detail we split
up the plots for positive and negative $\CBA$.

In \reffi{fig:couplings3split} we present the results where we have
divided the allowed region of 
benchmark plane 3 and the obtained cross sections and their sensitivity
in the couplings plane in positive and negative values of $\CBA$.
The left (right) column in \reffi{fig:couplings3split} shows the results
for $\CBA$ negative (positive). The upper row indicates the combination
of \kala\ and \lahhH\ that can be reached in the full benchmark
plane. The middle row shows $\sig_{\rm 2HDM}/\sig_{\rm SM}$, whereas the
lower row indicates the level of $\De\sig_{\rm SM}$ that can be
reached, where we have zoomed further into the interesting region.
The middle left plot demonstrates that for \kala\ close to~1 the cross
section can be strongly enhanced with the enhancement 
strongly depending on the BSM THC \lahhH. This behavior can be traced back to the
$H$~contribution in the $s$-channel for relatively small values of
$\MH$, as discussed in the previous subsections. Looking into the
(zoomed in) result for $\De\sig_{\rm SM}$ (lower left plot) one can
observe that for the smallest values of $\lahhH \sim -0.2$ a deviation
from the SM by up to $35\,\sig$ can be expected. 
More importantly, the
size of the deviation may give an indication of the value of \lahhH.
Going to positive values of $\CBA$ as presented in the right column, one
can observe that for large parts of the paramerter space a dependence
solely on \kala\ is found, as in the previous benchmark planes. However,
again for $\kala \sim 1$ strong enhancements are found due to the
presence of the heavy $\cp$-even Higgs in the resonance. This is better
visible in the lower right plot, demonstrating that for small, but
positive values of \lahhH\ deviations of up to $6\,\sig$ can be seen,
whereas for negative values of $\lahhH \sim -0.2$ even deviations of up
to $9\,\sig$ can be found. Here it should be kept in mind that this
analysis only demonstrates the possible dependences and effects of the
THCs on the di-Higgs production cross section. As will be discussed
below, an actual possibility for a determination of \lahhH\ is not
implied, as it depends on the precise knowledge of the other (free)
parameters.

Our final analysis in this section is done for benchmark \textbf{plane 4},
as shown in \reffi{fig:couplings8}. In this $\MA = \MHp$-$\MH$ plane the
value of \kala\ is always close to~1, varying only by about
$\sim 10\%$, so that large variations of the di-Higgs production cross
section can only be produced by resonant enhancement. The coupling
\lahhH\ varies between 0 to $\sim -1.5$ 
in the allowed region (and even down to $\sim -4.5$ for the largest
$\MH$ values). The cross section, as discussed already in the previous
subsection shows an interesting enhancement of up to $\sim 60\%$, where
the heavy Higgs is resonant and not too heavy, $\MH \sim 400 \gev$.
We will use the behavior of $\sig_{\rm 2HDM}$ in this  benchmark plane
for a more detailed analysis of the invariant $m_{hh}$ distribution in the
next section.

The projection into the \kala -\lahhH\ plane shows only a line, which
can be understood as follows.
Looking at \refeq{eq:hhh_phys} and \refeq{eq:hhH_phys} and
discarding all the terms proportional to constants (the angles $\al$ and
$\be$) we find the following relations: 

\begin{equation}
    \begin{split}
    \kala &= c_1 + c_2 \bar{m}^2,\\
    \lahhH  &= c_3 + c_4 \bar{m}^2,
    \end{split}
\end{equation}

\noindent where the $c_{1,2,3,4}$ are constant terms, and it is 
taken into account that according to \refeq{eq:m12special}
$\bar{m}^2 \propto \msq \propto \MH^2$. Consequently, one finds

\begin{equation}
\lahhH = c_3 - \frac{c_1 \,c_4}{c_2} + \frac{c_4}{c_2}\kala~,
\end{equation}
resulting in the linear dependence 
that is observed in the lower plots of \reffi{fig:couplings8}.

Now we proceed to analyze the values of the cross section that are
possible for the different values of the THCs, even though we
do not have a truly 2-dimensional plot in these cases. For the benchmark
\textbf{plane 4}, in the lower middle plot of
\reffi{fig:couplings8} the cross section is badly defined in the sense
that more than one value of the cross section corresponds to a
particular value of the THCs. This happens when we allow for a
change in the masses but fix the angles, as discussed above.
The THCs change in a
coherent way for different masses $\MH$ (see upper left and
middle plots in \reffi{fig:couplings8}), while the cross
section has different possible values. As an example, for $\MH$
in the range $\sim 220 \gev$ to $\sim 800 \gev$ it can vary 
within $\sim (0.8-1.6) \times \sig_{\rm SM}$,
as can be seen in the upper right  
plot. Therefore, in the lower middle plot we represented the mean
value of the cross section as a circle for a particular combination of
($\kala, \lahhH$).  We show maximum (upper triangle ``slightly
displaced above the circles'') and minimum (lower triangle
``slightly displaced below the circles'') values of the normalized cross
section for this same 
combination of ($\kala, \lahhH$). One can observe that 
the highest cross section is realized for a $\kala \sim 0.985$ and
$\lahhH \sim -0.2$. In this point the value of the cross section varies
roughly from 1 to 1.6 and the sensitivity that can be reached in the
most optimistic scenario (i.e.\ the largest deviation from the SM that
is realized) is almost 2.5 $\sig$, as can be seen in the lower
right plot. This enhancement is relatively small, but it demonstrates 
that in this plane the relevant role of the coupling $\lahhH$ is more significant
than $\kala$, which is very close to~1, where these effects are found.
The size of the deviation clearly depends on \lahhH\ in this case.

\clearpage
\newpage

\section{Analysis of \boldmath{\mhh}}
\label{sec:mhh}

In the next step of our analysis we will analyze 
the influence of the THCs on the di-Higgs production cross sections
by evaluating the di-Higgs invariant mass distribution, \mhh. 
We first demonstrate in a toy example the effect of the characteristic
properties of the resonant Higgs boson: its mass, its width, and the sign
of the coupling combination $(\lahhH \times \xi_H^t)$. Subsequently, 
the analysis will be performed for several benchmark points located in
the planes discussed in the previous section, where the effects of the
characteristics of the resonance will be demonstrated in a real model. 

The invariant mass distributions, $d\sig/d\mhh$, are also obtained with the code
\texttt{HPAIR}. We will use a grid of values for the invariant mass $\mhh$ that
range from $250 \gev$ to $1250 \gev$. As a default value we will use a
bin size of $20 \gev$ (where experimentally a bin size of $\sim 50 \gev$
appears more realistic, see the discussion below). This bin size is used
for demonstrative purposes. In a later step we will analyze the effect
of different bin sizes and other experimental effects to obtain a more
realistic picture of \mhh\ distributions. 


\subsection{General analysis of the effects}
\label{sec:toy}

In this subsection we will analyze a toy model for the resonance to
demonstrate the effects of the mass, width and couplings of the resonant
Higgs boson in the $s$-channel exchange.

\begin{figure}[ht!]
  \begin{center}
\includegraphics[width=0.44\textwidth]{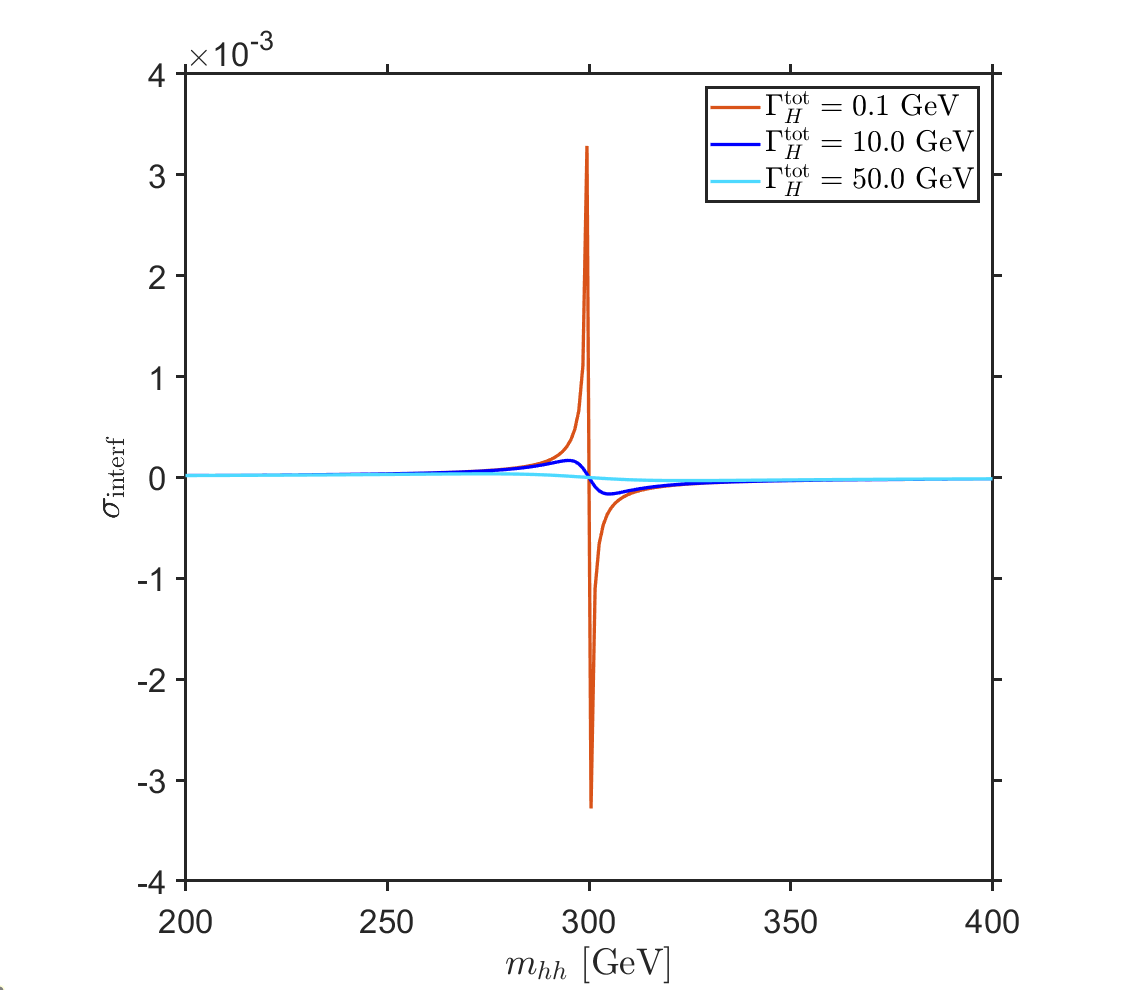}
\includegraphics[width=0.44\textwidth]{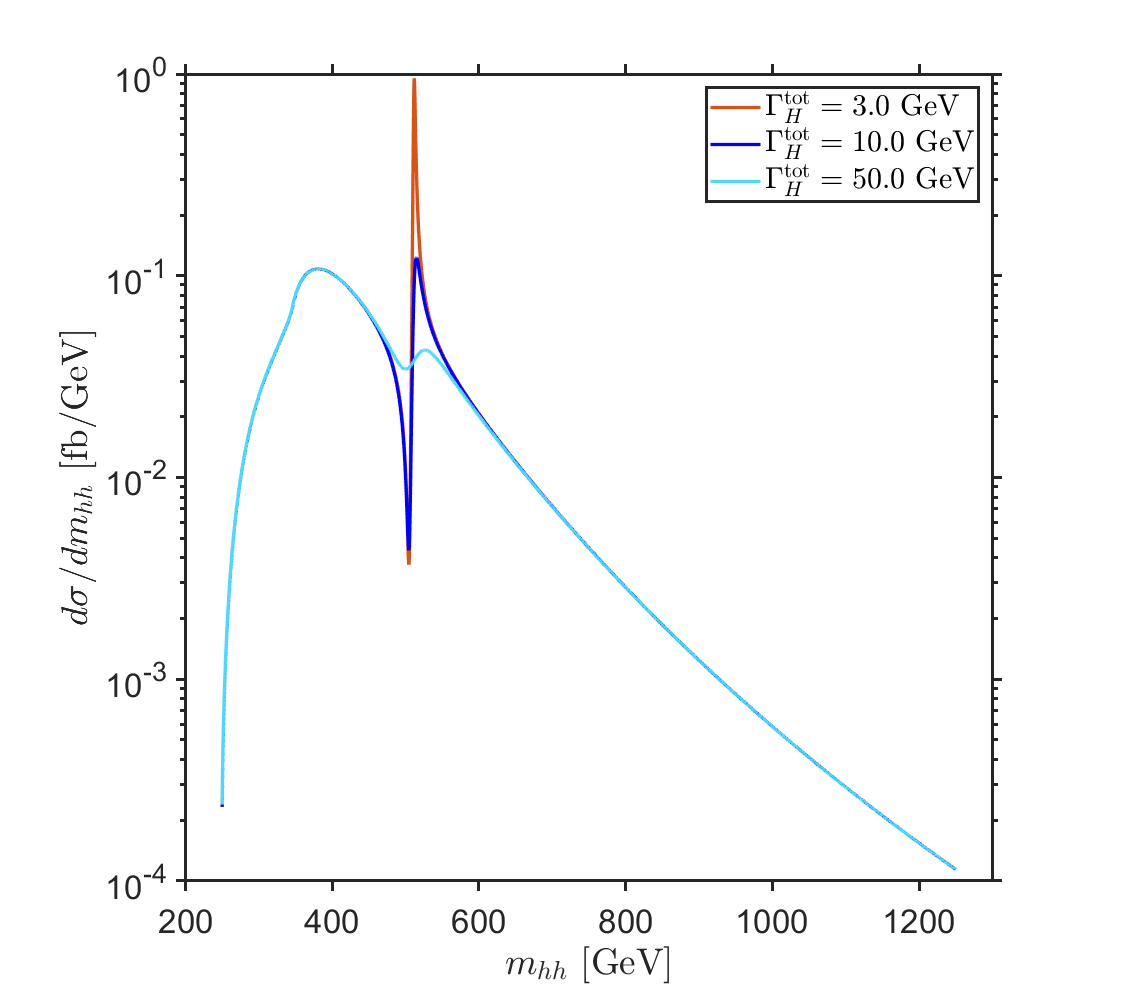}
  \end{center}
\caption{  \textit{Left:} Partial result $\sigma_{\rm interf}$
  \refeq{eq:sigmainterf} for 
  three different decay widths. \textit{Right:} Effect of \GaHtot\ on the 
  the invariant mass distribution for one benchmark point and three
  values of \GaHtot\ (see text).}
\vspace{-1em}
\label{fig:toydecays}
\end{figure}

The effect of the total decay width of the heavy Higgs boson,
$\Ga_H^{\rm tot}$, is important whenever the resonant diagram gains
significance in 
the calculation of the cross section. This happens close to the
resonance at $\MH \sim \mhh$, as discussed in the previous section.
For a correct treatment close to the resonance the total width has to be
included into the propagator, 
\begin{equation}
    \frac{1}{Q^2-\MH^2} \rightarrow \frac{1}{Q^2-\MH^2 + i\MH\Ga_H^{\rm tot}}.
    \label{eq:Breit-Wigner}
\end{equation}
From this expression one can clearly see that the dominant effect of
\GaHtot\ appears when the intermediate Higgs boson mass is equal to the
(reduced) center of mass energy $Q^2$.
In the Higgs pair production process, the total decay width of the heavy
Higgs becomes relevant near the resonant region where
the behavior of the cross section can be dominated by the interference
between the resonant and the non-resonant contributions, which is
proportional to
\begin{equation}
  {\sigma_{\rm interf} =} \frac{Q^2-\MH^2}
        {(Q^2-\MH^2)^2+\MH^2{\Ga^{\rm tot}_H}^2}\,.
    \label{eq:sigmainterf}
\end{equation}
We use this expression to investigate the resonant behavior of the
\mhh\ distribution. In \reffi{fig:toydecays} (left) we show
$\sig_{\rm interf}$ as a function
of $\mhh = Q$. We have chosen $\MH = 300 \gev$ and three exemplary
values of \GaHtot: $0.1 \gev$ (red), $10 \gev$ (dark blue) and $50 \gev$
(light blue). In all cases $\sig_{\rm interf}$ shows a peak-dip
structure, with the change exactly at $\mhh = \MH$, as expected from
\refeq{eq:sigmainterf}. Furthermore, one observes that the highest
(smallest) peak-dip structure is obtained for the smallest (highest)
value of \GaHtot, following the analytical behavior of
\refeq{eq:sigmainterf}. We  
furthermore observe that the ``total width of the effect'', given by the
width of the peak at half of its maximum value, increases with
increasing \GaHtot, as expected.

The features observed for $\sig_{\rm interf}$ are
also found in the calculation of the \mhh\ distribution of the
complete cross section, i.e.\ the
result from taking into account the complete resonant and the non-resonant
contributions, as shown in
the right plot of \reffi{fig:toydecays}. Here we depict $d\sig/d\mhh$ as a
function of \mhh\ for one benchmark point of benchmark \textbf{plane 4}
with $\MA = \MHp = 544.72 \gev$ and $\MH = 515.5 \gev$%
\footnote{This corresponds to the green
point of the model based analysis for plane 4 in the
\refse{sec:mhh-8}.}.%
~For the total
width of $H$ we find $\GaHtot \sim 3 \gev$, resulting in the red
curve. In order to illustrate the effects of the size of \GaHtot, as
seen in the left plot, we also show the results for ad-hoc set
values of $\GaHtot = 10 \gev$ (dark blue) and $50 \gev$ (light blue).
The main features of \GaHtot\ (height of the peak-dip
structure and the ``width of the effect'') are found in the full
calculation exactly as in $\sigma_{\rm interf}$.

\begin{figure}[ht!]
  \begin{center}
\includegraphics[width=0.41\textwidth]{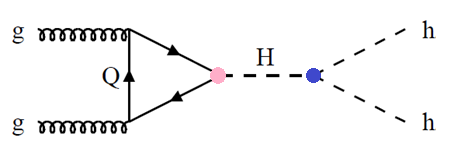}
\includegraphics[width=0.41\textwidth]{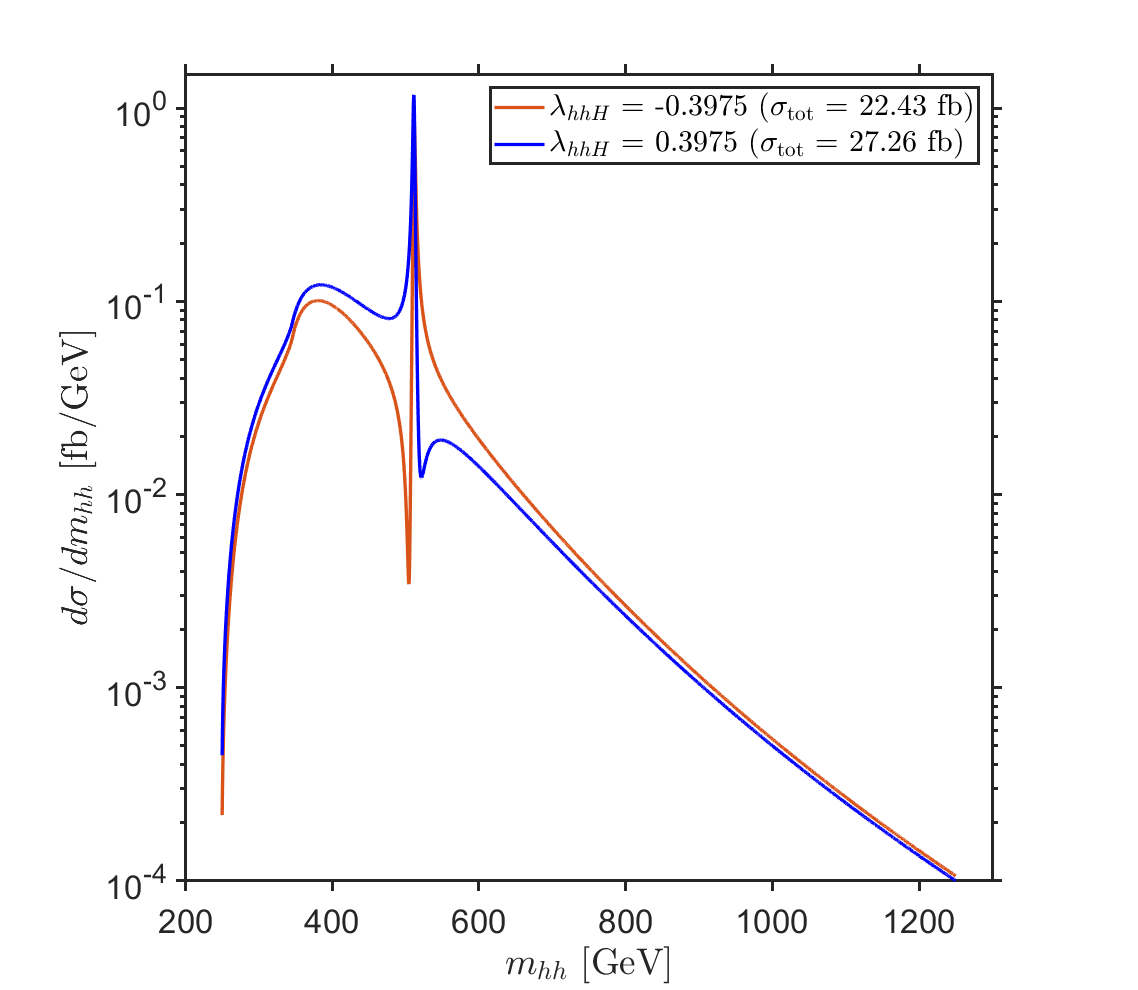}
  \end{center}
\caption{\textit{Left:} Couplings involved in the resonant diagram (top
  Yukawa, \textcolor{magenta}{$\xi_H^t$}, and BSM triple Higgs coupling,
  \textcolor{blue}{$\lahhH$}). \textit{Right:} Invariant mass
  distributions for different signs of ($\lahhH \times \xi_H^t$).}   
\vspace{-1em}
\label{fig:couplingsres}
\end{figure}

However, there is one important difference between the results for 
$\sigma_{\rm interf}$ 
and the full invariant mass distribution results which can be observed in
\reffi{fig:toydecays}. While for $\sigma_{\rm interf}$ a peak-dip
structure is found, in the full calculation for our chosen benchmark point
a dip-peak structure can be
observed. This difference can be traced back to the sign of
($\lahhH \times \xi_H^t$), which enters as prefactor in the
the resonant diagram.
In the left plot of \reffi{fig:couplingsres} the resonant $H$~diagram is
shown with the two coupling factors entering the amplitude:
the top-Yukawa coupling modification factor $\xi_H^t$ of the heavy
Higgs $H$ and the THC
\lahhH. The right plot in \reffi{fig:couplingsres} demonstrates the
effect of sign($\lahhH \times \xi_H^t$). The red curve is identical to
the red curve in the right plot of \reffi{fig:toydecays}, as obtained
for the value of $\lahhH = -0.3975$, with $\lahhH \times \xi_H^t < 0$.
The blue curve shows the \mhh\ distribution for the (ad hoc)
flipped sign, i.e.\ with $\lahhH = +0.3975$ (normalized to the
corresponding value of the full cross section obtained for this
trilinear Higgs coupling). As can be expected, the 
flip of the sign($\lahhH \times \xi_H^t$) also flips the dip-peak
structure to a peak-dip structure ($\sigma_{\rm interf}$ shown
above corresponds to 
a positive sign). The question arises whether an experimental analysis
can be sensitive to the difference between peak-dip and dip-peak, and
thus provide a handle on the sign of ($\lahhH \times \xi_H^t$). This question
will be analyzed below.

\subsection{Model based analysis: benchmark plane 3}
\label{sec:mhh-3}

We start our model based \mhh\ analysis for several points given in the
benchmark 
\textbf{plane 3}. First we will investigate the points that present the
largest enhancement of the cross section w.r.t.\ the SM within the
allowed region, as listed in \refta{tab:mhhBP3}.
Second we will look at points with $\CBA \sim 0.1$, as listed in
\refta{tab:mhhBP3pointscba01}. Finally we will analyze points 
with $\CBA$ $\sim 0.2$, i.e.\ a relatively large deviation from the
alignment limit, as listed in \refta{tab:mhhBP3pointscba02}. The aim of
this study is to extract the general behavior and the influence of specific
parameters on the experimental measurement of the cross section. 
This will allow us to track variations of the parameters that
we are mostly interested in ($\MH,\; \Ga_{H}^{\rm tot}$ and $\lahhH$).


\subsubsection{Benchmark plane 3: large di-Higgs production cross sections}
\label{sec:mhh-plane3}

\begin{figure}[ht!]
\vspace{-1em}
  \begin{center}
\includegraphics[width=0.49\textwidth]{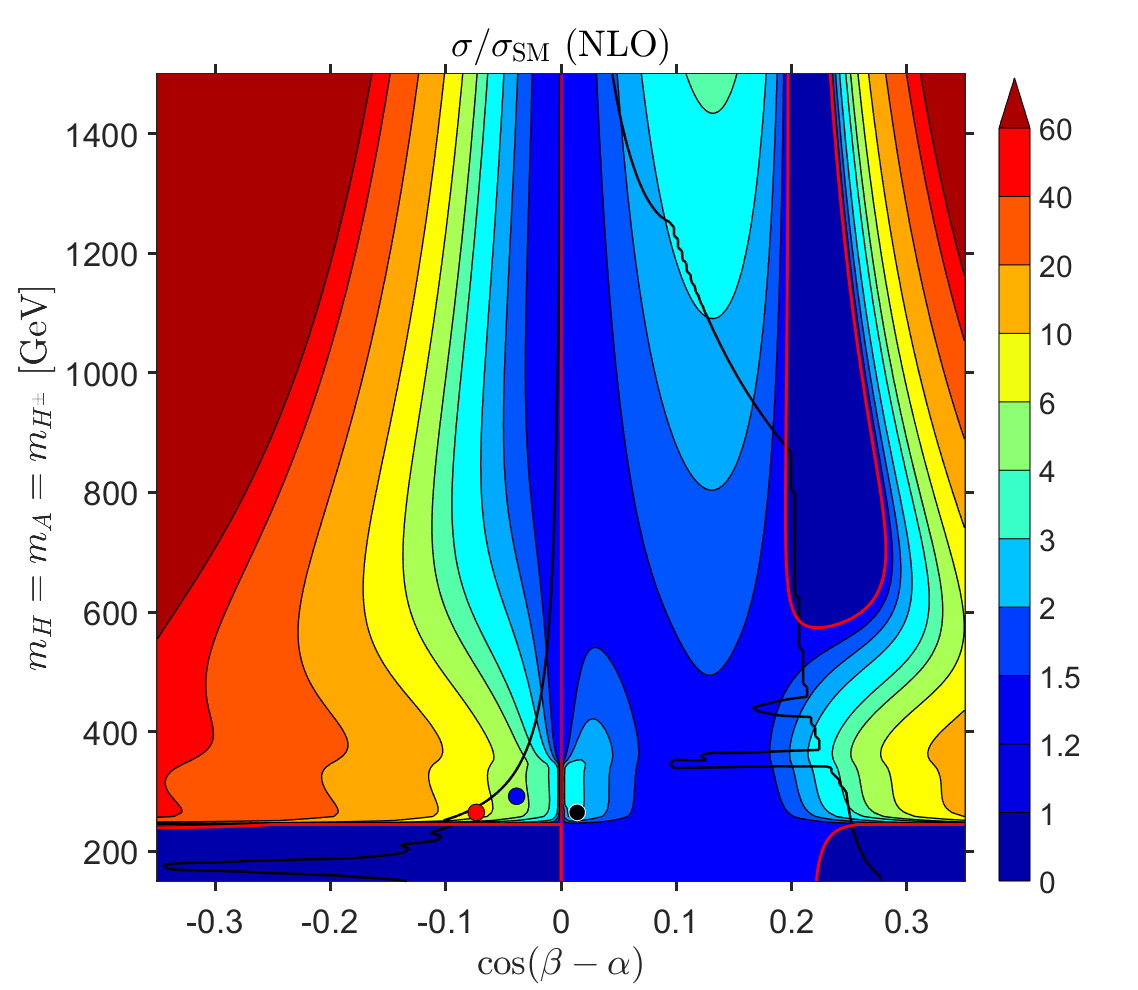}
\includegraphics[width=0.49\textwidth]{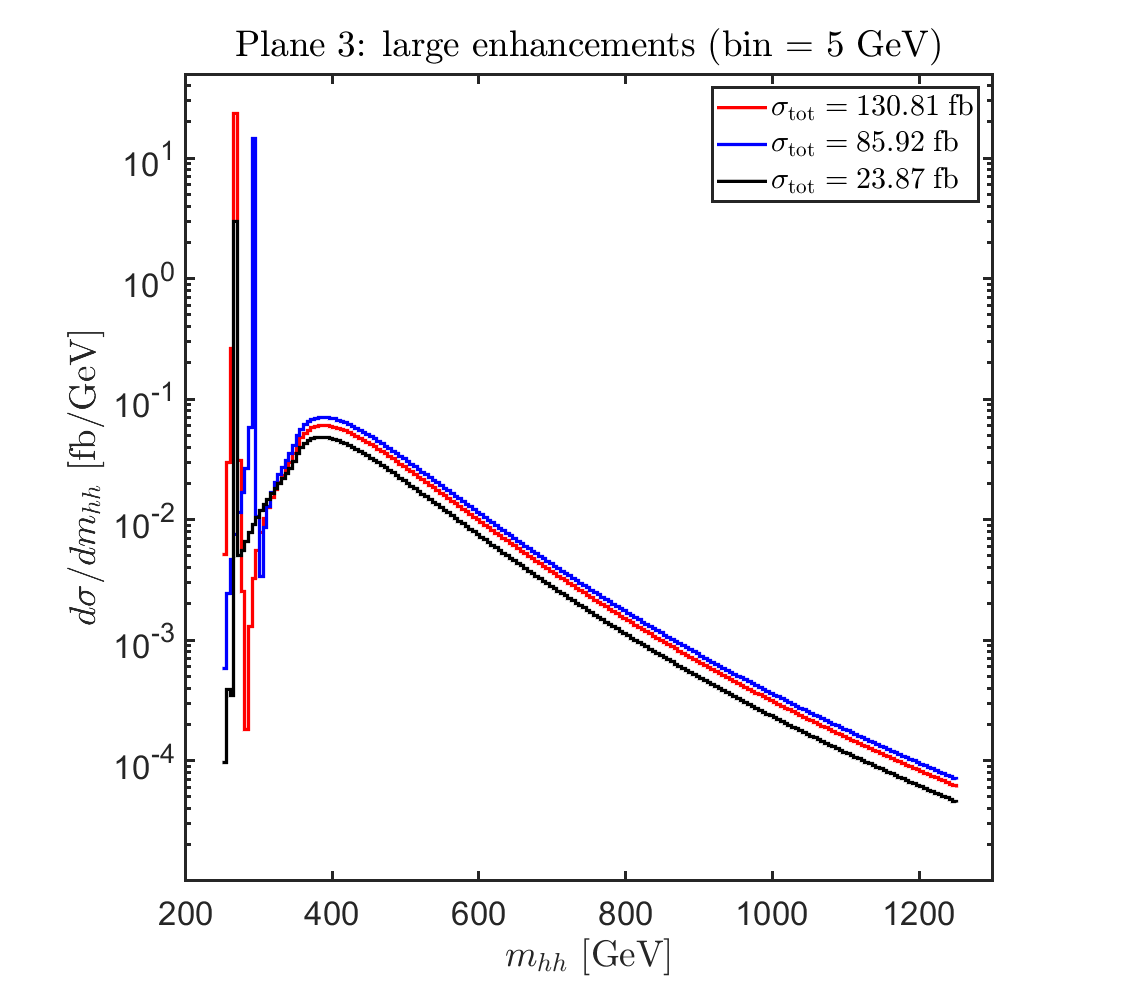}
  \end{center}
\caption{Analysis of points with large di-Higgs production cross
  sections in the   benchmark \textbf{plane 3}.
  \textit{Left.} Location of the benchmark
  points in the total production cross section plot. \textit{Right.}
  Invariant mass distribution for selected points (with a bin size of
  $5 \gev$). The colors of the points in the two figures are matching. The values of $\sig_{\rm tot}$ in the legend indicate the LO inclusive cross section prediction for each point.}  
\vspace{-1em}
\label{fig:mhh3}
\end{figure}

We first analyze three points in benchmark plane 3 with large
enhancements of the di-Higgs production cross section w.r.t.\ the SM.
They are located close to the alignment limit and can be seen in the left
part of \reffi{fig:mhh3} as red, blue and black dots.
In the first step of the analyses we choose a bin size of $5 \gev$ to
make the large resonant enhancement, which is very narrow, clearly visible.
The values of the parameters of each point are listed in
\refta{tab:mhhBP3}. 

\begin{table}[ht!]
\begin{center}
\begin{tabular}{|l||c|r|r|c|c|c|c|l|}
\hline
      & $\tb$ & $m^{2}_{12}$ [GeV$^2$]     &$\CBA$~~ & $\MH$ [GeV] & $\Ga^{\rm tot}_{H}$ [GeV] & $\kala$ & $\lahhH$ \\
      \hline\hline
red   & 10           & 6801.00                &-0.0735             & 264.75        & 0.09443   &  0.9475     &  -0.1505        \\
\hline
blue  & 10           & 8350.32                &-0.0385             & 291.75        & 0.04292   &   0.9838 & -0.0865           \\
\hline
black & 10           & 6957.94                &0.0140              & 264.75        & 0.02108   &    0.9986   &   0.0190  \\ \hline
\end{tabular}
\end{center}
\caption{Selected points in benchmark \textbf{plane 3} with large
  di-Higgs production cross section.}
\label{tab:mhhBP3}
\end{table}

The di-Higgs production process is kinematically forbidden for
$\mhh < 250 \gev = 2 \,\Mh$.
Once this threshold is surpassed, one can observe a resonant
enhancement for $\mhh \sim \MH$. This is clearly seen at the
location of the resonant peaks in the invariant mass
distribution in \reffi{fig:mhh3} (right). For the red and the black
points the resonant peak is found around $\sim 265 \gev$, while for the
blue point it is located at $\sim 292 \gev$, corresponding to the
respective $\MH$ value.
The ``height'' and ``width'' of the peaks is related to
the total decay width \GaHtot\ of the heavy Higgs boson,
which is largest for the red point  ($\sig/\sig_{\rm SM} \sim 8$) and
smallest for the black point 
($\sig/\sig_{\rm SM} \sim 2.5$), as shown in
\refta{tab:mhhBP3}. Furthermore, the resonant heavy Higgs 
contribution yields the already observed typical
pattern, a peak-dip or dip-peak structure, depending on the
parameter point. The peak-dip
structure is observed in the blue and red points, whereas in the case of
the black point one can only see the peak, see the discussion below. 
Moreover, we observe for all three points an enhancement at an
invariant mass of $350-400 \gev$, which is  
related to the top pair production threshold
in the resonant diagram%
\footnote{The mass of the
  top quark used in the calculation is $m_t = 173.2
  \gev$, therefore $2m_t = 346.4 \gev$.}. 

The three points have different values of $\CBA$, which 
change the Yukawa coupling of the top quark according to the expression:
\begin{equation}
    \xi_H^t = \CBA -\SBA\cot\be,
    \label{eq:topyukawa}
\end{equation}
resulting in a positive (negative) sign of $(\lahhH \times \xi_H^t)$
for the red and blue (black) points, as shown in \refta{tab:coupling}. 
This in turn results in a peak-dip (dip-peak) structure, cf.~\refta{tab:yukawas}. The
pattern of the differential distribution changes according to the sign of
$(\lahhH \times \xi_H^t)$. For the black curve only a
peak is visible, because the dip before it 
cannot be produced for masses below $250 \gev$. 

\begin{table}[ht!]
\begin{center}
    \begin{tabular}{|l||c|c|c|c|}
    \hline
     & $\lahhH$ & $\xi_H^t$  & sign & structure \\ \hline\hline
      red   &  -0.1505 & -0.1738  & +    & peak-dip\\ \hline
      blue  &  -0.0865 & -0.1384  & +    & peak-dip\\ \hline
      black &   0.0190 & -0.0810  & -    & dip-peak\\ \hline
     \end{tabular}
     \caption{Structure of the resonance.}
     \label{tab:yukawas}
\end{center}
\end{table}


\begin{figure}[ht!]
\vspace{-1em}
  \begin{center}
\includegraphics[width=0.41\textwidth]{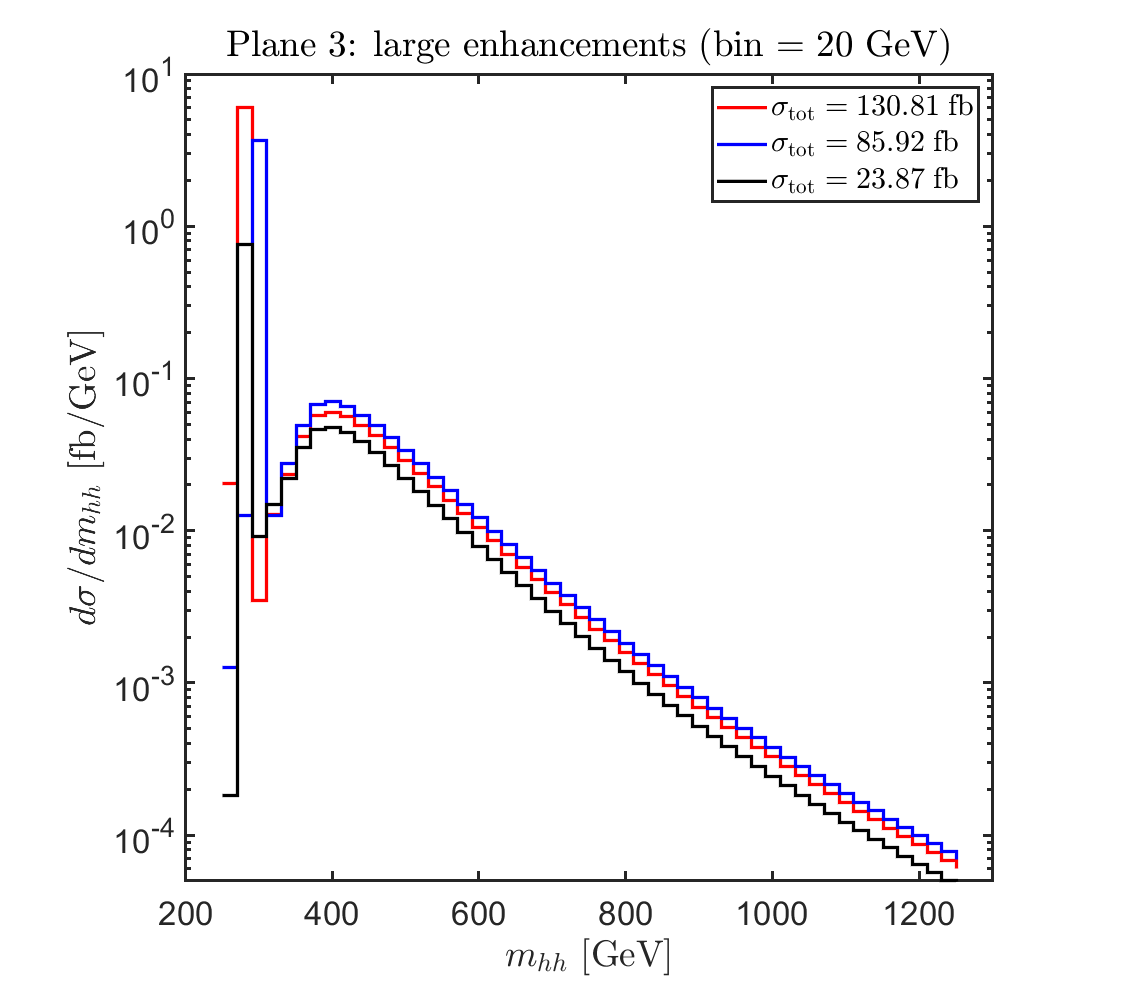}
\includegraphics[width=0.41\textwidth]{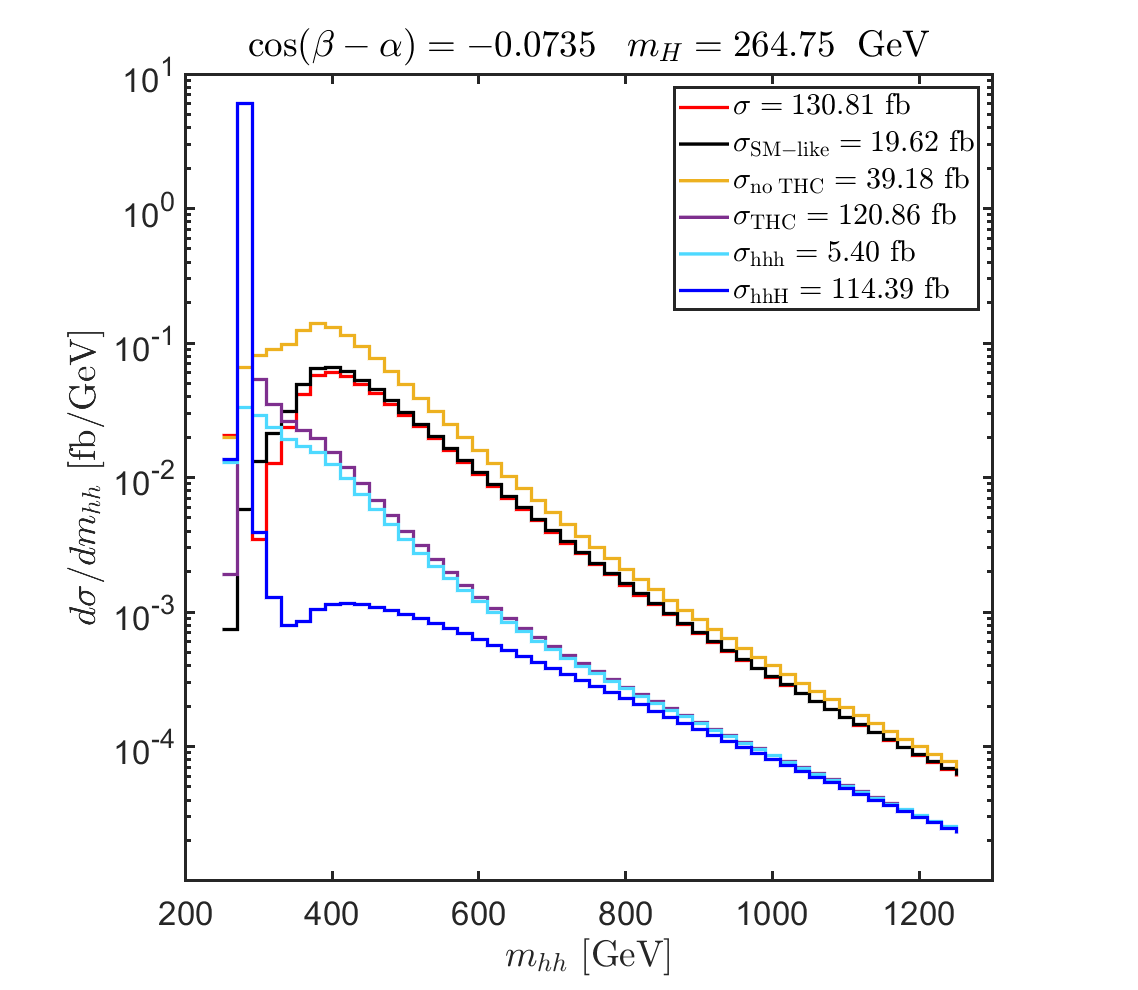}
\includegraphics[width=0.41\textwidth]{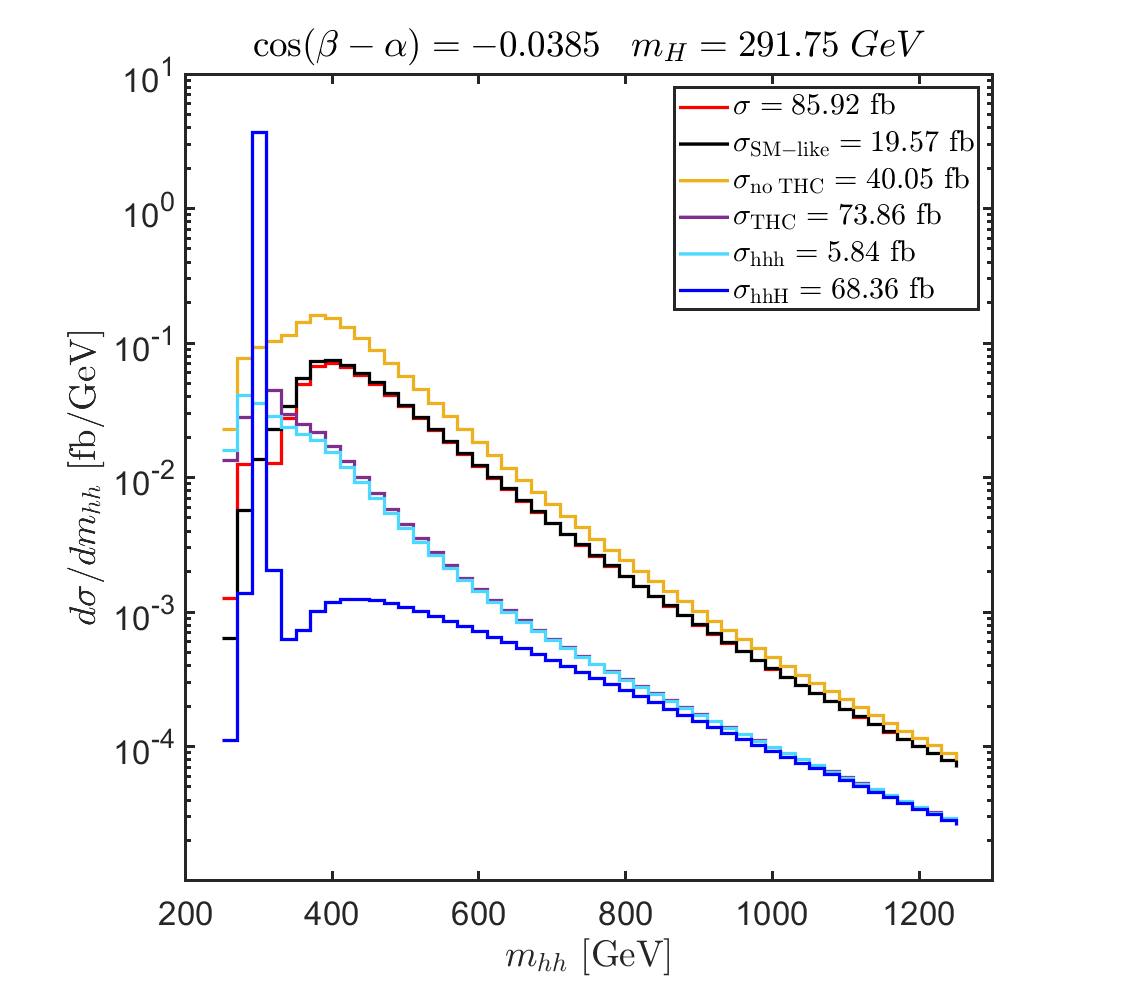}
\includegraphics[width=0.41\textwidth]{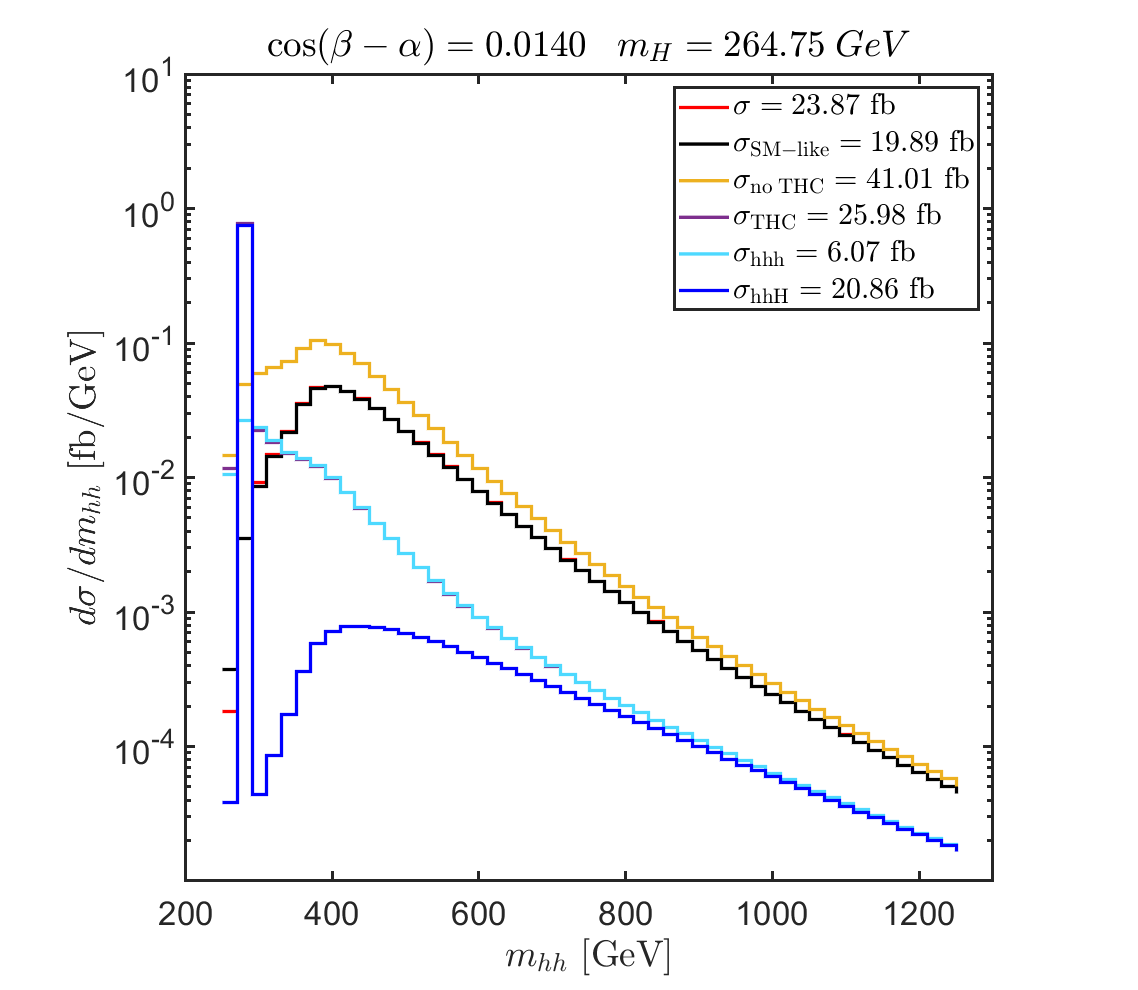}
  \end{center}
\caption{\mhh\ distributions for selected points in
  \textbf{plane 3}. \textit{Upper lef}t: Total \mhh\ distributions for the points
of \refta{tab:mhhBP3}; \textit{upper right} (red), \textit{lower left} (blue), \textit{lower
right} (black): Individual contributions to the 
\mhh\ distribution for the three points: total cross section (red), 
continuum cross section (black), \mhh\ involving only \lahhh\ (\lahhH) in
light (dark) blue, \mhh\ involving both (no) THCs in purple (yellow).}
\vspace{-1em}
\label{fig:mhh3largepoints}
\end{figure}

A more detailed analysis of the three points is presented in
\reffi{fig:mhh3largepoints}, where we analyze the contribution of
individual (groups of) diagrams.
In the upper left plot we show
the total, i.e.~including all diagrams, distribution for the three
points (as in the right plot 
of \reffi{fig:mhh3}) but changing the bin size to 20 GeV in order to represent a 
more realistic experimental set up.
One can observe already in the example of the black point that the bin size
(and location) is important for the observation or non-observation of an
enhancement.

In the upper right, and lower row of \reffi{fig:mhh3largepoints} we
disentangle the contributions of the individual diagrams for the red,
black and blue points, respectively. We have calculated the
total differential distribution including all three diagrams for
di-Higgs production (red), the SM-like cross section, called continuum, -
including only the box and the SM-like Higgs boson in the 
$s$-channel diagrams - (black), the contribution of the diagram with no
triple Higgs couplings involved, i.e.\ the box diagram (yellow), the
contribution of the two diagrams with the triple Higgs couplings,
which includes the $h$ and 
$H$ in the $s$-channel (purple), and the contribution with only $h$
(light blue) and with only $H$ (dark blue), respectively, in the ${s{\rm- channel}}$.  
For all three points the same pattern can be observed.
The 2HDM cross section (red) follows largely the SM-like distribution
(black), apart from the strong resonant enhancement at $\mhh \sim \MH$.
This is caused by the $H$ $s$-channel contribution (blue), potentially
providing sensitivity to \lahhH. Furthermore, the destructive
interference of the box-diagram (yellow) and the SM-like $h$-exchange
contribution (light blue) is clearly visible in the
continuum, i.e. SM-like, contribution of the distribution (black).


\subsubsection{Plane 3: $\CBA \sim$ 0.1}

Next we proceed to study a set of points that are all located at the
same value of $\CBA \sim 0.1$. The exact value
of $\CBA = 0.1015$ results from the grid used to scan this plane.
In this case the only change
between the benchmark points, as listed in
\refta{tab:mhhBP3pointscba01}, is the common mass $m_H$ of the heavy Higgses,
with correspondingly modified couplings and decay widths. 

\begin{table}[ht!]
\begin{center}
\begin{tabular}{|l||c|r|c|r|r|c|c|}
   \hline
        & $\tb$ & $m^{2}_{12}$ [GeV$^2$] & $\al$ & $\MH$ [GeV] & $\Ga^{\rm tot}_{H}$ [GeV] & $\kala$ & $\lahhH$ \\
      \hline   \hline
orange  & 10           & 5978.00            & 0.00201  & 244.50        & 0.03648                  & 0.9658           & 0.0490          \\   \hline
yellow  & 10           & 23111.96           & 0.00201  & 480.75        & 0.61080                  & 0.8536           & 0.1880          \\   \hline
purple  & 10           & 51408.69           & 0.00201  & 717.00        & 2.2380                   & 0.6684           & 0.4175          \\   \hline
garnet  & 10           & 90868.19           & 0.00201  & 953.25        & 5.4440                   & 0.3338           & 0.5481          \\   \hline
green   & 10           & 141490.45          & 0.00201  & 1189.5        & 10.750                   & -0.0397          & 0.8501          \\   \hline
\end{tabular}
\end{center}
\caption{Selected points in benchmark \textbf{plane 3} for $\CBA$ $\sim$ 0.1.}
\label{tab:mhhBP3pointscba01}
\end{table}

\begin{figure}[ht!]
\vspace{-1em}
  \begin{center}
\includegraphics[width=0.41\textwidth]{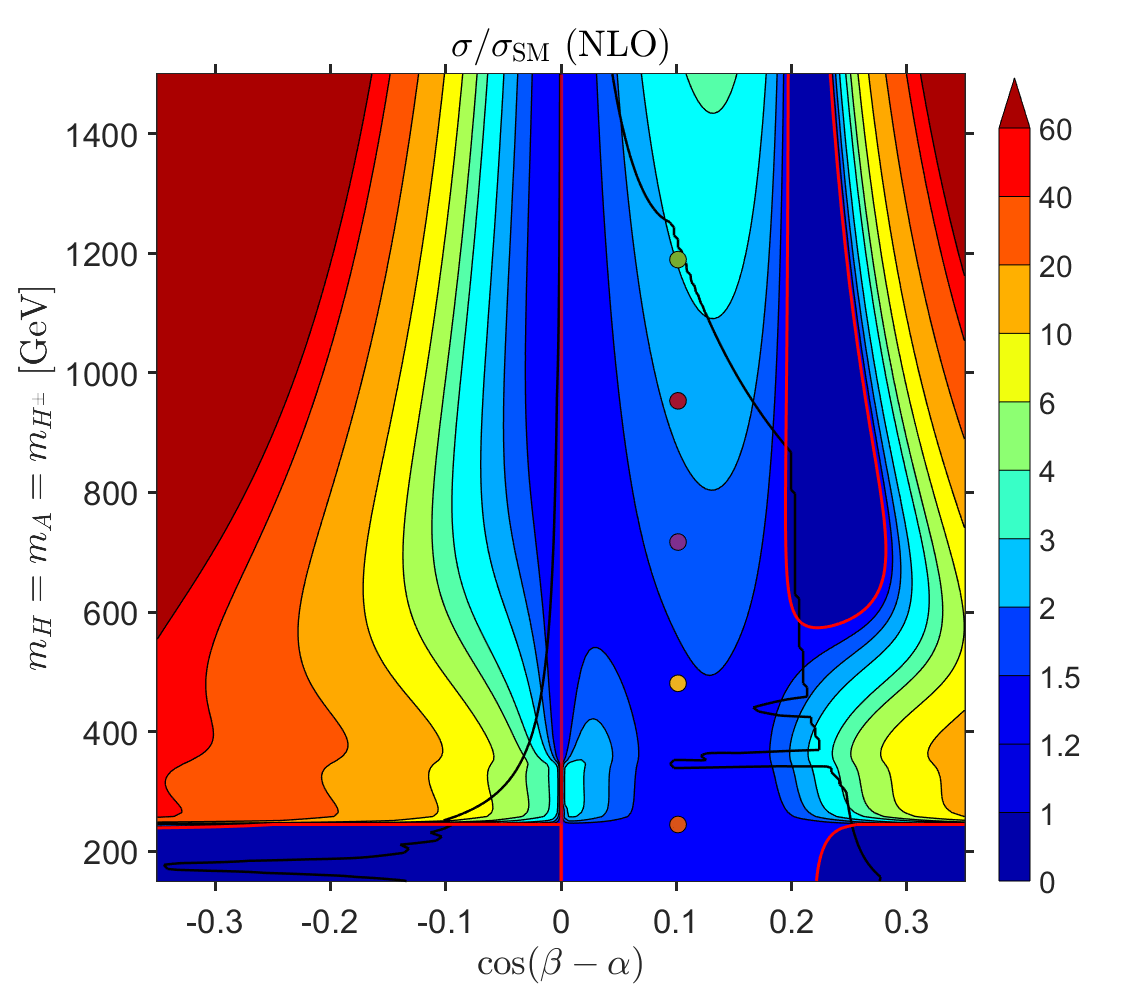}
\includegraphics[width=0.41\textwidth]{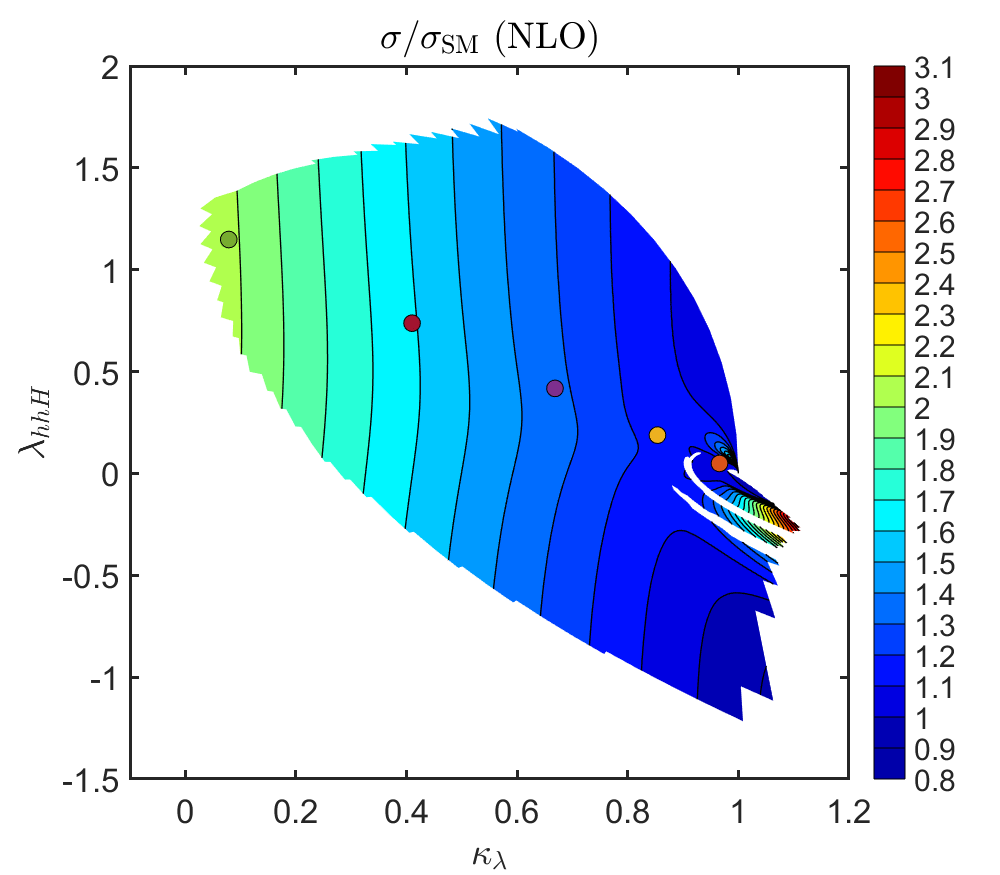}
\includegraphics[width=0.55\textwidth]{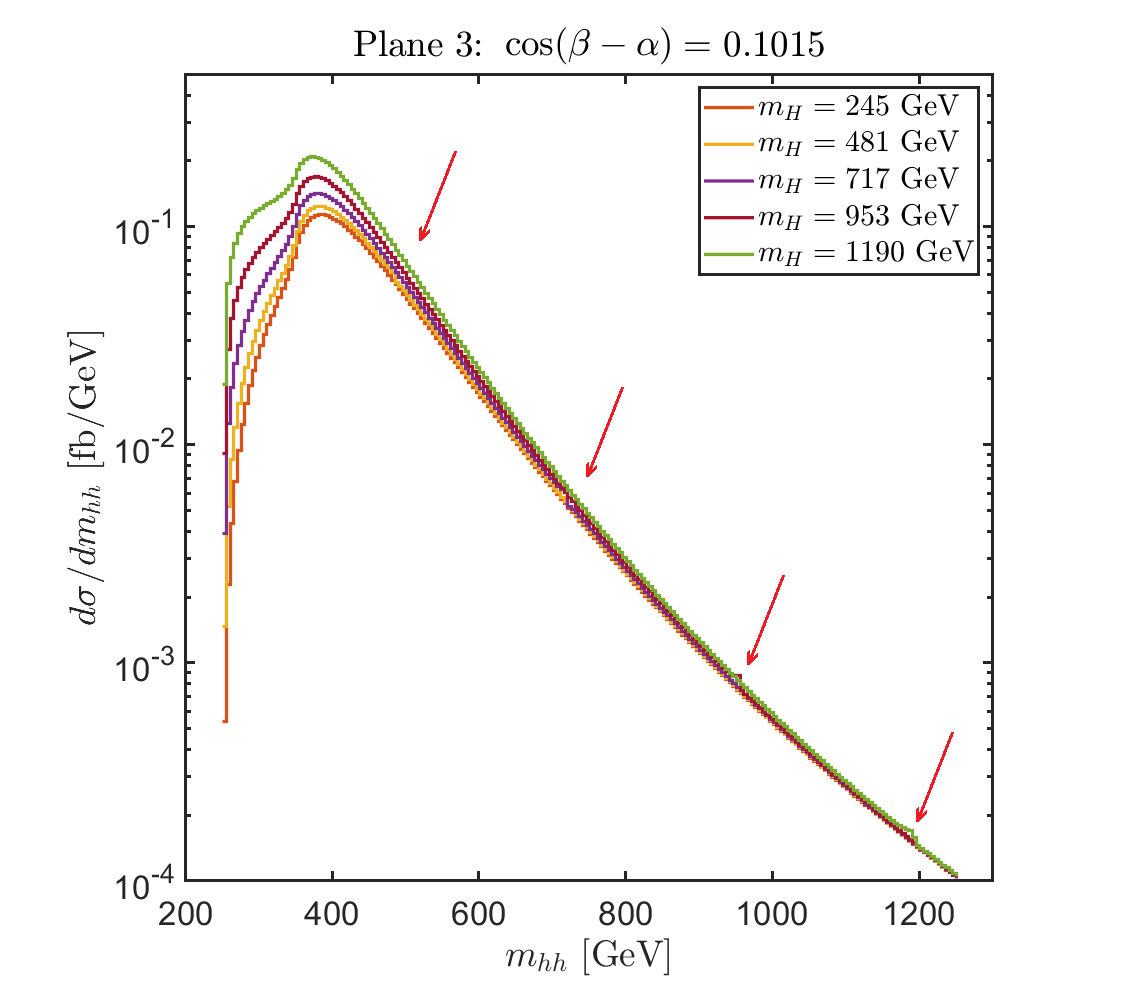}
  \end{center}
  \caption{Sensitivity to triple Higgs couplings for points with
    $\CBA = 0.1015$ in benchmark \textbf{plane~3}.
    \textit{Upper left}: The common mass $m \equiv \MH = \MA = \MHp$ and correspondingly
    $\msq$ change, as 
    marked with orange, yellow,
    purple, garnet and green dots (in ascending mass order,
    color coding indicating $\sig/\sig_{\rm SM}$). 
    \textit{Upper right}: Location of the benchmark points in the
    \kala-\lahhH\ plane (color coding indicating $\sig/\sig_{\rm SM}$).
  \textit{Lower plot}: \mhh\ distribution for the five benchmark points with a
  $5 \gev$ bin size. The colors indicate $m_H$, the red arrows show the
  location of the ``resonant peak''.}  
\vspace{-.5em}
\label{fig:plane3_mhh_cba0.1}
\end{figure}

The points are also shown as orange, yellow,
purple, garnet and green dots (in ascending mass order)
in the upper left plot of \reffi{fig:plane3_mhh_cba0.1} (repeating the
upper left plot of \reffi{fig:couplings3}).  The upper right 
plot of \reffi{fig:plane3_mhh_cba0.1} (repeating the middle right plot
of \reffi{fig:couplings3split}) indicates the location of these points
in the \kala-\lahhH\ plane. One can observe that with increasing $m_H$ the
points are decreasing in \kala (from $\kala \sim 1$ down to zero) and
are increasing in \lahhH\ (from \lahhH\ close to zero to $\lahhH \sim 1$). 

The \mhh\ distributions for the five benchmark points are presented in
the lower plot or \reffi{fig:plane3_mhh_cba0.1}, with a bin size of $5 \gev$.
The color indicates the value of $m_H$ (as defined in the upper plots).
The red arrows indicate the location of the ``resonant peaks'' (for
$m_H > 250 \gev$). The resonant enhancement is found to be tiny, despite
the non-negligible values of \lahhH. The reason for the unrealistically
small bin size of $5 \gev$ is to see any peak at all.
It is clear that for these points the enhancement of $\sig_{\rm 2HDM}$
w.r.t.\ $\sig_{\rm SM}$ is caused purely by the reduced \kala\ values
which alleviates the destructive interference between triangle and box
contributions present in the SM.
The resonant $H$~exchange hardly contributes to
the total cross section.
The structure of the enhancement in this case is hard to infer from
the plot, but looking closely one can see a peak-dip structure. The reason for
this small resonant contribution can be found in the top Yukawa
coupling value of the heavy $\cp$-even Higgs. 
Following \refeq{eq:topyukawa} we obtain a value of
$\xi_H^t = 5 \times 10^{-4} >0$, thus rendering the resonant contribution
negligible. The triple Higgs couplings $\lahhH$ listed in
\refta{tab:mhhBP3pointscba01} are all positive, so that the overall
sign for the coupling factor of the triangle contribution is
positive and the (hardly visible) structure is a peak-dip one as expected.
Furthermore, we have seen in \reffi{fig:mhh3largepoints} that the largest
contribution in the lower mass spectrum comes from the diagram with a
light Higgs $h$ exchange, i.e.\ from the diagram involving
$\kala$. This diagram hence drives the behavior of the
distribution. In the lower plot of \reffi{fig:plane3_mhh_cba0.1} this
trend is clearly visible in the lower part of the spectrum,
$\mhh \lsim 350 \gev$. The smaller
the value of $\kala$ for a particular point (as seen in upper right plot),
the larger the enhancement in the invariant mass distribution at lower
$\mhh$. The most extreme point is the green one, for which $\kala$ is
close to zero, 
and the $\mhh$ spectrum shows a clear bump at $\mhh \sim 350 \gev$. It should 
be recalled that the gluon fusion di-Higgs production cross section has
a minimum for $\kala \sim 2.5$ and is higher for small (or very large
values) of $\kala$.


\subsubsection{Plane 3: $\CBA \sim$ 0.2}

\begin{figure}[ht!]
  \begin{center}
\includegraphics[width=0.41\textwidth]{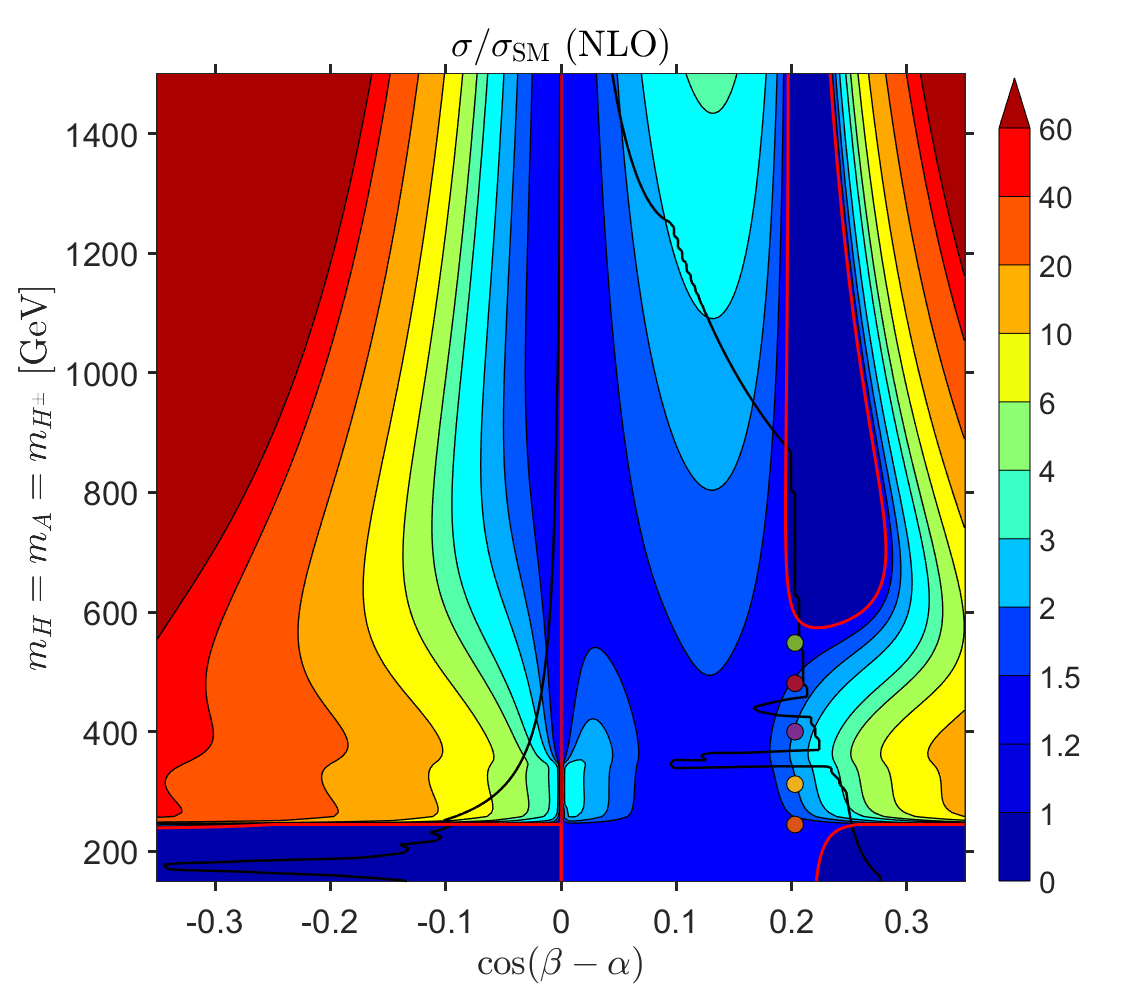}
\includegraphics[width=0.41\textwidth]{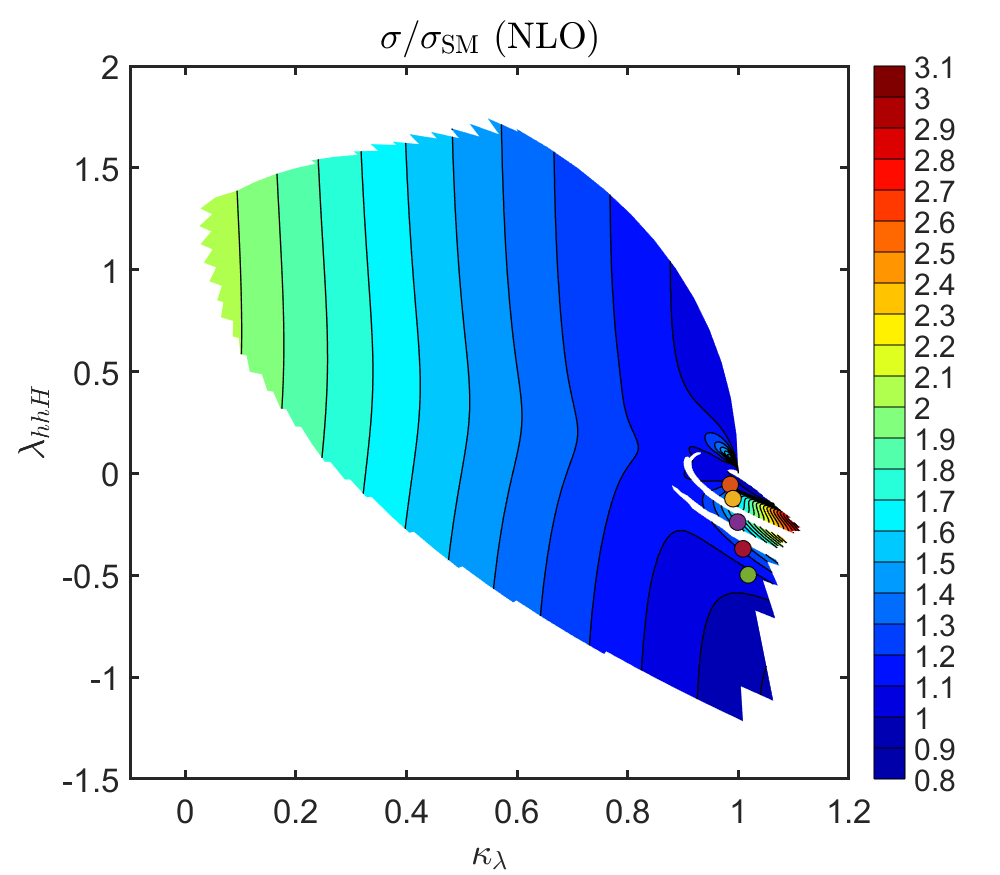}
\includegraphics[width=0.55\textwidth]{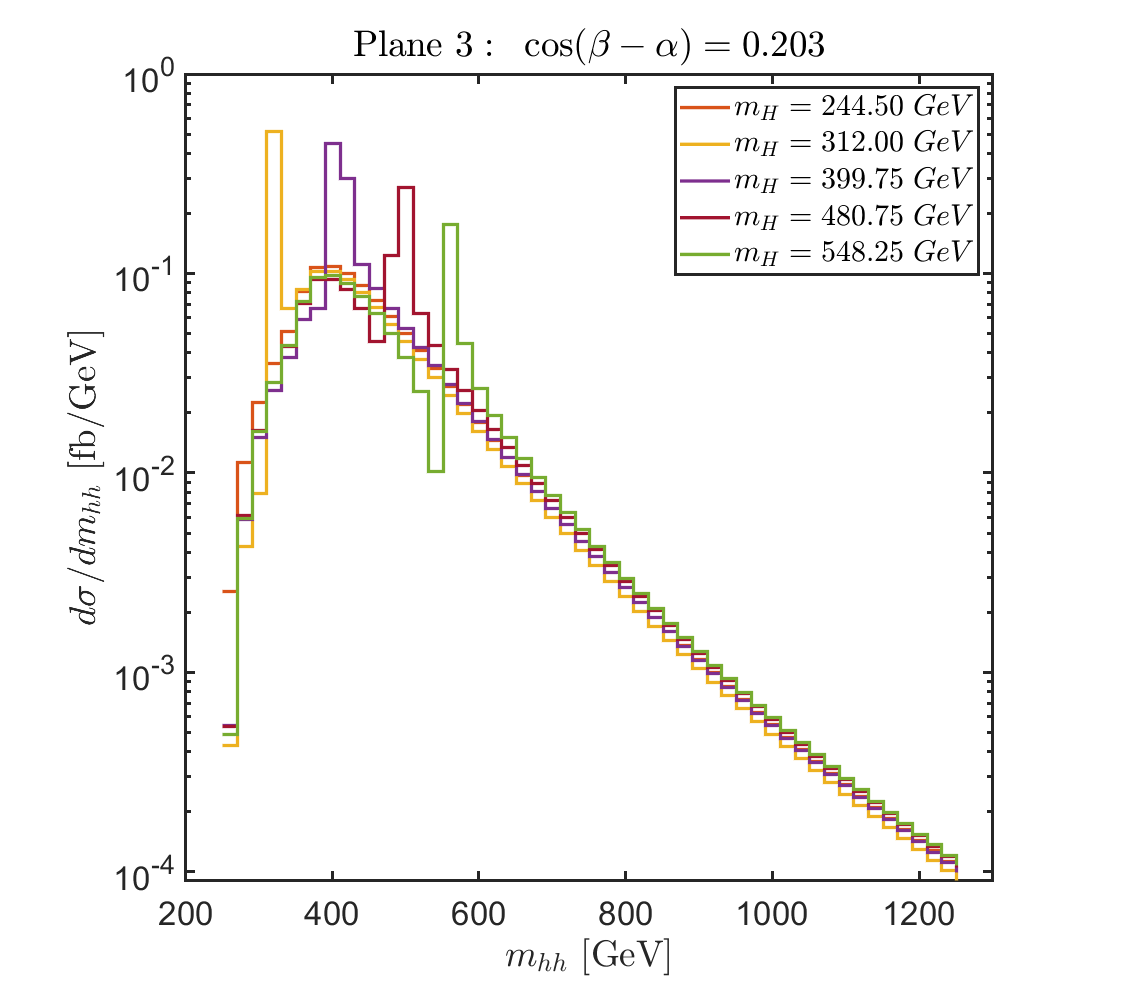}
  \end{center}
  \caption{Sensitivity to triple Higgs couplings for points with
    $\CBA = 0.203$ in benchmark \textbf{plane~3}. \textit{Upper left}:
    $m \equiv \MH = \MA = \MHp$ and correspondingly $\msq$ changes, as
    marked with orange, yellow,
    purple, garnet and green dots (in ascending mass order,
    color coding indicating $\sig/\sig_{\rm SM}$).~\textit{Upper right}: Location of the benchmark points in the
    \kala-\lahhH\ plane (color coding indicating $\sig/\sig_{\rm SM}$).
  \textit{Lower plot}: \mhh\ distribution for the five benchmark points.
  The colors indicate $m_H$.}
\vspace{-1em}
\label{fig:plane3_mhh_cba0.2}
\end{figure}

We finish our analysis of benchmark points in \textbf{plane 3} with five
points with a relatively large value of $\CBA \sim 0.2$,
i.e.\ relatively far away from the alignment limit, as given in 
\refta{tab:mhhBP3pointscba02}. 
As above, the exact value of $\CBA = 0.203$ is given by the scanned
grid.
The points are shown in the upper left and upper right plot of
\reffi{fig:plane3_mhh_cba0.2} as colored dots in orange, yellow,
purple, garnet and green dots (in ascending mass order),
color coding indicating $\sig/\sig_{\rm SM}$ in the 
$\CBA$-$m_H$ and \kala-\lahhH\ plane in the upper left and right plot,
respectively. As can be observed in the upper right plot, all points
have $\kala \sim 1$, i.e.\ no relevant change in the cross section can
be expected from the contribution of the $h$-exchange diagram, so the lower part of the $\mhh$ spectrum is very similar for all the points.
On the other hand, the values of \lahhH\ decrease from around zero down
to $\lahhH \sim -0.5$. However, as can be seen in the two upper plots of
\reffi{fig:plane3_mhh_cba0.2}, the variation of the total cross section
is relatively small. The largest cross sections are found for
$\MH = 312.0 \gev$ (yellow) and $\MH = 399.75 \gev$ (purple),
i.e.\ around $\sim 350 \gev$, see the discussion in
\refse{sec:mhh-plane3}. 
In the lower plot of \reffi{fig:plane3_mhh_cba0.2} we show the
\mhh\ distribution for the five points. For four of them with
$\MH > 2 \Mh$ a clear resonance dip-peak structure can be observed at
$\MH \sim \mhh$, as expected. 

\begin{table}[ht!]
\begin{center}
\begin{tabular}{|l||c|r|c|c|c|c|c|}
\hline
        & $\tb$ & $m^{2}_{12}$ [GeV$^2$] & $\al$ & $\MH$ [GeV] & $\Ga^{\rm tot}_{H}$ [GeV] & $\kala$ & $\lahhH$ \\
      \hline\hline
orange  & 10           & 5912.66            & 0.10475  & 244.50        & 0.146                    & 0.9851           & -0.0549          \\
\hline
yellow  & 10           & 9627.97            & 0.10475  & 312.00        & 0.463                    & 0.9902           & -0.1240          \\
\hline
purple  & 10           & 15805.30           & 0.10475  & 399.75        & 1.291                    & 0.9864           & -0.2388          \\
\hline
garnet  & 10           & 22859.37           & 0.10475  & 489.75        & 2.550                    & 1.0083           & -0.3699          \\
\hline
green   & 10           & 29729.19           & 0.10475  & 548.25        & 3.995                    & 1.0177          & -0.4977          \\
\hline
\end{tabular}
\end{center}
\caption{Selected points in benchmark \textbf{plane 3} for $\CBA$ $\sim 0.2$.}
\label{tab:mhhBP3pointscba02}
\end{table}

\begin{figure}[ht!]
  \begin{center}
\includegraphics[width=0.32\textwidth]{figs/plot3cba02.png}
\includegraphics[width=0.32\textwidth]{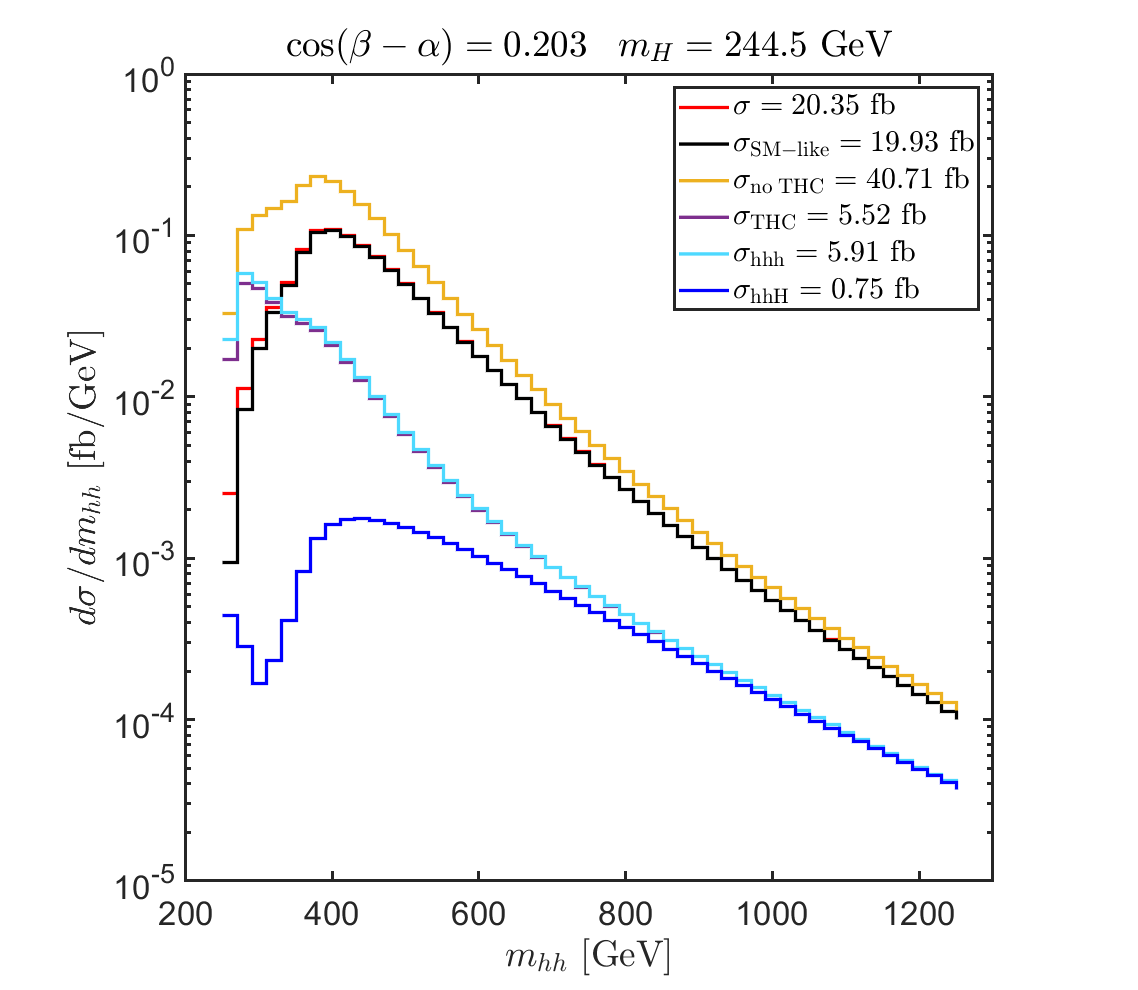}
\includegraphics[width=0.32\textwidth]{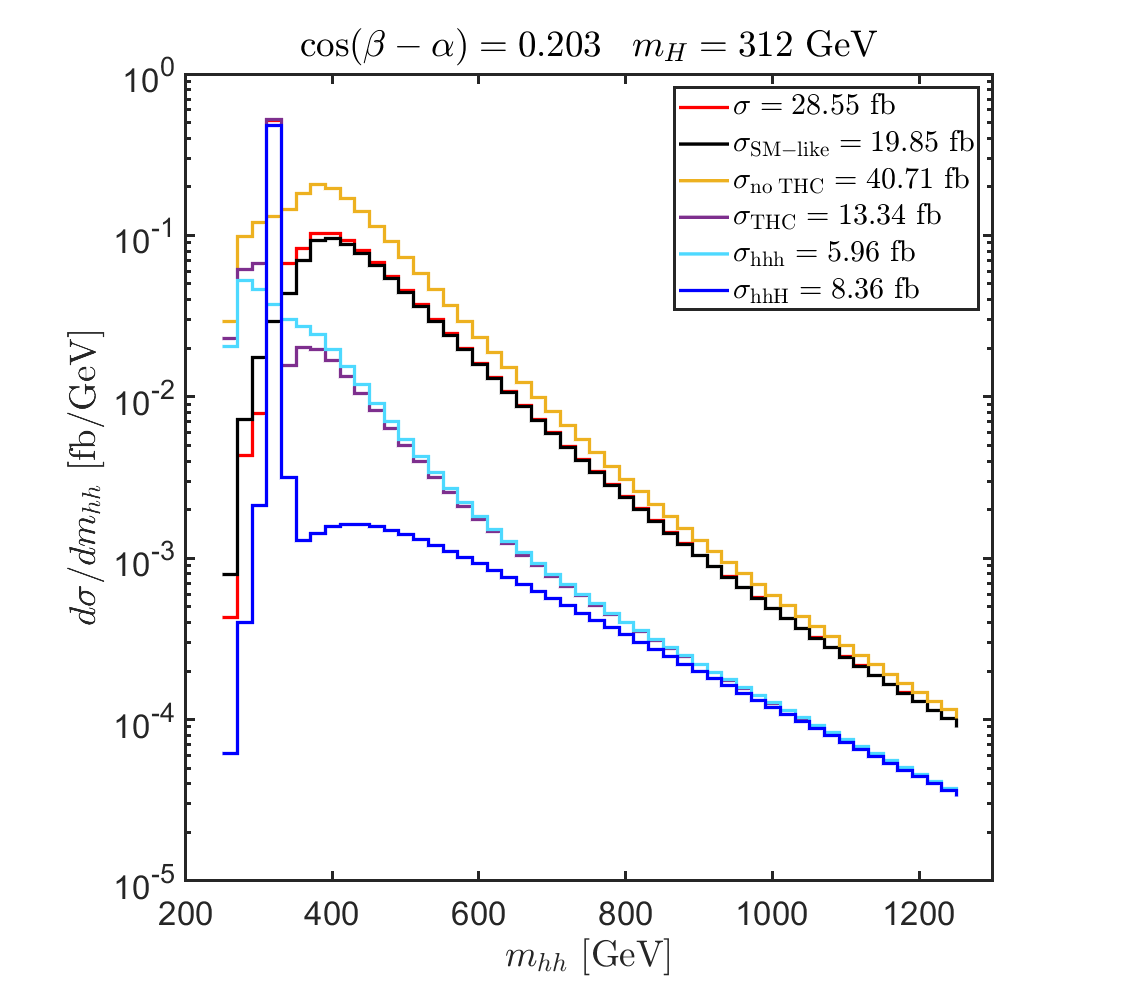}
\includegraphics[width=0.32\textwidth]{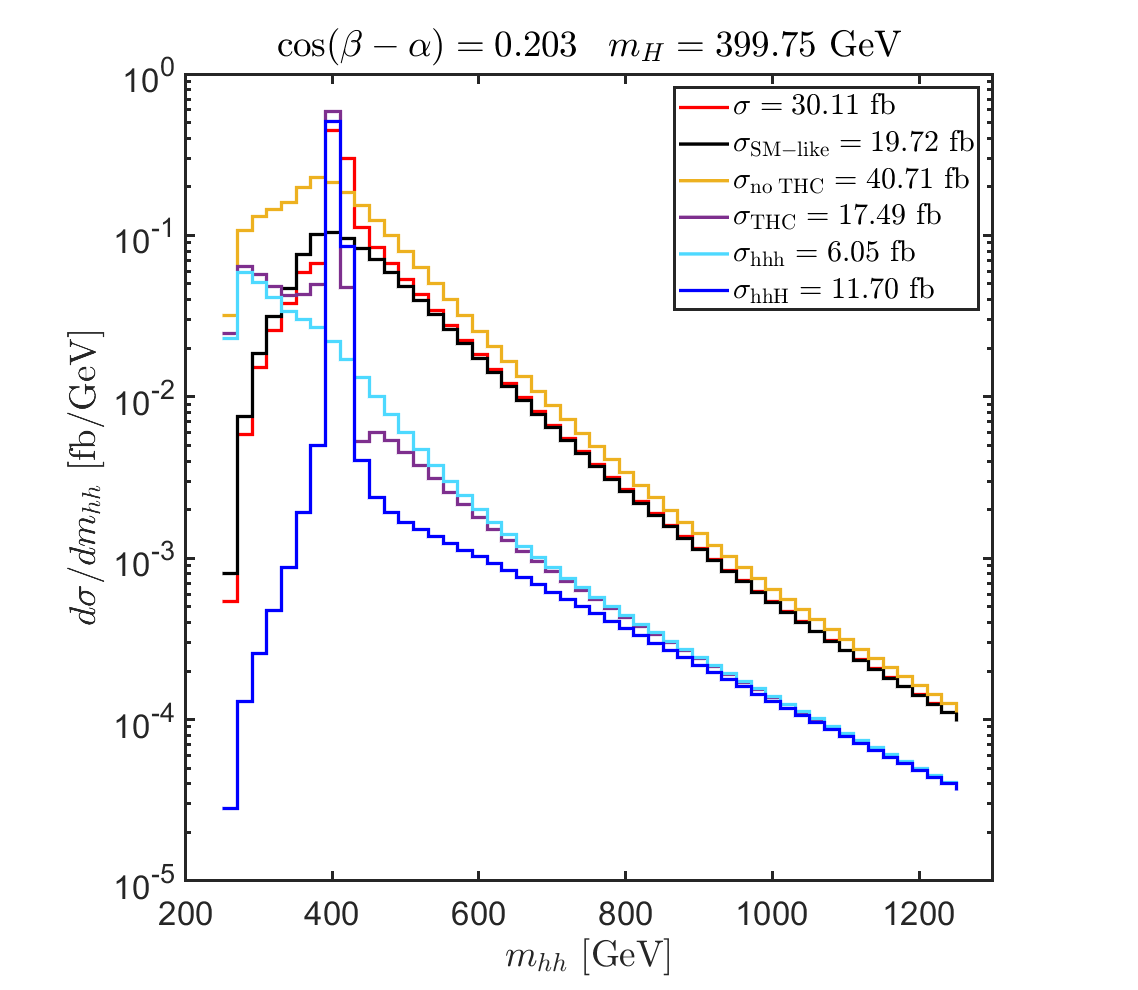}
\includegraphics[width=0.32\textwidth]{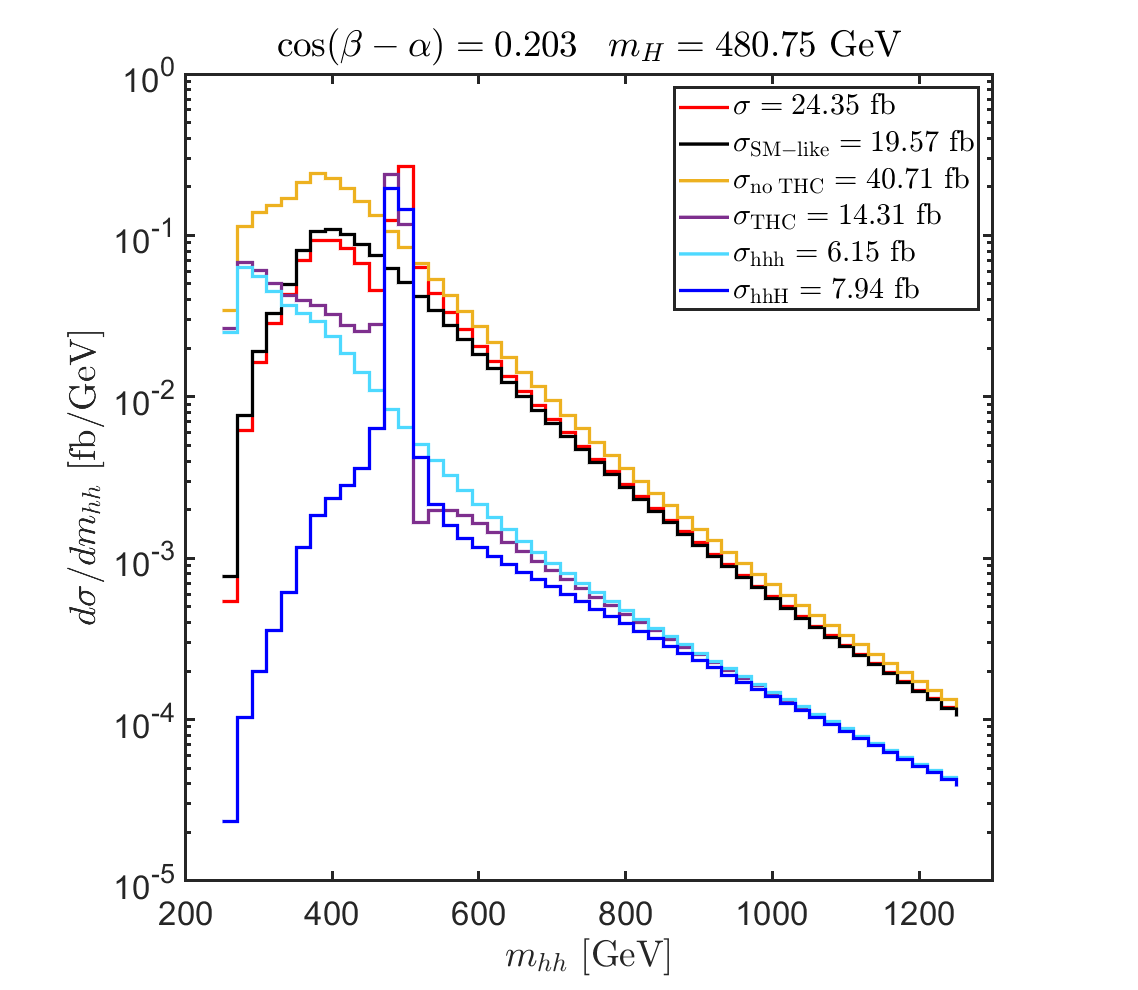}
\includegraphics[width=0.32\textwidth]{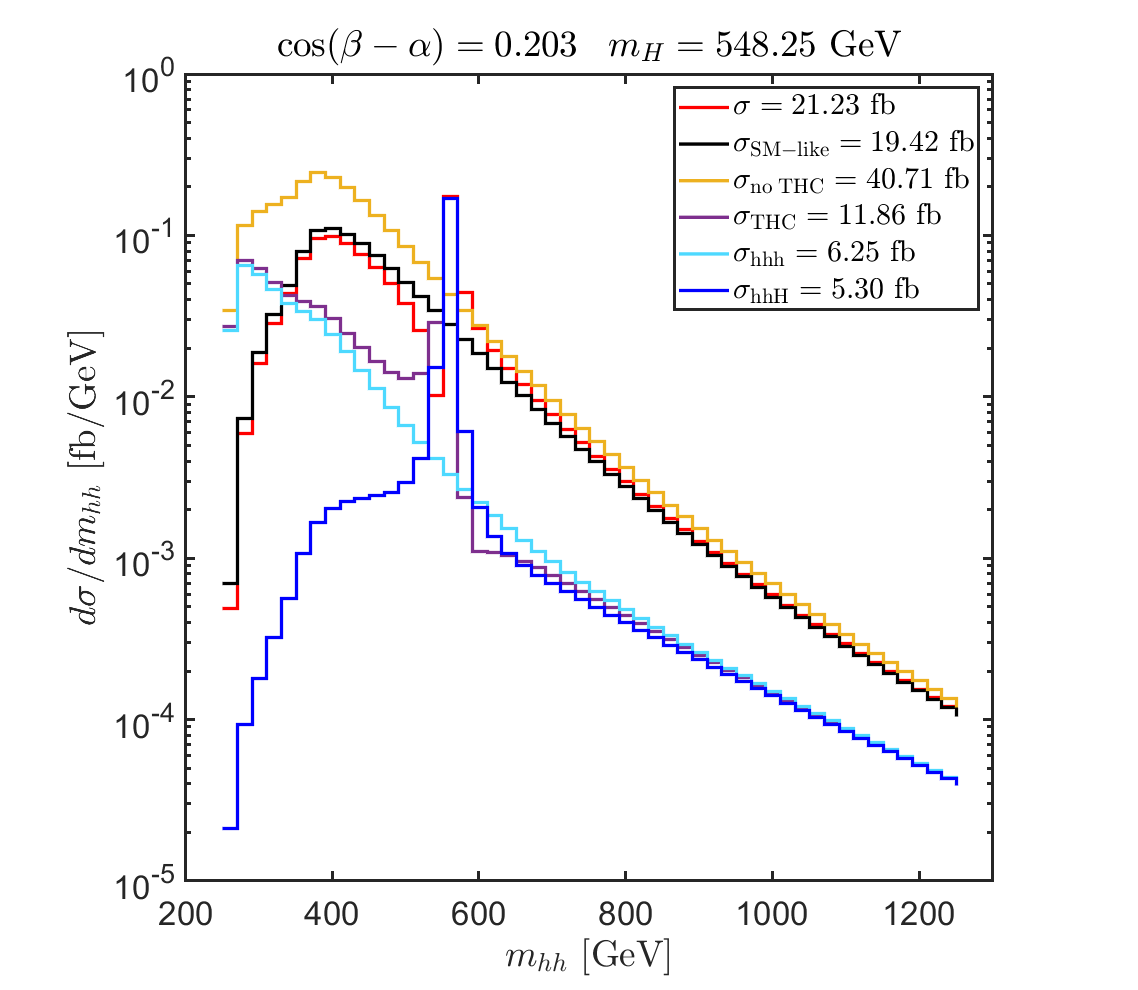}
  \end{center}
\caption{\mhh\ distributions for selected points in
  \textbf{plane~3} with $\CBA \sim 0.2$.
  \textit{Upper left}: Total \mhh\ distributions for the points
of \refta{tab:mhhBP3pointscba02}; \textit{upper middle, left} and \textit{lower left,
middle} and \textit{right}: Individual contributions to the
\mhh\ distribution for the five points: total SM-like cross
section in black, 
total cross section in red, \mhh\ involving only \lahhh\ (\lahhH) in
light (dark) blue, \mhh\ involving both (no) THCs in purple (yellow).}
\label{fig:mhh3cba02points}
\end{figure}

In \reffi{fig:mhh3cba02points} we present a more detailed analysis of the
sensitivity to the triple Higgs couplings in the invariant mass
distribution following the same notation as in
\reffi{fig:mhh3largepoints}. We show the invariant mass distribution
from all the diagrams in the upper right plot and then split the
individual contributions for each particular mass point in the rest of the
plots.
The first one for $\MH = 244.5 \gev$ is shown in the upper middle
plot, which has $\MH$ 
below the di-Higgs production threshold. Consequently, no enhancement 
due to the diagram containing $\lahhH$ can be observed, and the total
cross section is almost indistinguishable from the SM-like result
in this case. For the other masses we
find a similar result as in \reffi{fig:mhh3largepoints}.
One can observe that the $s$-channel contribution
involving the heavy Higgs with its trilinear coupling
$\lahhH$ is responsible for the  
enhancement close to the mass of the intermediate Higgs boson, while the
effect of the $\kala$ is mostly significant in the low mass region of
the plot. The contribution of the diagrams involving THCs (purple)
interferes
with the continuum (box diagram) shown in yellow and creates the
dip-peak structure that can be observed in the total distributions, see \reffi{fig:mhh3cba02points} (upper left).

The top Yukawa coupling for this value of $\CBA$ is $\xi_H^t = 0.1020 > 0$, and
thus the sign of $(\lahhH \times \xi_H^t)$ is negative, resulting in the
dip-peak structure observed. In principle, this type of distributions
can yield a handle on the size and sign of \lahhH, as will be discussed
in more detail below.

\begin{figure}[ht!]
  \begin{center}
\includegraphics[width=0.32\textwidth]{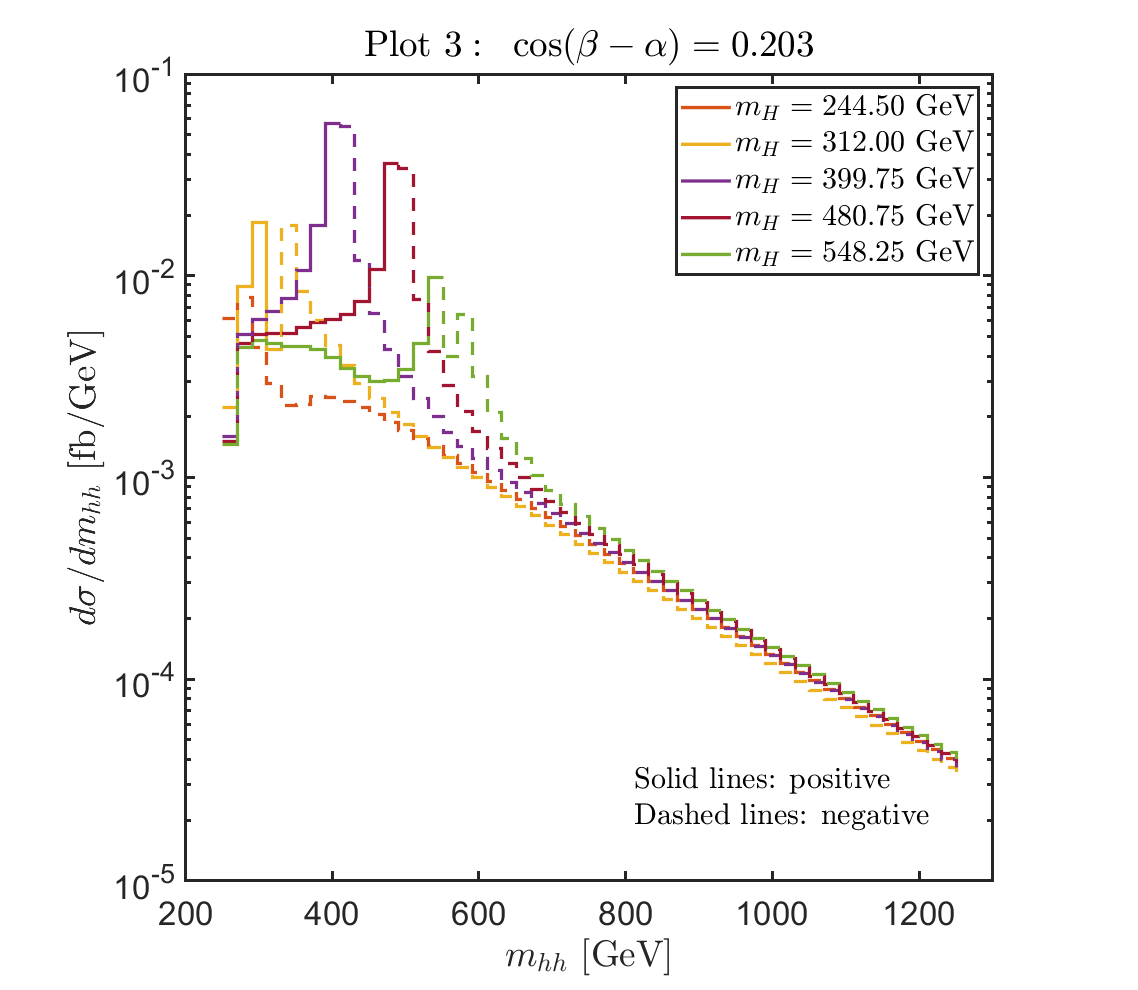}
\includegraphics[width=0.32\textwidth]{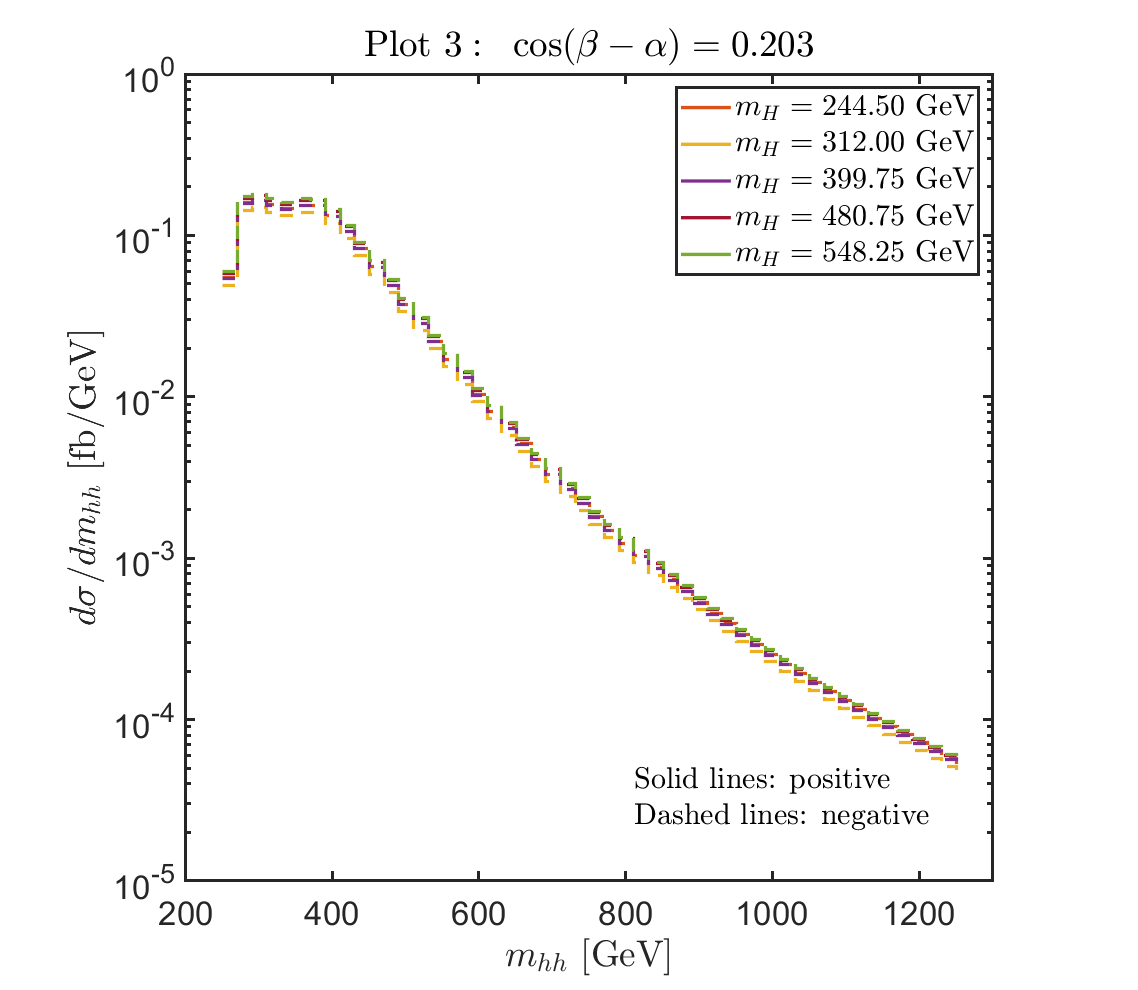}
\includegraphics[width=0.32\textwidth]{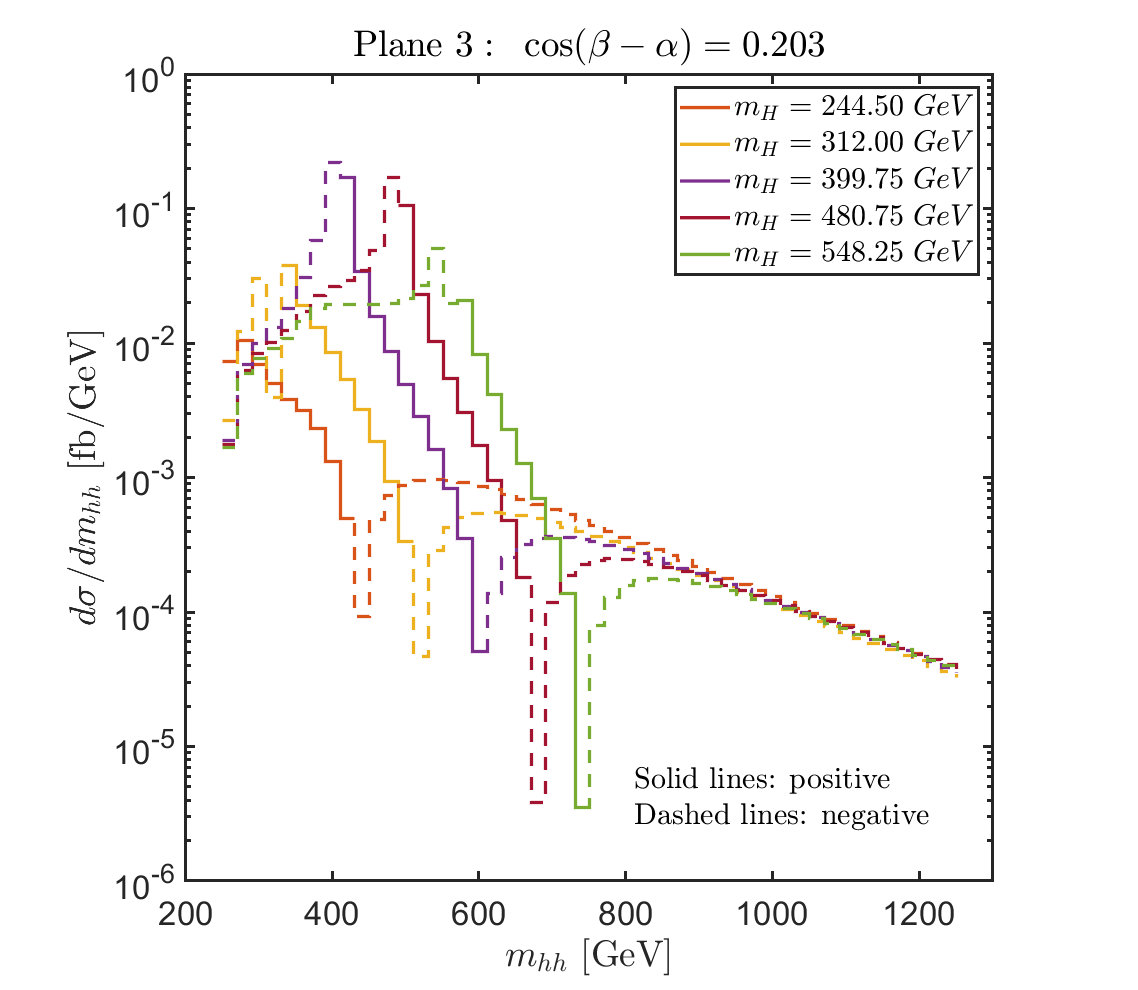}
  \end{center}
\caption{Interference contributions for the benchmark points in
  \textbf{plane~3} with $\CBA$ = 0.203. \textit{Left}:
    Interference of the two triangle diagrams. \textit{Middle}:
    Interference of the SM-like diagrams. \textit{Right}: Interference between the box and the resonant diagram. Solid lines
    indicate that the interference is positive and dashed lines indicate
    that it is negative.}
\vspace{-1em}
\label{fig:plane3_mhh_cba0.2_interf}
\end{figure}

In \reffi{fig:plane3_mhh_cba0.2_interf} we further analyze the
interference contributions of the $s$-channel
diagrams in \reffi{fig:gghhdiagrams}. 
In the left plot we show the interference term of
the diagram with the $s$-channel $h$ exchange (A) and diagram with the
$s$-channel $H$ 
exchange (B). The interference term in this case is defined as
$|A+B|^2-|A|^2-|B|^2$. The solid (dashed) lines indicate positive
(negative) interference. One can observe a similar behavior to the
above discussed interference between resonant and box contributions,
i.e.\ 
that these two diagrams interfere constructively up to $\mhh \le \MH$, and
destructively for larger \mhh\ values, i.e.\ the interference term
changes its sign. 
The diagram A and the box diagram (C) interfere negatively accross the
whole invariant mass range as shown in the middle plot of
\reffi{fig:plane3_mhh_cba0.2_interf}. This behavior corresponds to the
result found for the SM di-Higgs production, where only these two
diagrams are present 
Finally, the interference of B and C, shown in the right plot of
\reffi{fig:plane3_mhh_cba0.2_interf}, has two sign changes. Up to
$\mhh \le \MH$ the interference is negative, leading to the dip in the
total \mhh\ distribution. For larger values it turns positive, leading
the subsequent peak in the total distribution, see the discussion in
\refse{sec:toy}. 
The second sign change happens because the interference approaches
zero at an \mhh\ value not related to $\MH$. In the plot the
interference lines in principle go down to zero, which, however, is not
visible due to the log scale and the finite bin width.


\subsection{Model based analysis: benchmark plane 4}
\label{sec:mhh-8}

To complete our \mhh\ analysis we investigate benchmark points in
\textbf{plane 4}. In this plane the heavy Higgs boson masses are free
parameters and are in agreement with the applied constraints over a
relatively large interval, i.e.\ the effects of a mass variation (as
well as a variation of \lahhH) can be readily analyzed. The selected
points are listed in \refta{tab:plane8_mhh_cba0.2}. They are all located
at the same mass of $\MA = \MHp = 544.75 \gev$. As was demonstated in
\refse{sec:sensitivity}, \kala\ and \lahhH\ are proportional to each
other, and allowed values of \lahhh\ are within $\sim 2\%$ of $\kala = 1$,
whereas \lahhH\ is found for our benchmark points in the interval
$[\sim 0, -0.7]$. Similarly, 
also $\Ga_H^{\rm tot}$ grows with increasing $\MH$. All corresponding
numerical values can be found in \refta{tab:plane8_mhh_cba0.2}.

\begin{table}[ht!]
\begin{centering}
\begin{tabular}{|l||c|r|c|c|c|c|c|}
\hline
        & $\tb$ & $m^{2}_{12}$ [GeV$^2$] & $c_{\beta-\al}$& $\MH$ [GeV]  & $\Ga^{\rm tot}_{H}$ [GeV] & $\kala$ & $\lahhH$ \\
        \hline\hline
purple  & 10           & 4899.61            & 0.2                & 222.50         & 0.09306                  & 0.9816           & -0.0321       \\
\hline
orange   & 10           & 8612.82            & 0.2                & 295.00         & 0.3464                   & 0.9835           & -0.0964       \\
\hline
blue    & 10           & 13366.45           & 0.2                & 367.50         & 0.8694                   & 0.9860           & -0.1788       \\
\hline
light blue    & 10           & 19160.49           & 0.2                & 440.00         & 1.7540                   & 0.9891           & -0.2791       \\
\hline
green   & 10           & 25994.95           & 0.2                & 512.50         & 3.0020                   & 0.9927           & -0.3975       \\
\hline
yellow  & 10           & 33869.83           & 0.2                & 585.00         & 4.6720                   & 0.9968           & -0.5339       \\
\hline
garnet  & 10           & 42785.13           & 0.2                & 657.50         & 9.2810                   & 1.0015           & -0.6883       \\
\hline
\end{tabular}
\par\end{centering}
\caption{Selected points in benchmark \textbf{plane 4} for $\MA = \MHp
  \sim 545 \gev$.}
\label{tab:plane8_mhh_cba0.2}
\end{table}

\begin{figure}[ht!]
  \begin{center}
\includegraphics[width=0.41\textwidth]{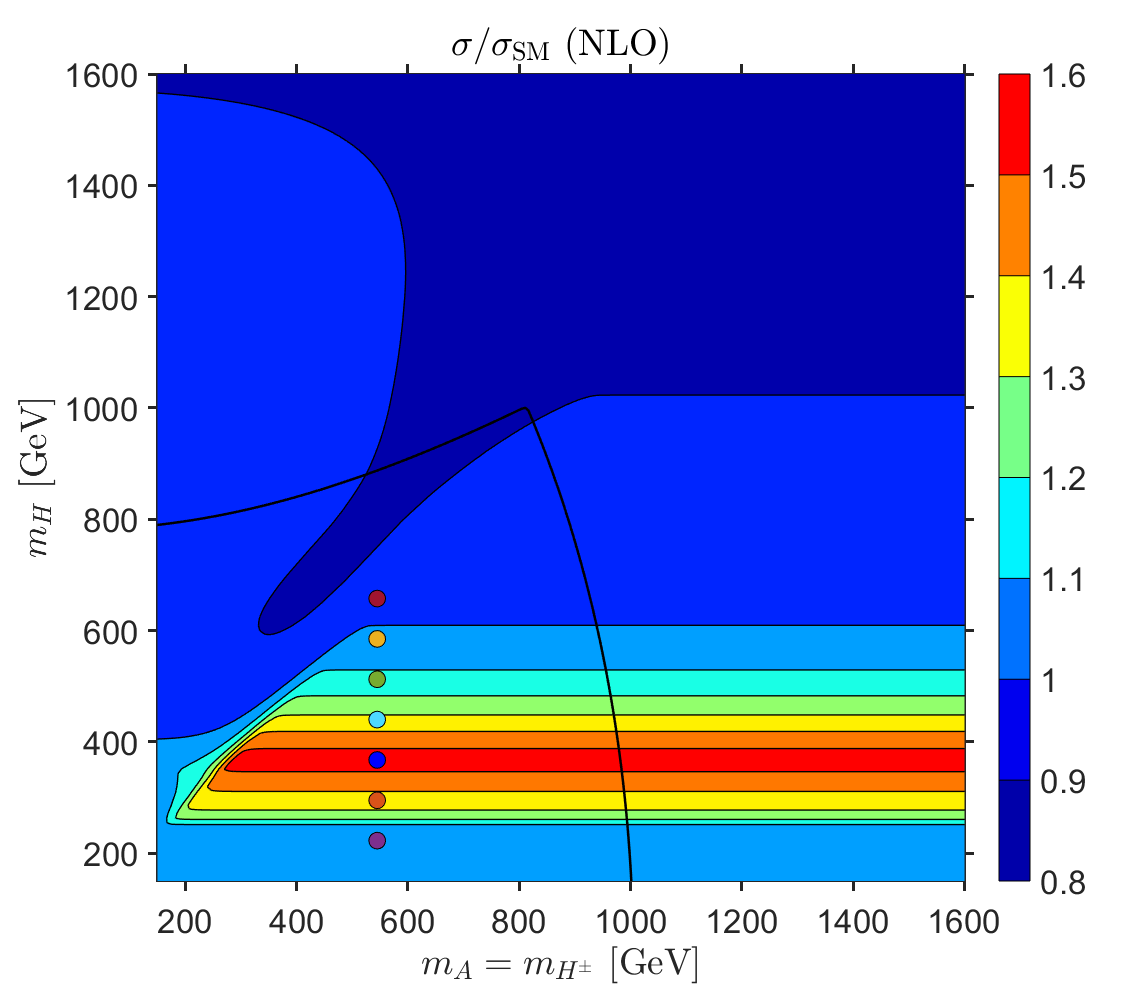}
\includegraphics[width=0.41\textwidth]{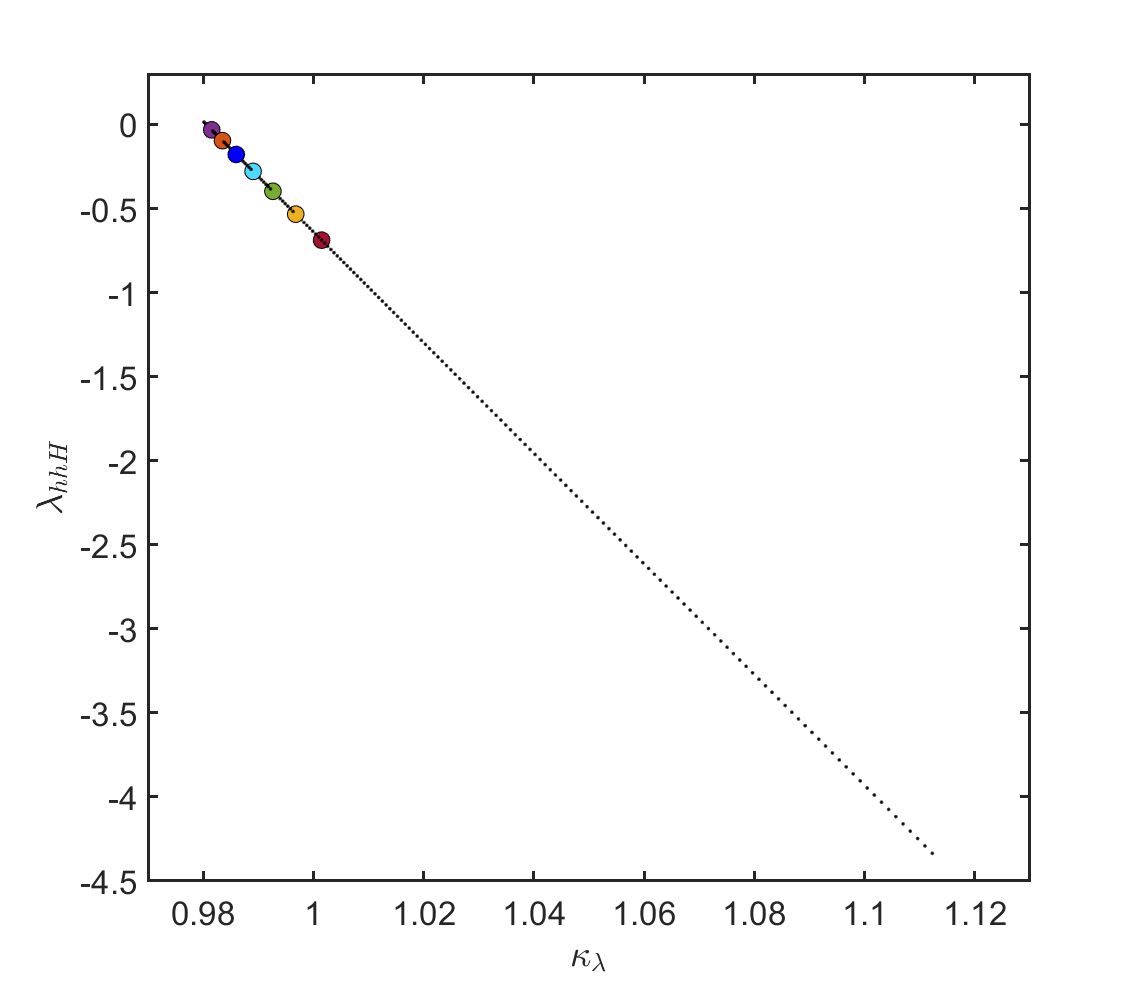}
\includegraphics[width=0.41\textwidth]{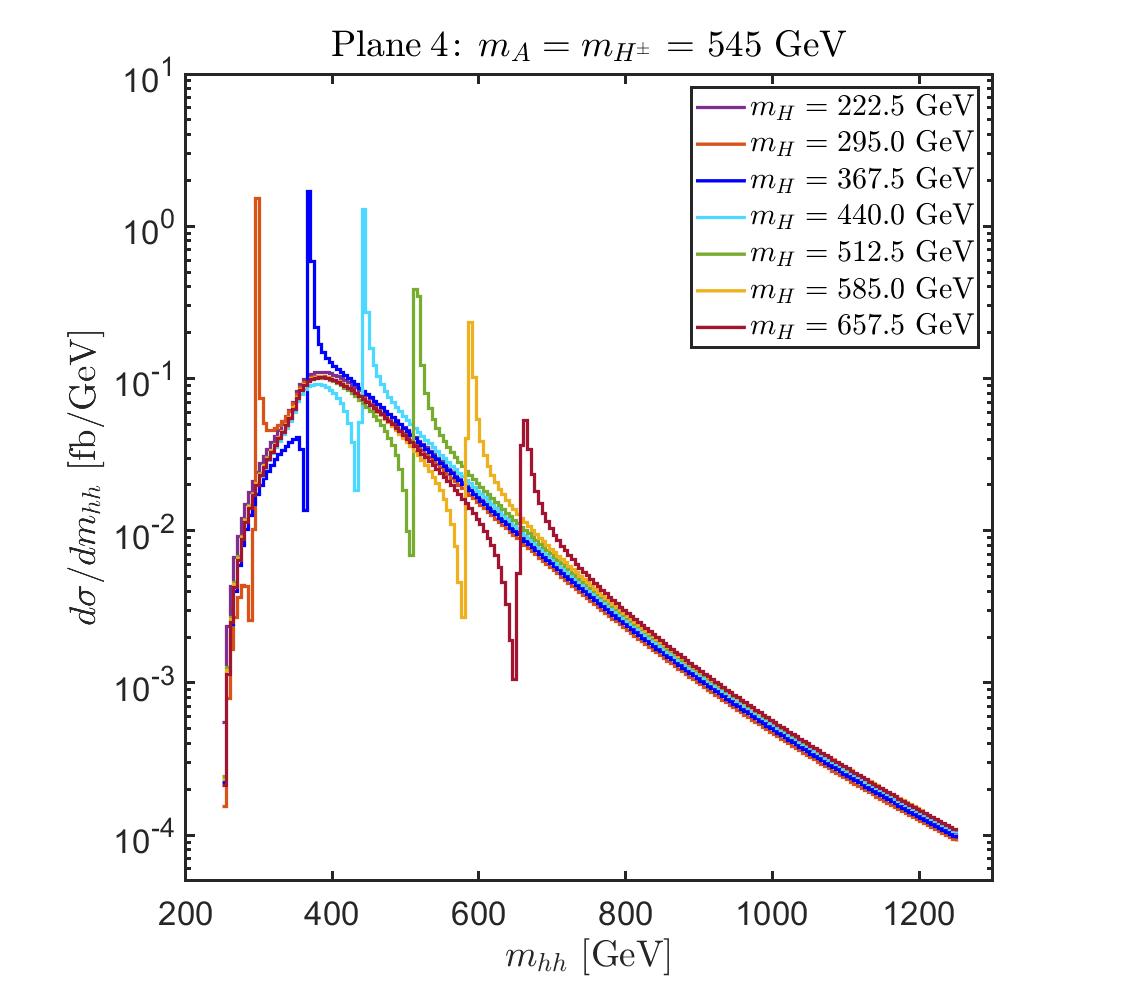}
\includegraphics[width=0.41\textwidth]{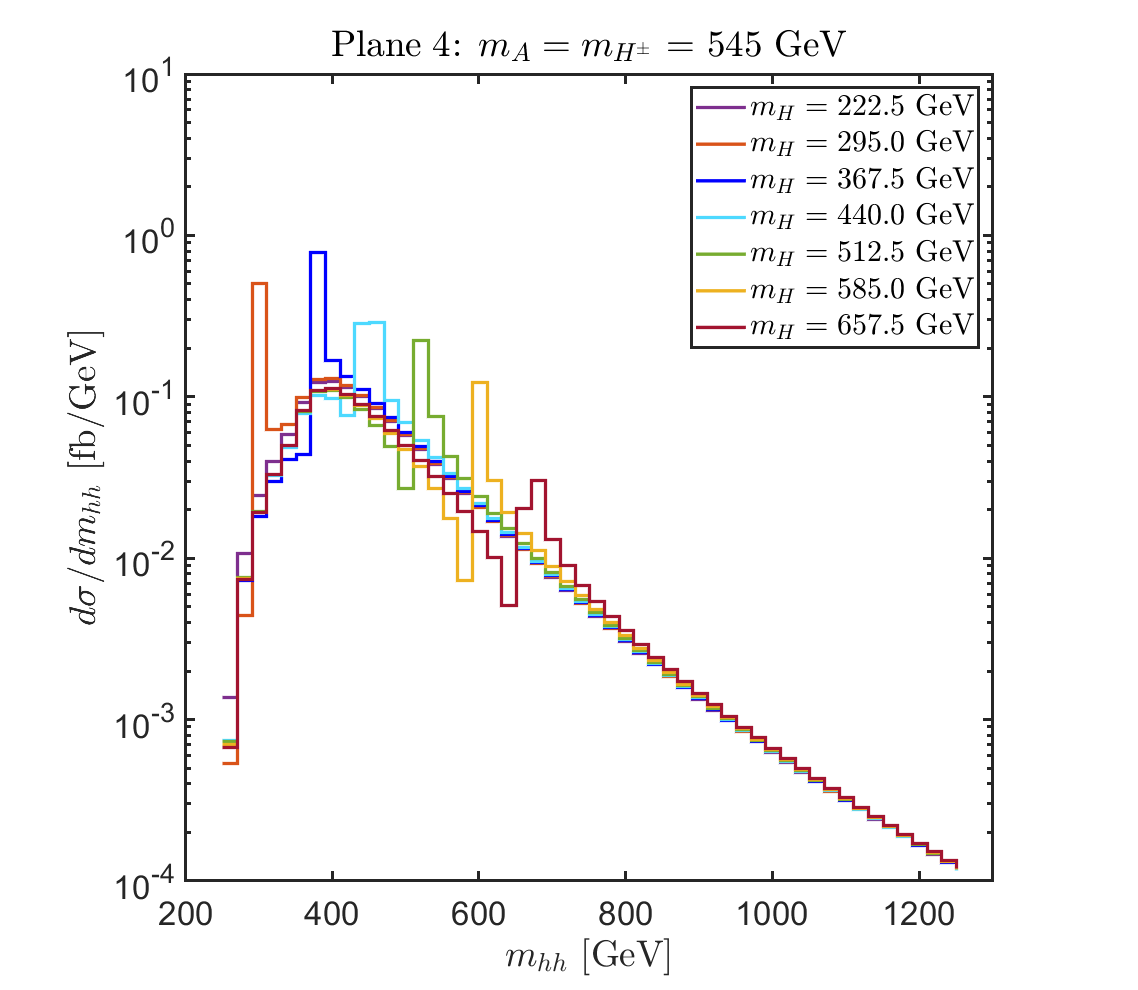}
  \end{center}
  \caption{Sensitivity to triple Higgs couplings for points within
    benchmark \textbf{plane 4}. \textit{Upper left}: The points have fixed
    $\MA = \MHp = 544.75 \gev$, $\CBA = 0.2$, whereas
    $\MH$ and correspondingly $\msq$ change, as
    marked with purple, orange, blue, light blue, green,
    yellow and garnet dots (in ascending mass order,
    color coding indicating $\sig/\sig_{\rm SM}$). 
    \textit{Upper right}: Location of the benchmark points in the
    \kala-\lahhH\ plane.
  \textit{Lower line}: \mhh\ distribution for the seven benchmark points with a
  bin size of $5 \, (20) \gev$ in the left (right) plot.  
  The colors indicate $\MH$.}
\vspace{-1em}
\label{fig:plane8_mhh_cba0.2}
\end{figure}

The location of the points in the benchmark plane is shown in the upper
left plot of \reffi{fig:plane8_mhh_cba0.2}, and in the
\kala-\lahhH\ plane in the upper right plot. 
The lower row shows the corresponding \mhh\ distributions with the color
indicating the $\MH$ value. In the left plot, for better visibility,  we
show the idealized case of a bin size of $5 \gev$, whereas the right
plot shows the more realistic case of $20 \gev$. 

One can observe that the resonant
structure in this case is dip-peak since (as in the previous case) 
the top Yukawa coupling is positive and the BSM THC is negative.
For the purple point no enhancement can be observed, since 
its mass is below the di-Higgs production threshold,
$\MH^{\rm purple} = 222.5 \gev$.  
The remaining points show a resonant enhancement as the invariant
mass approaches the mass of the heavy Higgs, i.e.\ one expects to
have the highest sensitivity to the THC $\lahhH$ in this region. The
difference between the heights of the peaks is found to be largest for small $\MH$, but
naturally also decreases with increasing bin size, which will be further
discussed below. The increasing bin size, in particular, decreases the
visibility of the dip, an effect that is in particular visible for low
$\MH$. This demonstrates the relevance of the bin size in a realistic
experimental analysis (and will also be further discussed below).



\section{Impact of experimental uncertainties \label{sec:expuncert}}

In this section we will analyze the impact of experimental uncertainties
on the possible sensitivity to \lahhH. These effects are the
experimental smearing, i.e.\ the uncertainty in the \mhh\ measurement,
the experimental resolution, i.e.\ the size of the bin width, as well as
the arbitrary location of the bin. We will also demonstrate how the
experimental results for \lahhH\ change with a variation of
sign$(\lahhH \times \xi_H^t)$.

In order to estimate the sensitivity to the BSM coupling $\lahhH$,
following \citere{Arco:2021bvf}, we define a theoretical quantity that
aims to quantify the ``sensitivity to \lahhH'' (but is not meant as a
determination of its precision, which requires a detailed experimental
analysis, which is beyond the scope of our paper). We define
\begin{equation}
    R := \frac{\sum_i |N_i^R - N_i^C|}{\sqrt{\sum_i N_i^C}},
    \label{eq:R}
\end{equation}
where $N^{R}$ is the number of events of the resonant contribution,
and $N^C$ is the number of events of the continuum. The window in
which the events are counted is defined by
\begin{align}
|N^R - N^C| &> {\rm bin\;size} \times 20 \gev~.
\label{eq:R-window}
\end{align}
The sum over $i$ in \refeq{eq:R} runs over all the bins
that fulfill this condition.
The chosen condition in \refeq{eq:R-window} starts with a minimum of 
1000~excess events due to the resonance when the bin
size is 50 GeV and 200 events when the bin size is $10 \gev$, 
i.e.\ smaller bin sizes are not ``punished''. 
Using the absolute value in the definition of $R$ in \refeq{eq:R},
  as well as in the definition of the window in \refeq{eq:R-window}
  effectively makes use of both the dip and the peak of the smeared
distribution. This constitutes a simplified theory definition, where in
a realistic experimental analysis the dip-peak structure would be taken
into account via a template fitting, see e.g.\ the analysis in
\citere{CMS:2019pzc}. 
The numbers of events are in turn obtained using the relation between
the cross section and the integrated luminosity of the collider, 
\begin{equation}
    N = \sig \cdot \mathcal{L},
\end{equation} 
where we have used $\mathcal{L} = 6000\; {\rm fb^{-1}}$, i.e.\ the sum
of the anticipated luminosity of ATLAS and CMS combined at the end of
the HL-LHC run. This constitutes the most optimistic case.


\subsection{Smearing}
\label{sec:smearing}

Differential cross section measurements are affected by the finite
resolution of the detectors. This translates into a blurred or
``smeared'' spectrum that can be observed in such experiments. We try to
mimic this effect by artificially smearing the theoretical prediction
for the invariant mass distributions of the chosen benchmark points. To
do this we introduce a statistical error to our prediction of the
invariant mass. We apply the uncertainties in \mhh\ by allowing the
value of an event to shift to the left or to the right in the spectrum
according to 
a Gaussian probability distribution. The amount of smearing is defined
in terms of a percentage of smearing $p$ that indicates that the
predicted value $x$ should stay within the interval
$[(1-p) \cdot x, (1+p) \cdot x]$ with a 78 \% probability,
which corresponds to the
definition of the full width half maximum of the Gaussian distribution
(as given for the experimental analyses). 

\begin{figure}[ht!]
  \begin{center}
\includegraphics[width=0.41\textwidth]{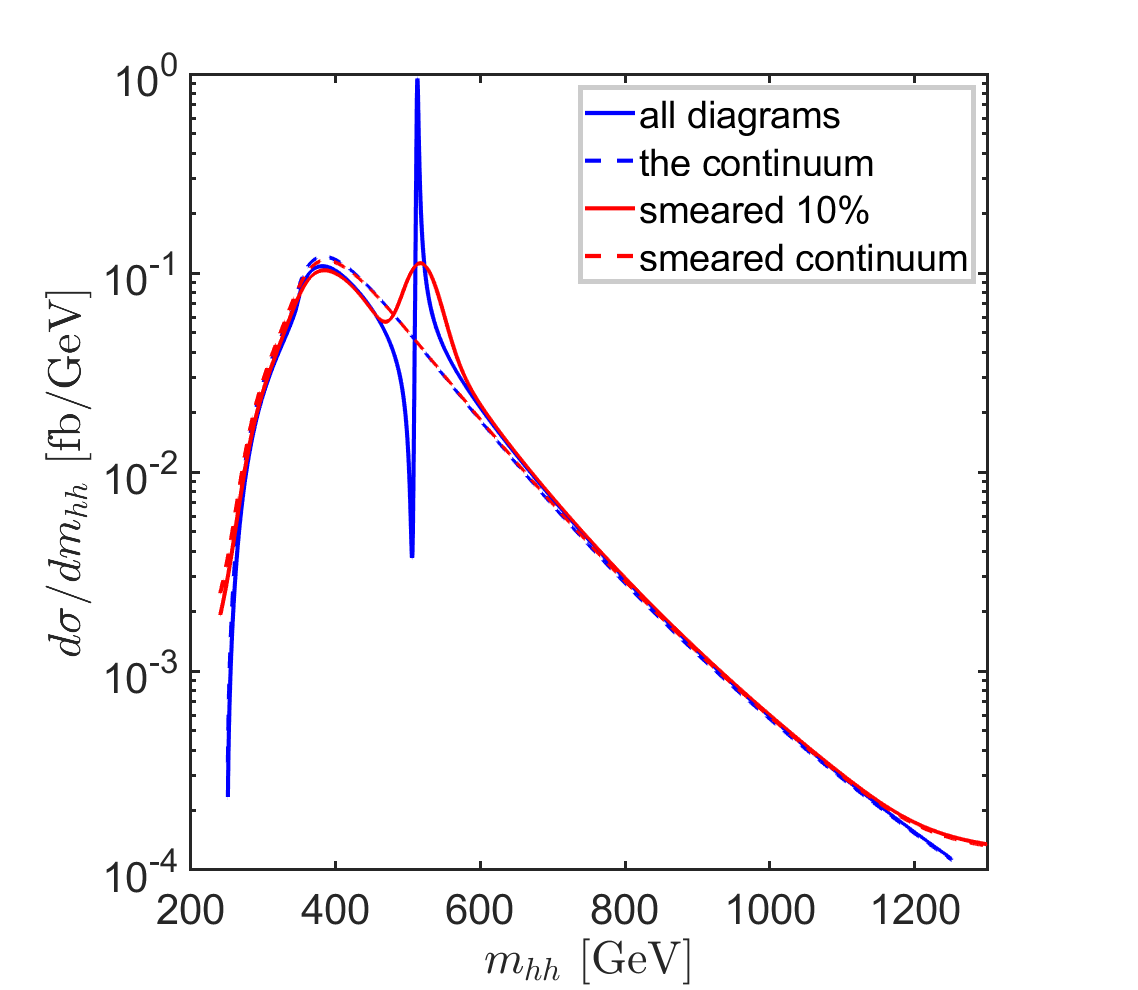}
\includegraphics[width=0.41\textwidth]{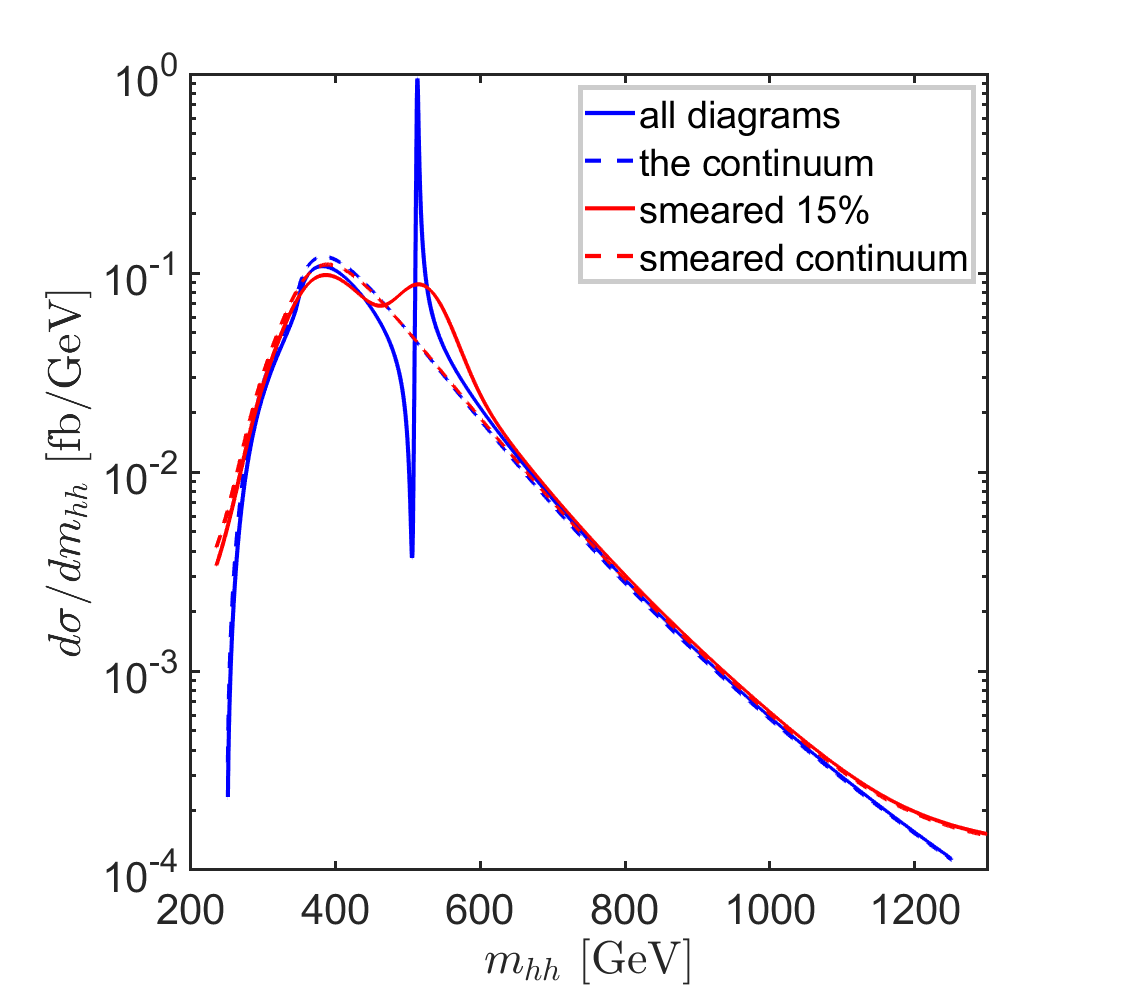}
  \end{center}
\caption{Theoretical (blue) and smeared (red) invariant mass
  distributions for the selected benchmark point (see text). Solid (dashed) lines
  show the contribution of the total (continuum) differential cross
  section. Left (right) plot has a 10\% (15\%) smearing.}
\label{fig:smear}
\end{figure}

We illustrate this effect in \reffi{fig:smear} for one particular
example of a benchmark point 
taken from the benchmark plane 4 with the masses fixed to $\MA = \MHp = 544.72
\gev$ and $\MH = 515.5 \gev$\footnote{This corresponds to the green
point of the model based analysis for plane 4 in 
\refse{sec:mhh-8}, which is also the point used in
\refse{sec:toy}.}. In
this figure we show in blue the 
\mhh\ distribution without smearing (the ideal case). The solid line
depicts the full distribution, whereas the dashed line shows the result
for the continuum (non-resonant) diagrams. The
red lines demonstrate the effect of applying a 10\% (left plot) and 15\%
(right plot) smearing on the theoretical prediction of the
\mhh\ distributions, where the solid (dashed) lines indicates the full
(continuum) result. 
While a 15\% smearing was given as a realistic future estimate, the 10\%
smearing indicates a potential optimistic improvement. 
One can observe that from the original dip-peak structure as seen in the
solid blue line effectively only a peak or bump around the original
peak remains. The original dip is visible only as a very small reduction
of the unsmeared distribution, as the relative weight of the points
below the continuum is smaller than those above the continuum (note the
logarithmic scale). 

\subsection{Bin width}
\label{sec:bin}

As a further step in the evaluation of the experimental challenges, we
analyze the effect of the bin width. The binning means that the
data in a particular interval in $\mhh$ is presented as the mean value
of the differential cross section of all the points that fall in that
interval. Assuming that at least one of the Higgs bosons analyzed
will decay in a $b\bar b$ pair\footnote{The most promising decay mode
for the other Higgs is $\ga \ga$ because of the excellent di-photon
mass resolution.}, the bin size will eventually be determined by the
$b$-jet mass 
resolution from the reconstruction of the $h \to b\bar b$ decay mode.
This affects the visualization of the results in a
realistic experimental set up, but also the counting of events for the
evaluation of the experimental sensitivity, see \refeq{eq:R}.
The binning is applied after the smearing discussed in the previous
subsection.

\begin{figure}[ht!]
  \begin{center}
\includegraphics[width=0.41\textwidth]{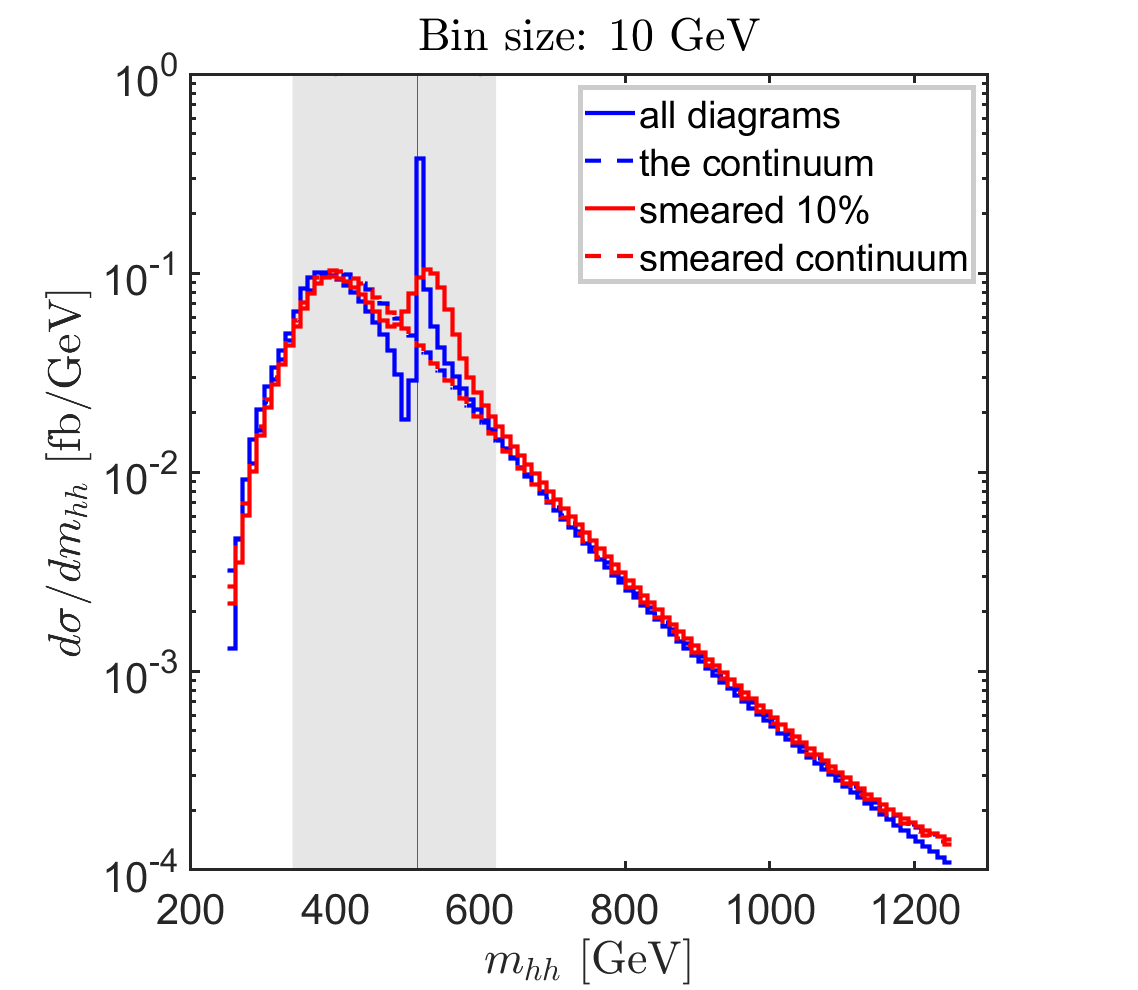}
\includegraphics[width=0.41\textwidth]{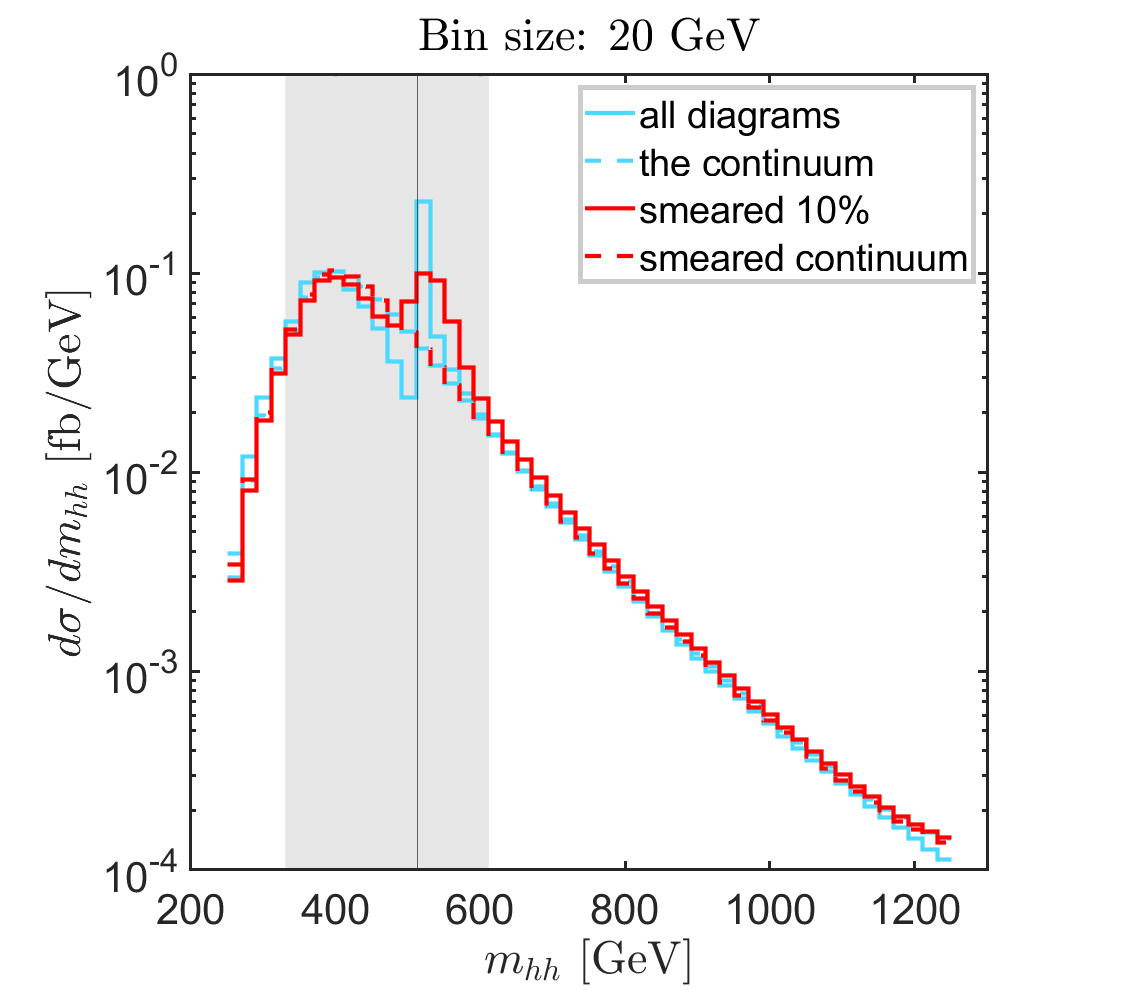}
\includegraphics[width=0.41\textwidth]{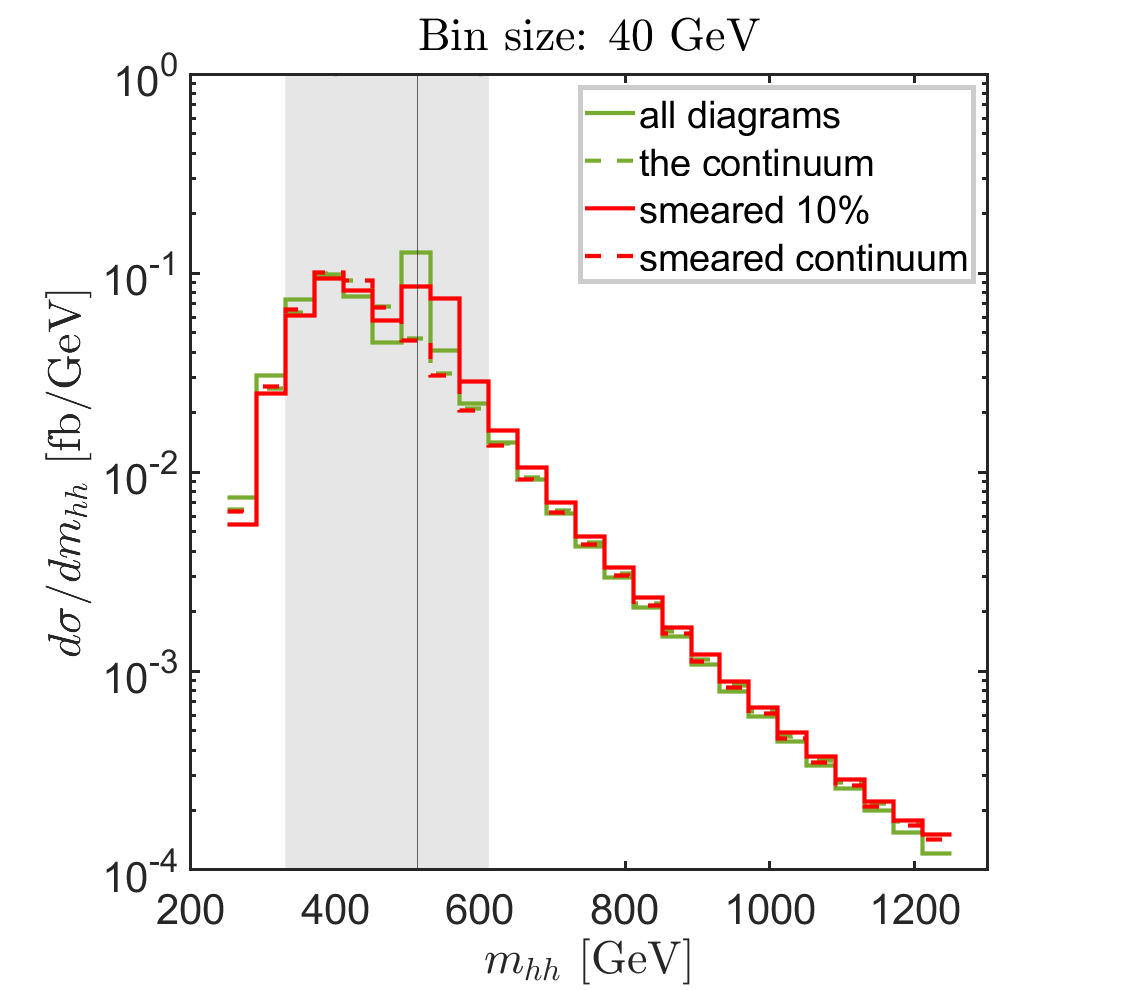}
\includegraphics[width=0.41\textwidth]{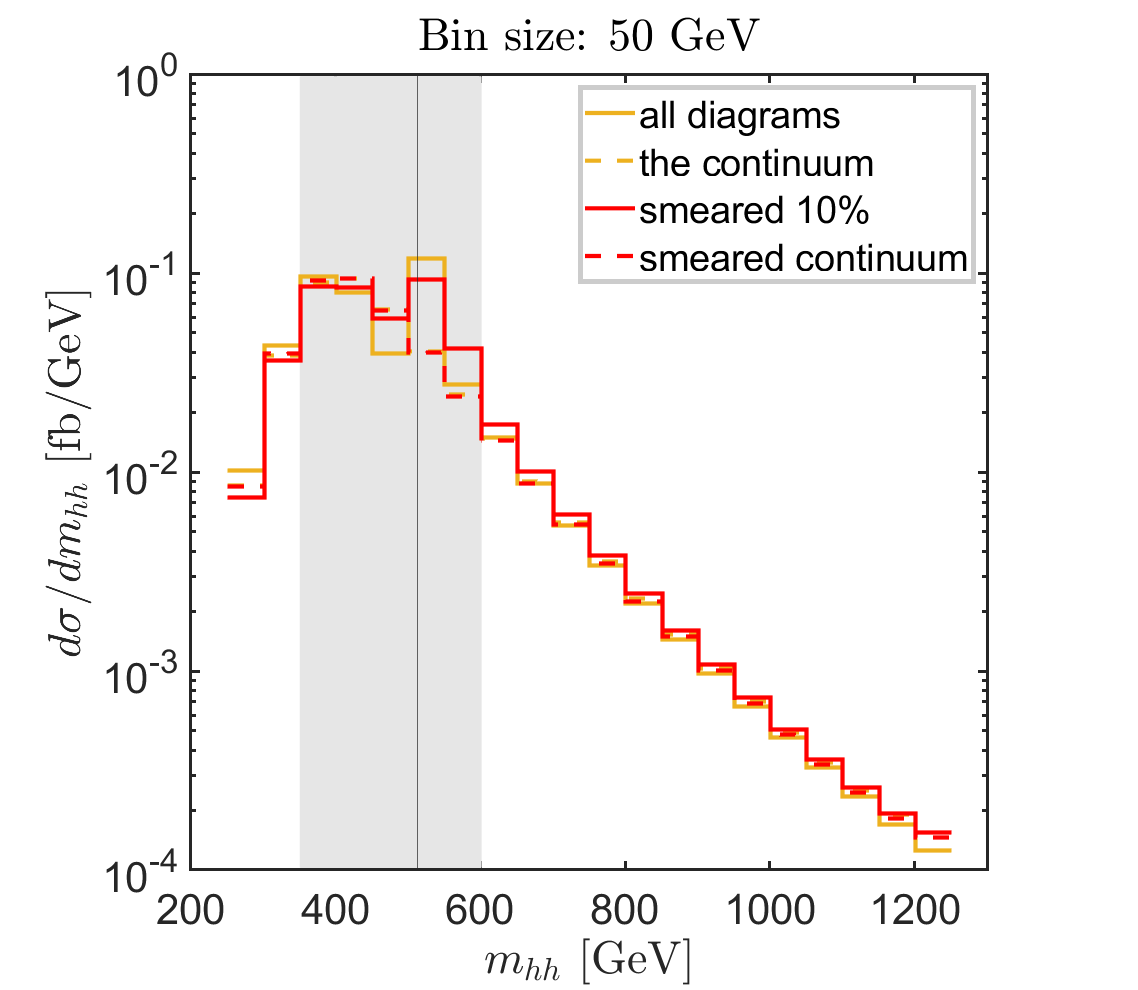}
  \end{center}
\caption{Different bin sizes for a 10\% smeared distribution in the
  example benchmark point (10, 20, 40 and 50 $\gev$). The red lines
  correspond to the true (smeared and binned) prediction of the
  \mhh\ distribution. The other color indicates the corresponding binned,
  but unsmeared distribution. Solid (dashed) lines represent the total
  (continuum) contribution to the cross section. The grey region
  represents the region that falls into the window defined to compute
  the variable $R$. The black vertical line indicates the value of
  the resonant mass, i.e.\ $512.5 \gev$.} 
\label{fig:smear10bin}
\end{figure}

\begin{figure}[ht!]
  \begin{center}
\includegraphics[width=0.41\textwidth]{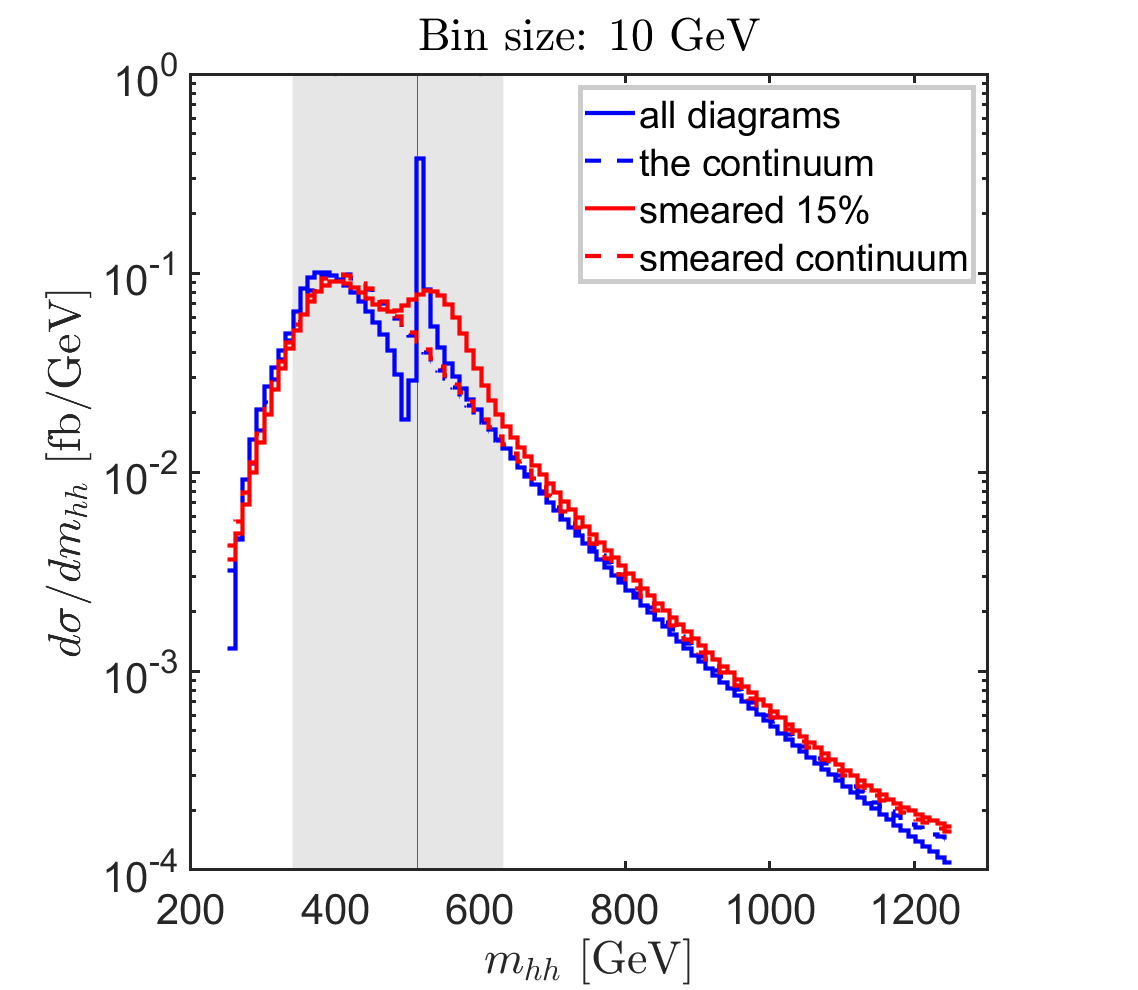}
\includegraphics[width=0.41\textwidth]{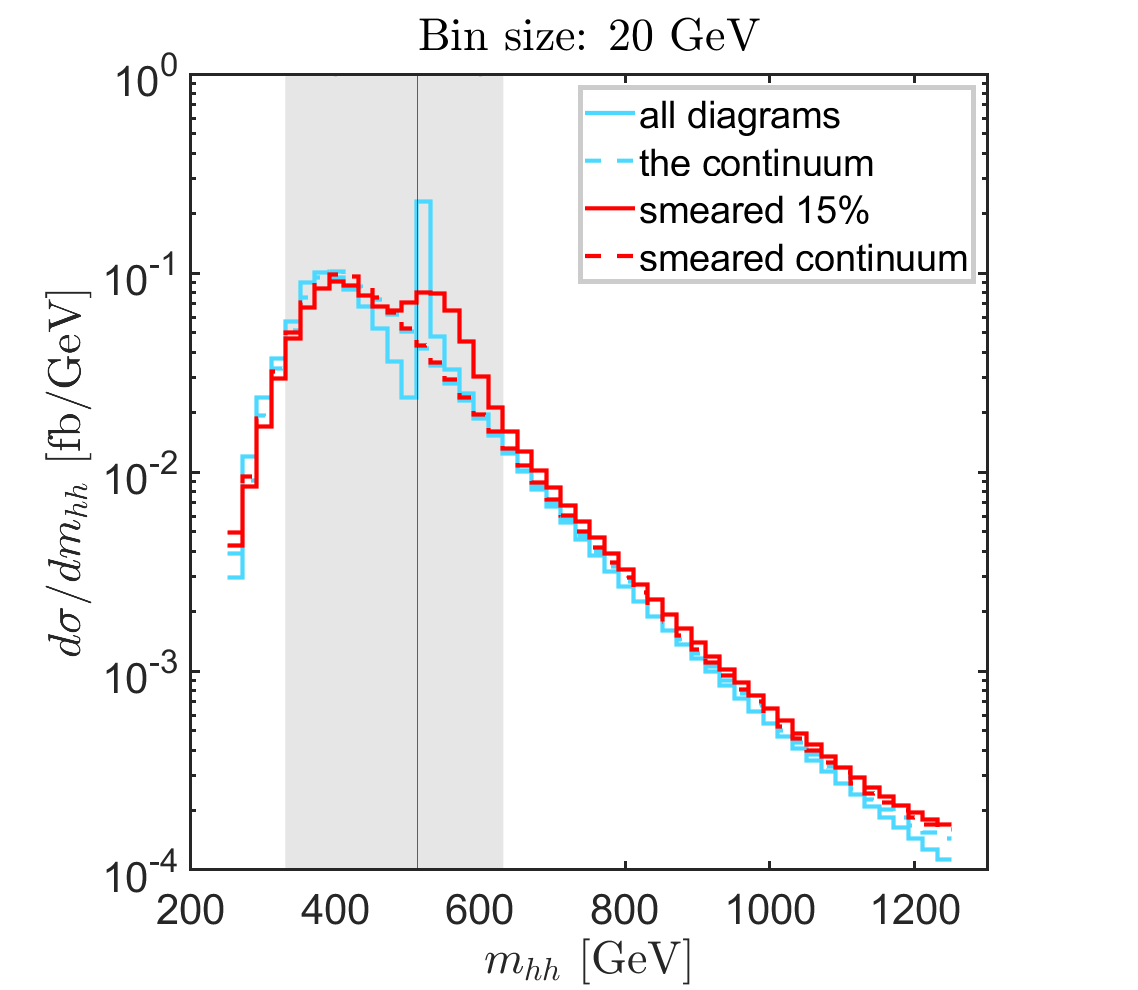}
\includegraphics[width=0.41\textwidth]{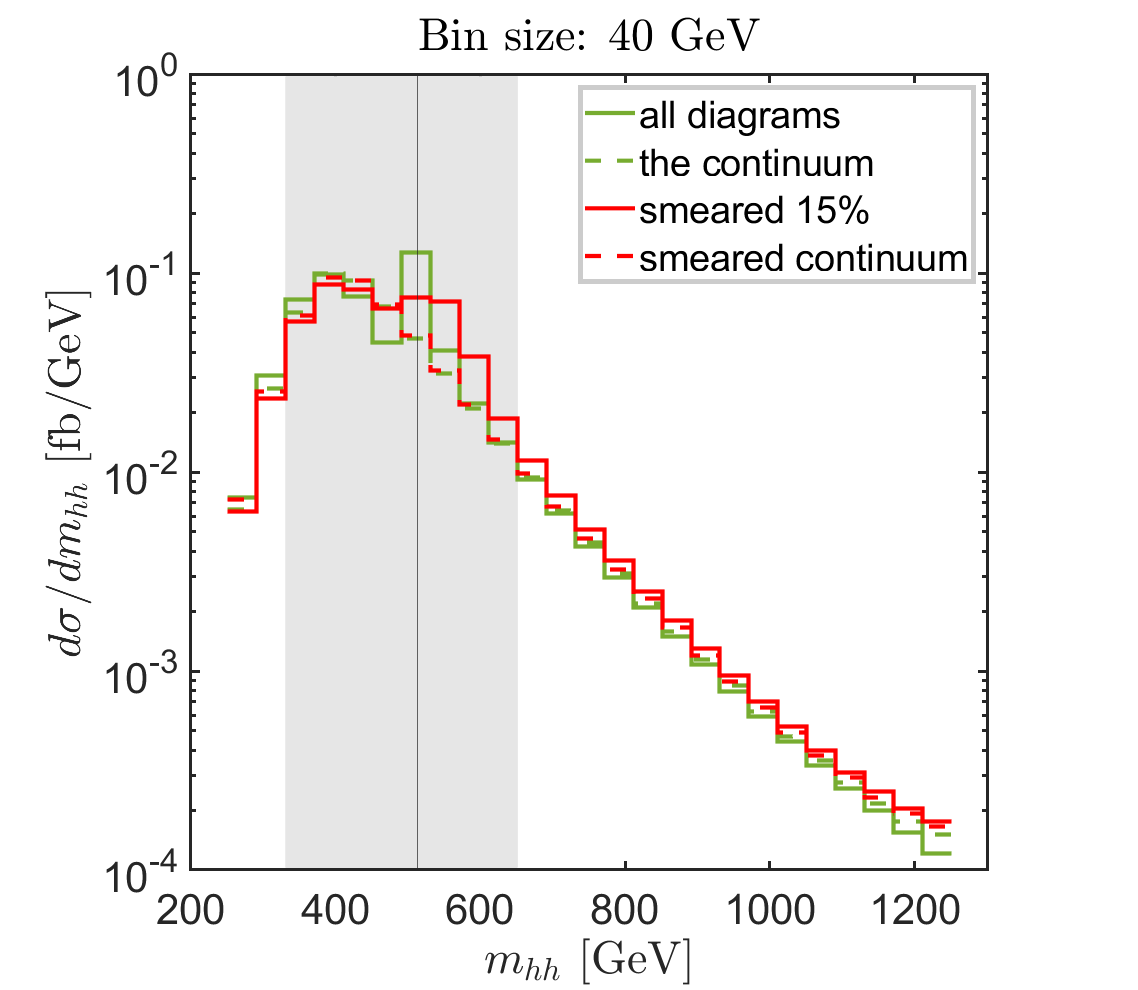}
\includegraphics[width=0.41\textwidth]{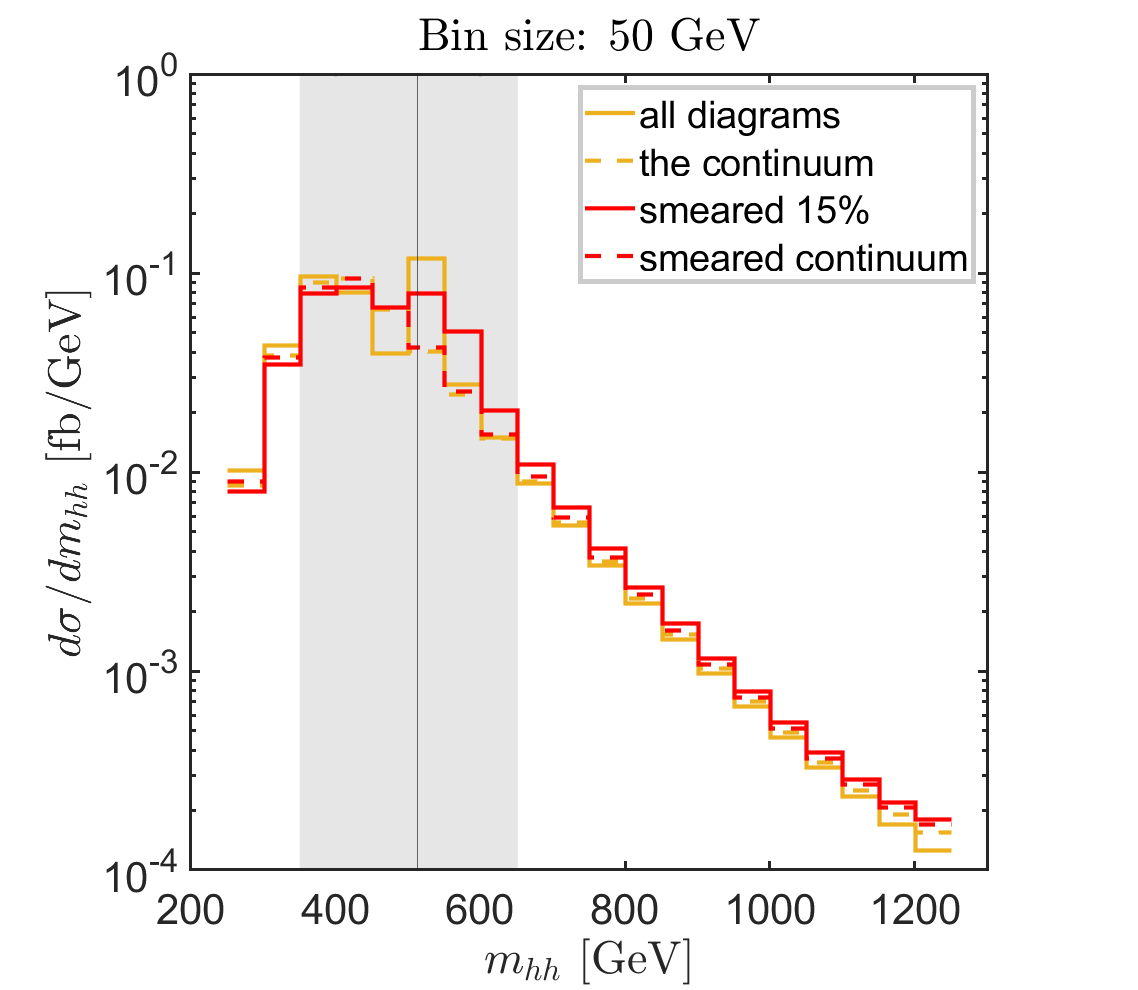}
  \end{center}
\caption{Same as Fig.~\ref{fig:smear10bin} but for 15\% smearing.} 
\label{fig:smear15bin}
\end{figure}

The analysis is performed for the same benchmark point as in 
\refse{sec:smearing}. In \reffis{fig:smear10bin} and \ref{fig:smear15bin}
we show the same spectrum but for a different bin size in the $\mhh$
variable: $10 \gev$ (upper left), $20 \gev$ (upper right), $40 \gev$
(lower left) and $50 \gev$ (lower right).
\reffi{fig:smear10bin} assumes a 10\% smearing, whereas in
\reffi{fig:smear15bin} we show the more realistic result with 15\%
smearing. The red lines show the true
(smeared and binned) prediction, whereas the other colors indicate
the unsmeared, but binned results for comparison. 
One can observe that the effect of the smearing
becomes less significant in the region of resonant production for a
larger bin size. The resonance is already partially diluted by the
smearing, and the effect of the binning becomes less visible, as can be
observed best in the lower right plots of \reffis{fig:smear10bin} and 
\ref{fig:smear15bin}. The effect of the binning is less important once
the smearing of the experimental data is taken into account.
After the binning the ``dip'' is effectively 
indistinguishable from the continuum contribution. The peak is still
persistent and for larger bin size approaches the same height as the
bump at $\sim 400 \gev$ before binning.

In the most conservative result the expected experimental resolution
should have a bin size of $50 \gev$ and a smearing of $\sim 15\%$. The
expected results in this case would possibly give access to the location
of the resonance (the mass of the $\cp$-even~$H$ should be know via
single production by the time the di-Higgs cross section is measured)
and partially to the height, and thus possibly to the size of \lahhH.
In order to make a quantitative estimate of the sensitivity of the
signal produced by the resonant diagram we have calculated the value of
the variable $R$ defined in \refeq{eq:R} that is obtained from
\reffis{fig:smear10bin} - \ref{fig:smear15bin}, shown in
\refta{tab:Rbinsize}. 
Overall, one can see that the values of $R$ are significant,
i.e.\ the signal could possibly be distinguished at the HL-LHC if our
assumptions on the experimental uncertainties are met. It should
be noted that we are not taking into 
account the efficiency of the particle detectors, which could reduce
signifficantly the estimate of $R$. 
Comparing the two columns in \refta{tab:Rbinsize} one observes 
$R$ is roughly 10\% worse as the 
assumed percentage of smearing increases by 5\%. However, $R$ is
is somewhat more stable w.r.t.\ the bin size, where deviations
within $\sim$ 5\% are found. This would constitute a rather
positive feature for an experimental set-up.

\begin{table}[ht!]
\centering
\begin{tabular}{|l|c|c|}
        \hline
         Bin size & R(10\% smear)  & R(15\% smear)\\
        \hline
        \hline
        10 GeV (blue) &  94.7 & 84.9   \\
        \hline 
        20 GeV (light blue) &  92.6 & 86.5  \\
        \hline
        40 GeV (green) &  92.6 & 86.8   \\
        \hline
        50 GeV (yellow) &  89.6 & 87.5   \\
        \hline
\end{tabular}
\caption{Values of the variable $R$ for the significance of the signal
  for different bin sizes for a 10\% and 15\% smeared distribution, see
  \protect\reffis{fig:smear10bin}-\ref{fig:smear15bin}.}
\label{tab:Rbinsize}
\end{table}


\subsection{Bin location}
\label{sec:location}

The next part of the analysis concerns the arbitrary choice of the
location of the bin. This choice can also affect the pattern of
the invariant mass distribution. The concrete 
value of $\mhh^{\min}$ (the value of $\mhh$ at the bin start) and
$\mhh^{\max}$ (the value of $\mhh$ at the bin end)
affects the number of events that fall into that bin and thus can have
an impact on the evaluation of the sensitivity~$R$.
For the previously used benchmark point we change 
the location of the bin for a 10\% and 15\% smeared
distribution and a $50 \gev$ bin size. 

\begin{figure}[ht!]
  \begin{center}
\includegraphics[width=0.41\textwidth]{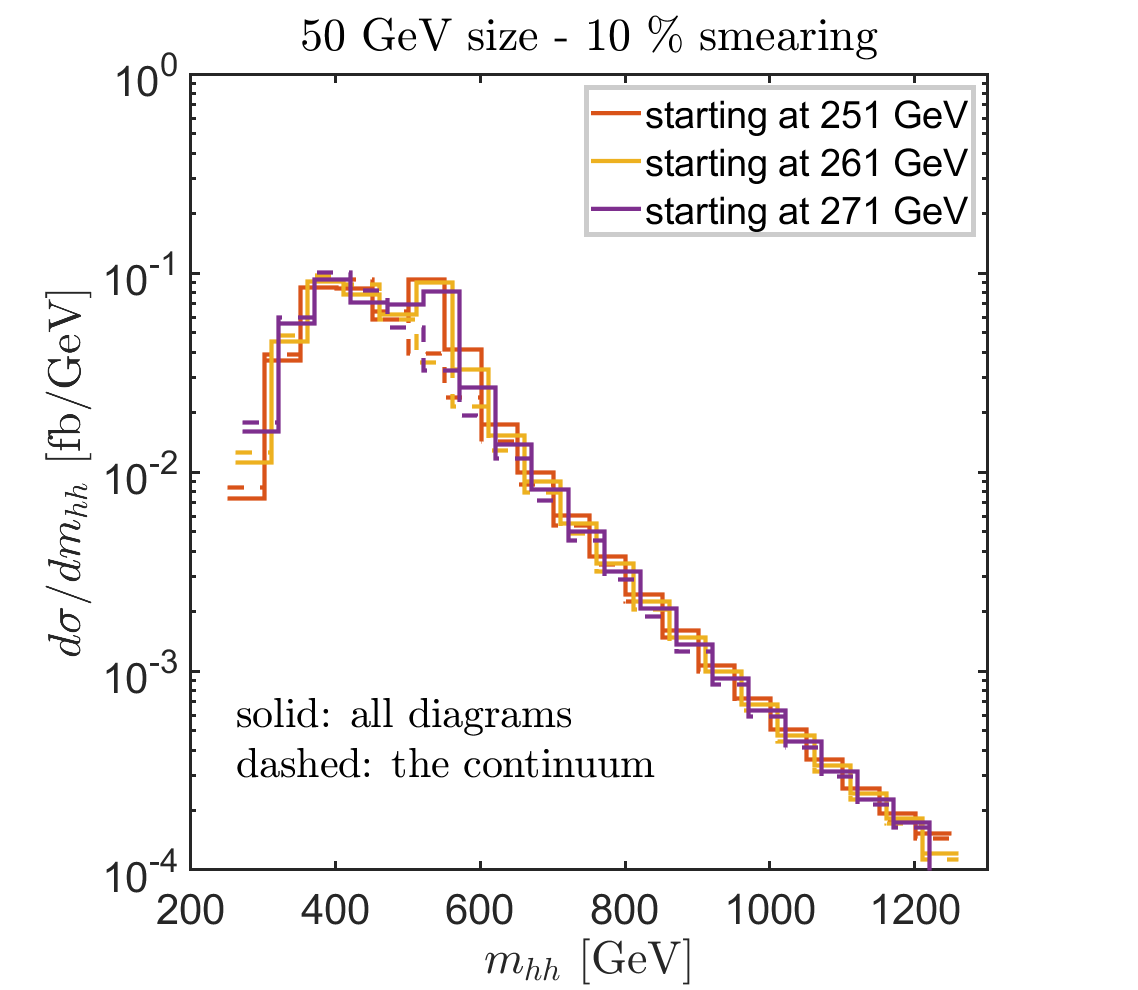}
\includegraphics[width=0.41\textwidth]{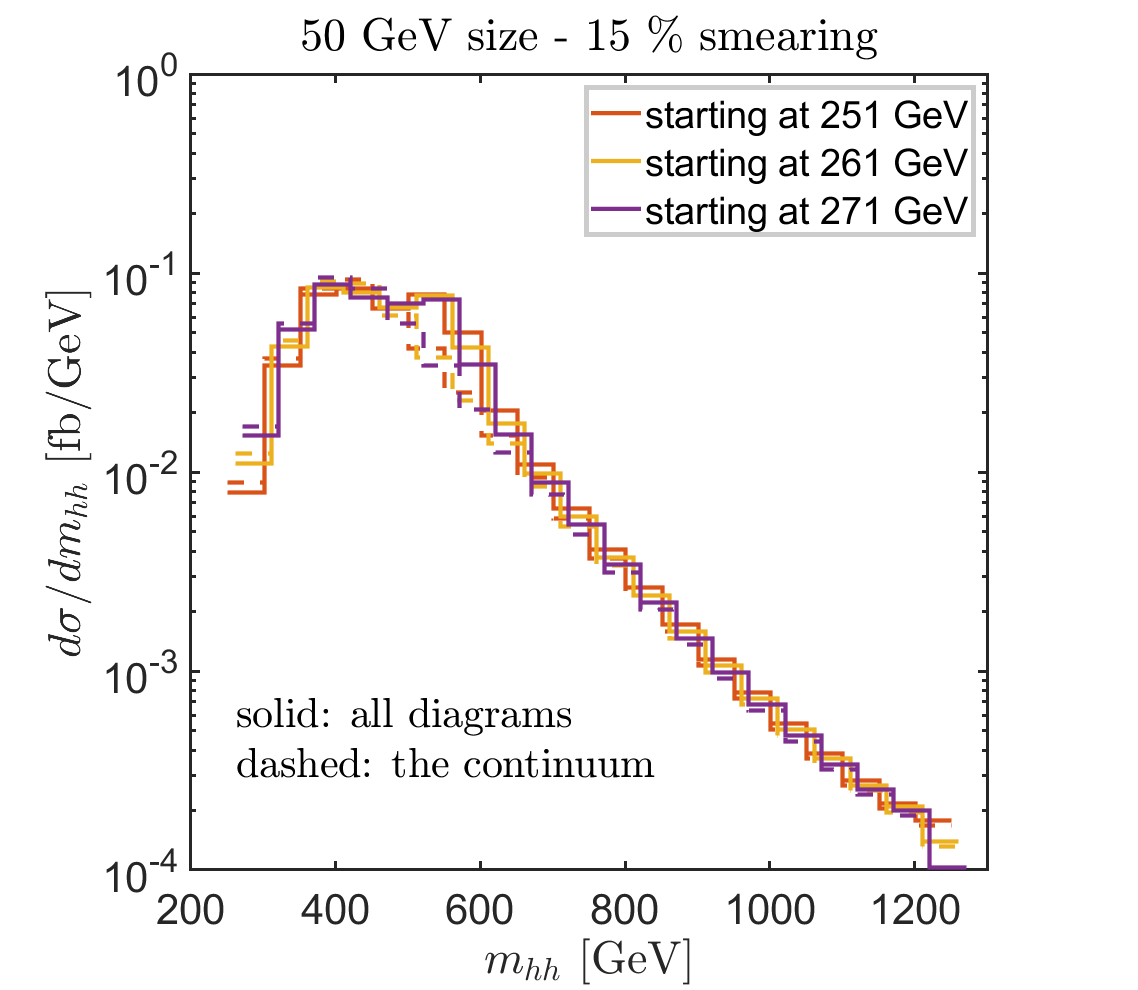}
  \end{center}
\caption{Invariant mass distributions for different bin locations
  assuming a bin size of 50 GeV and a smearing of 10\% (left) and 15\%
  (right). Solid (dashed) curves show the full (continuum) result.}
\label{fig:location}
\end{figure}

In \reffi{fig:location} we show the difference in the invariant mass
distribution created by a change in the location of the first bin
by $10$ or $20 \gev$,  
i.e.\ we start the distribution at $251, 261, 271 \gev$ as
orange, yellow and purple lines, respectively.
In both plots we show the difference between the total differential
cross section (solid lines) and the continuum contribution (dashed
lines). The left (right) plot uses a smearing of 10\% (15\%). 
One can observe that for all three choices of bin locations the peak
structure remains similarly visible (the dip is strongly diluted from
the smearing and the binning as discussed in the previous subsections). 
To quantitatively evaluate the significance of the signal of the
resonant enhancement we list the values of $R$ for the two plots discussed
above in \refta{tab:Rlocation}.

\begin{table}[ht!]
\centering
\begin{tabular}{|l|c|c|}
        \hline
         Bin location & R(10\% smear)  & R(15\% smear)\\
        \hline
        \hline
        start at 251 GeV (orange) &  89.6 & 87.5   \\
        \hline 
        start at 261 GeV (yellow) & 85.8 & 82.6   \\
        \hline
        start at 271 GeV (purple) &  87.8 & 81.6   \\
        \hline
\end{tabular}
\caption{Values of the variable $R$ for the sensitivity of the signal
  for different bin locations for a 10\% and 15\% smeared distribution
  and a bin size of $50 \gev$.}
\label{tab:Rlocation}
\end{table}

In \refta{tab:Rlocation} one can observe that the variation in $R$
stays within 5\% when we modify the location of the bins. That means
that the uncertainties associated to the location of the bin are
smaller than the ones associated to the smearing
and about the same as for the bin size.
Therefore, overall we find that the experimental resolution
of the particle detector, which we tried to mimic by smearing the
data, has a larger impact on the resonance, and the width and location
of the binning has a smaller effect in diluting the resonance. 


\subsection{Effect of  \boldmath{sign($\lahhH \times \xi_H^t$)}}
\label{sec:sign}

\begin{figure}[ht!]
  \begin{center}
\includegraphics[width=0.41\textwidth]{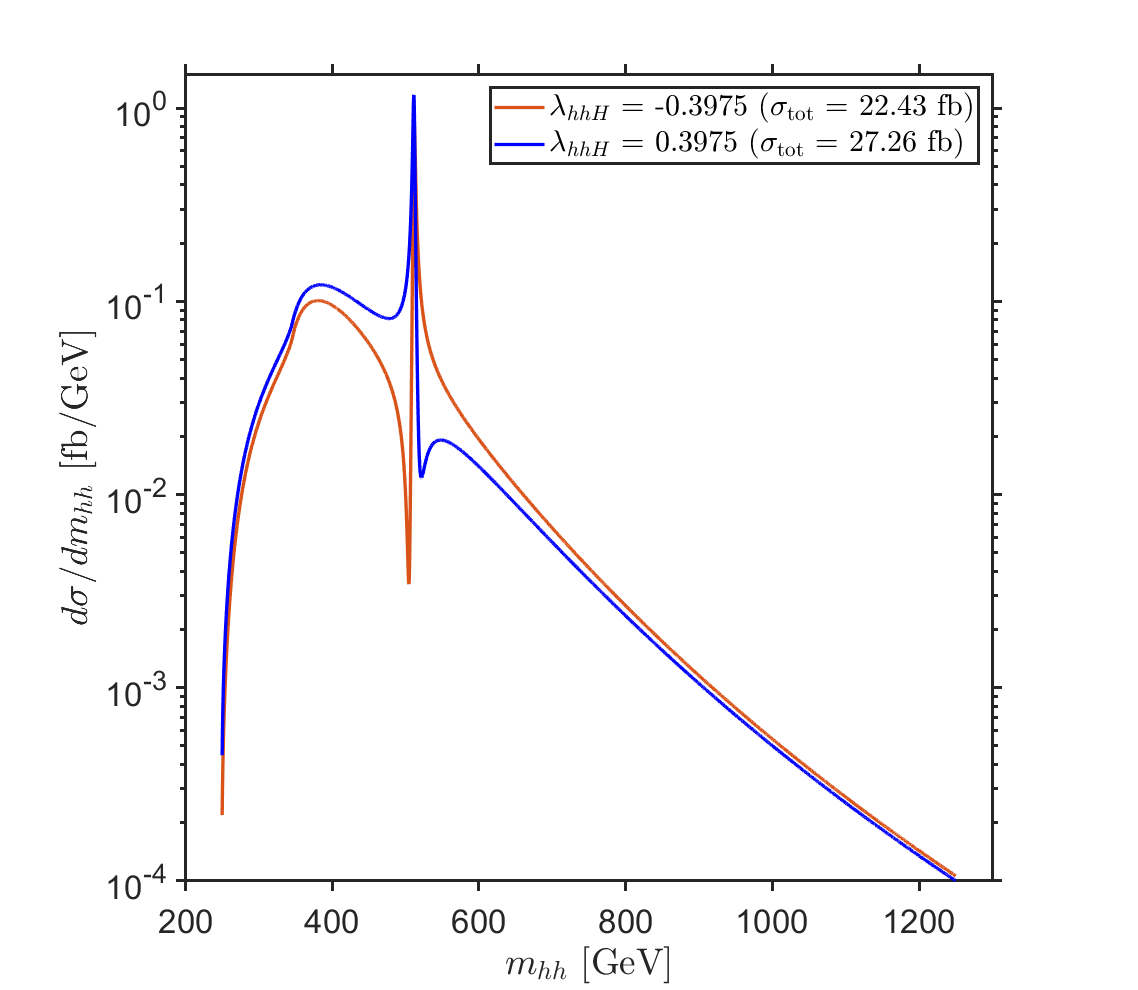}
  \end{center}
\vspace{-1em}
  \caption{Invariant mass distributions for different signs of
    ($\lahhH \times \xi_H^t$) (see the right plot of
    \protect\reffi{fig:couplingsres}).}  
\label{fig:mhhsigns}
\end{figure}

Finally, taking into account the above discussed experimental
uncertainties, we analyze the possibility to access experimentally
the relative sign of the involved couplings in the resonant
diagram. For this analysis we will use the same benchmark point as
above, where the effect of an ad-hoc change of
sign($\lahhH \times \xi_H^t$) has been shown already in the right plot of
\reffi{fig:couplingsres} and is reproduced for completeness of this
analysis in \reffi{fig:mhhsigns}. On this ``ideal'' result 
we will apply a smearing and a binning in order to see whether the
difference in the peak-dip vs.\ dip-peak structure, and thus in the
sign($\lahhH \times \xi_H^t$), persists in the experimental analysis.
This will demonstrate whether the access to this parameter is a realistic
goal at the HL-LHC.  

The invariant mass distribution of the benchmark point is shown in red
for the original value of $\lahhH = -0.3975$ and a positive sign of
($\lahhH \times \xi_H^t$), whereas the blue line gives the result for an
ad-hoc changed sign of $\lahhH = +0.3975$ and a negative sign of
($\lahhH \times \xi_H^t$). Both distributions are normalized to
the corresponding value of the total cross section, which is indicated in
the legend of \reffi{fig:mhhsigns}. The change in 
the structure is clearly visible in this plot. We remind that the
value of the top Yukawa coefficient in this case was, as obtained from
\refeq{eq:topyukawa}, $\xi_H^t = 0.104 > 0$. Therefore we see that for
an overall minus (plus) sign the imprinted structure of the resonance in
the $\mhh$ spectrum is a peak-dip (dip-peak) one. 

\begin{figure}[ht!]
  \begin{center}
\includegraphics[width=0.41\textwidth]{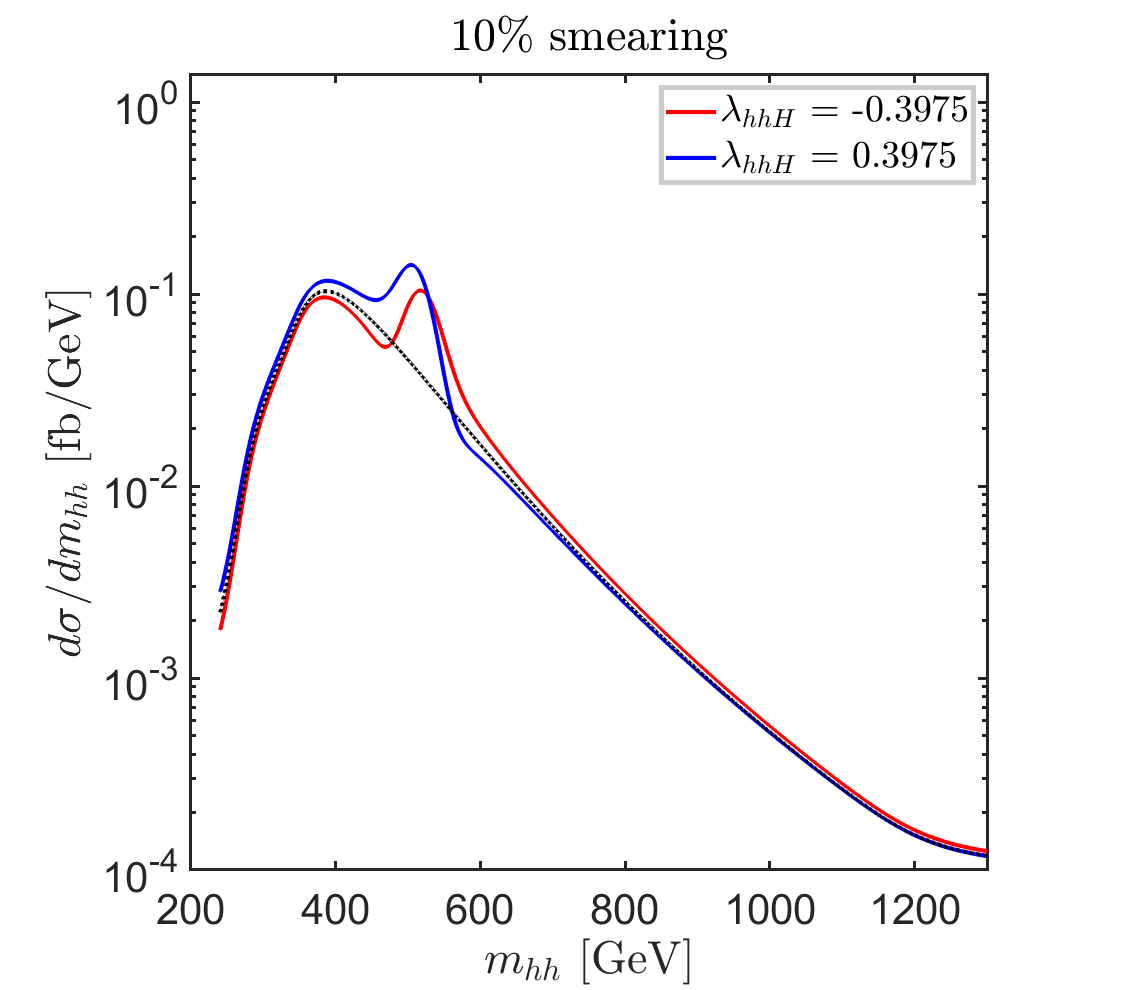}
\includegraphics[width=0.41\textwidth]{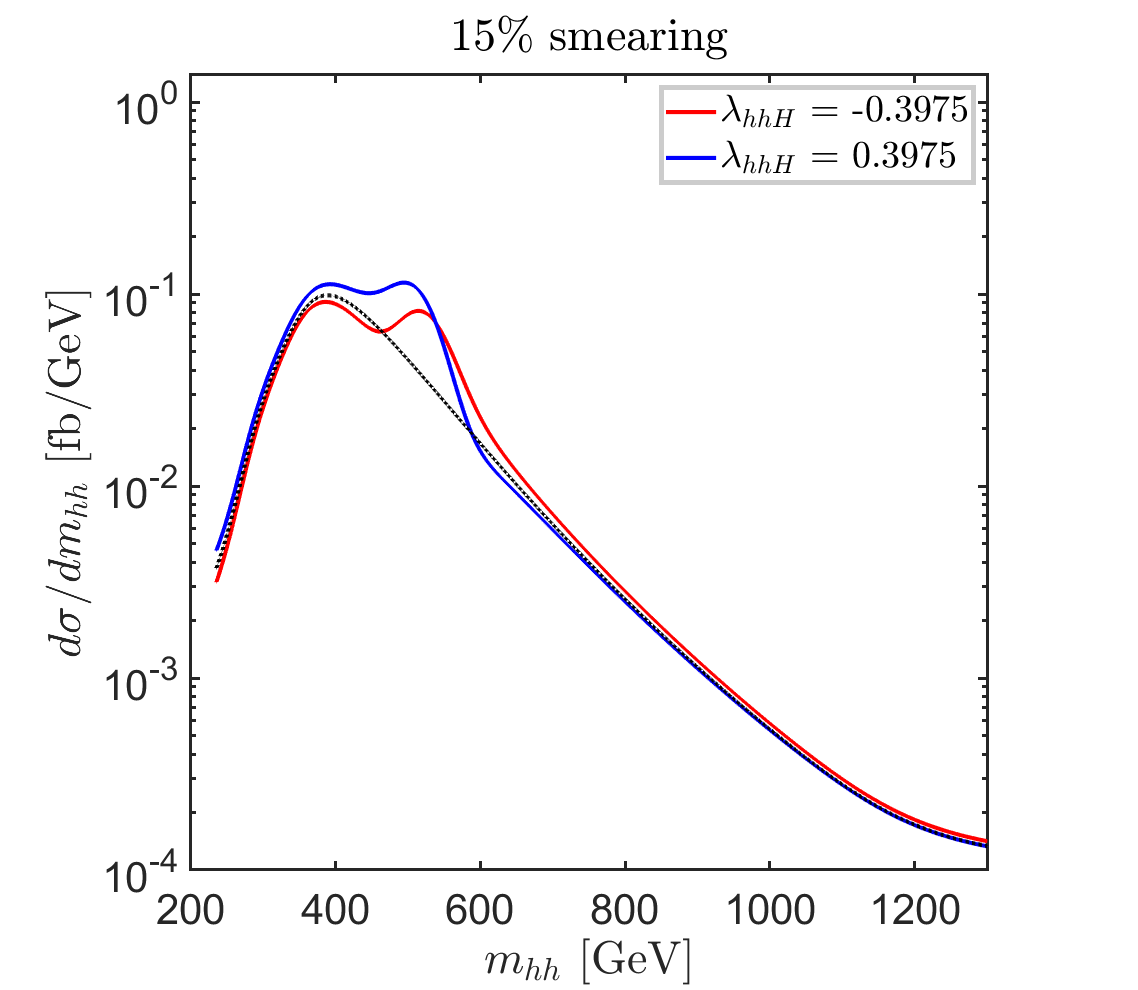}
  \end{center}
\vspace{-1em}
\caption{Invariant mass distributions for different signs of ($\lahhH
  \times \xi_H^t$) with a 10\% (left) and 15\% (right) smearing
  applied. The black line shows the continuum distribution.}
\label{fig:mhhsigns-smeared}
\end{figure}

We now impose the various experimental uncertainties on the ``ideal''
result shown in \reffi{fig:mhhsigns}. We start with the application of
smearing as presented in \reffi{fig:mhhsigns-smeared}.
In the left (right) plot a 10\% (15\%) smearing is applied. The red and
blue curves correspond, as before, to the negative and positive sign of
\lahhH. The black curves indicate the continuum distribution.
One can observe that the blue curve is above (below) the red one right
before (after) the bump. However, the overall structure cannot be
resolved once the data is smeared.

\begin{figure}[ht!]
\vspace{-1em}
  \begin{center}
\includegraphics[width=0.41\textwidth]{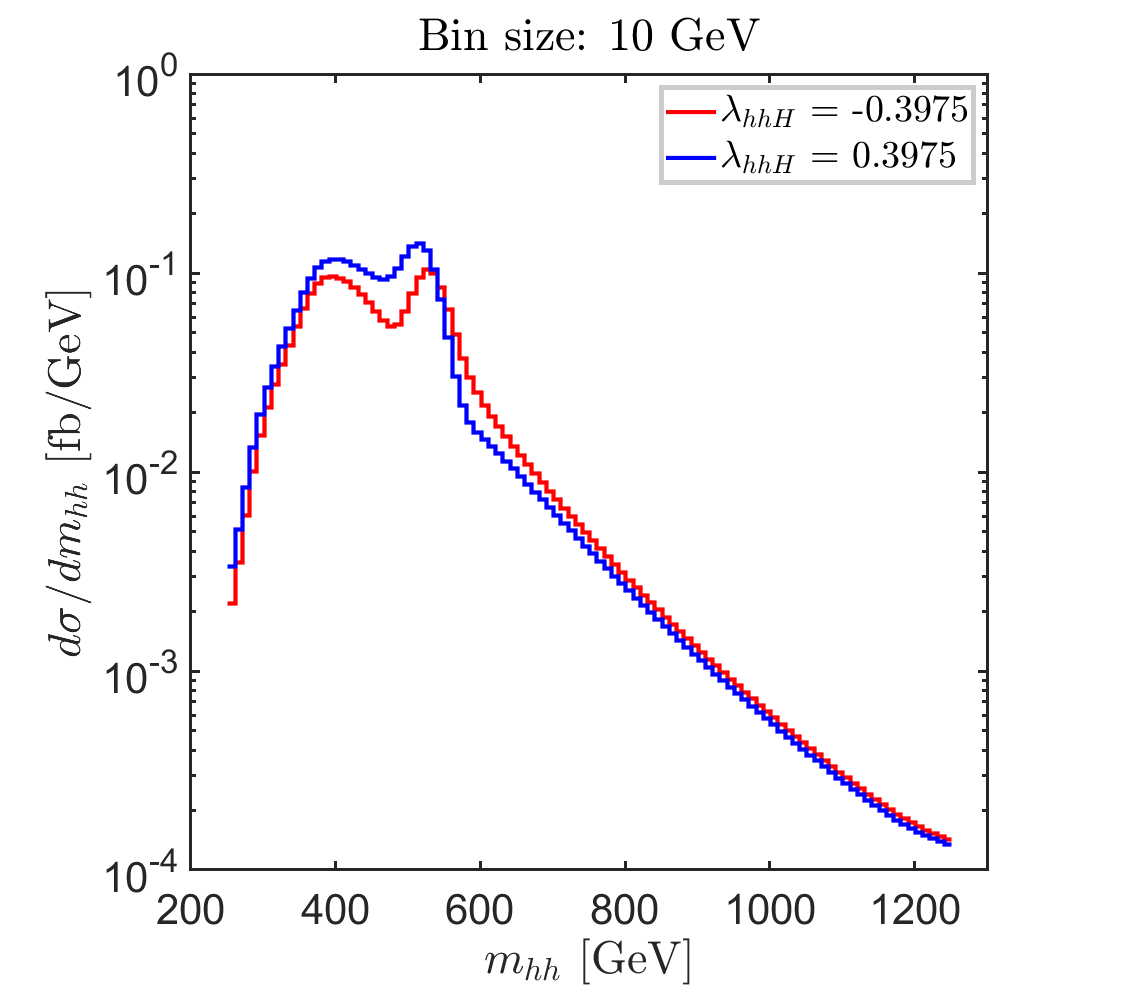}
\includegraphics[width=0.41\textwidth]{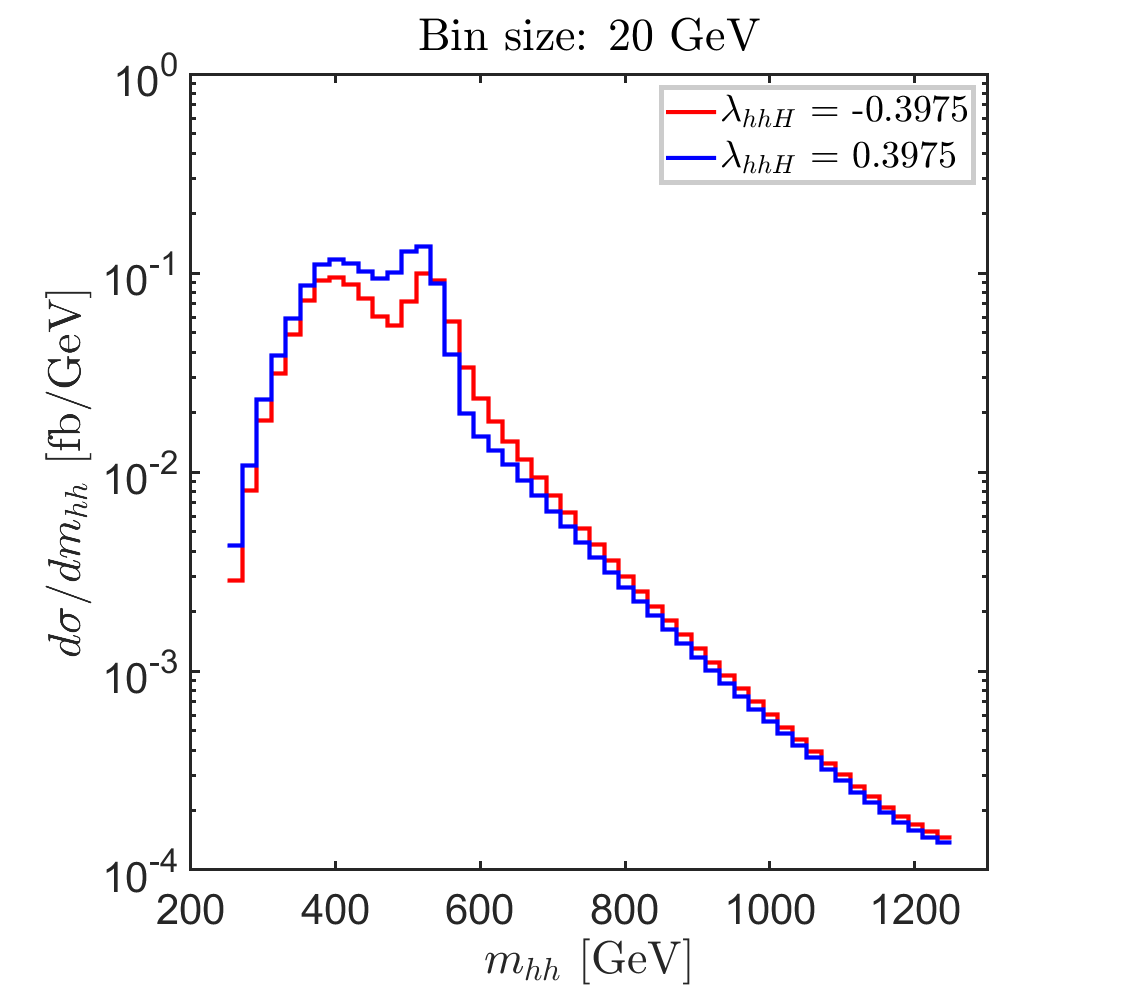}
\includegraphics[width=0.41\textwidth]{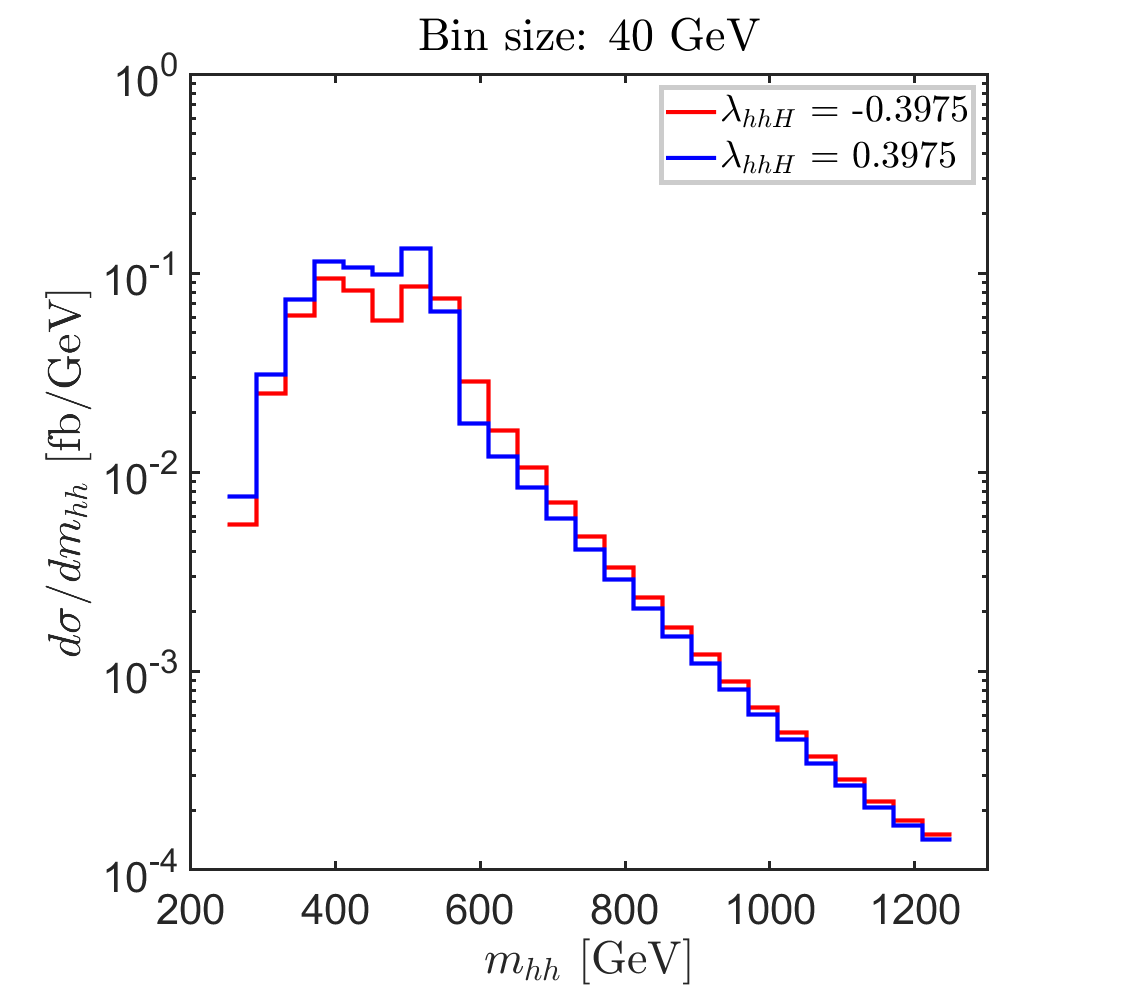}
\includegraphics[width=0.41\textwidth]{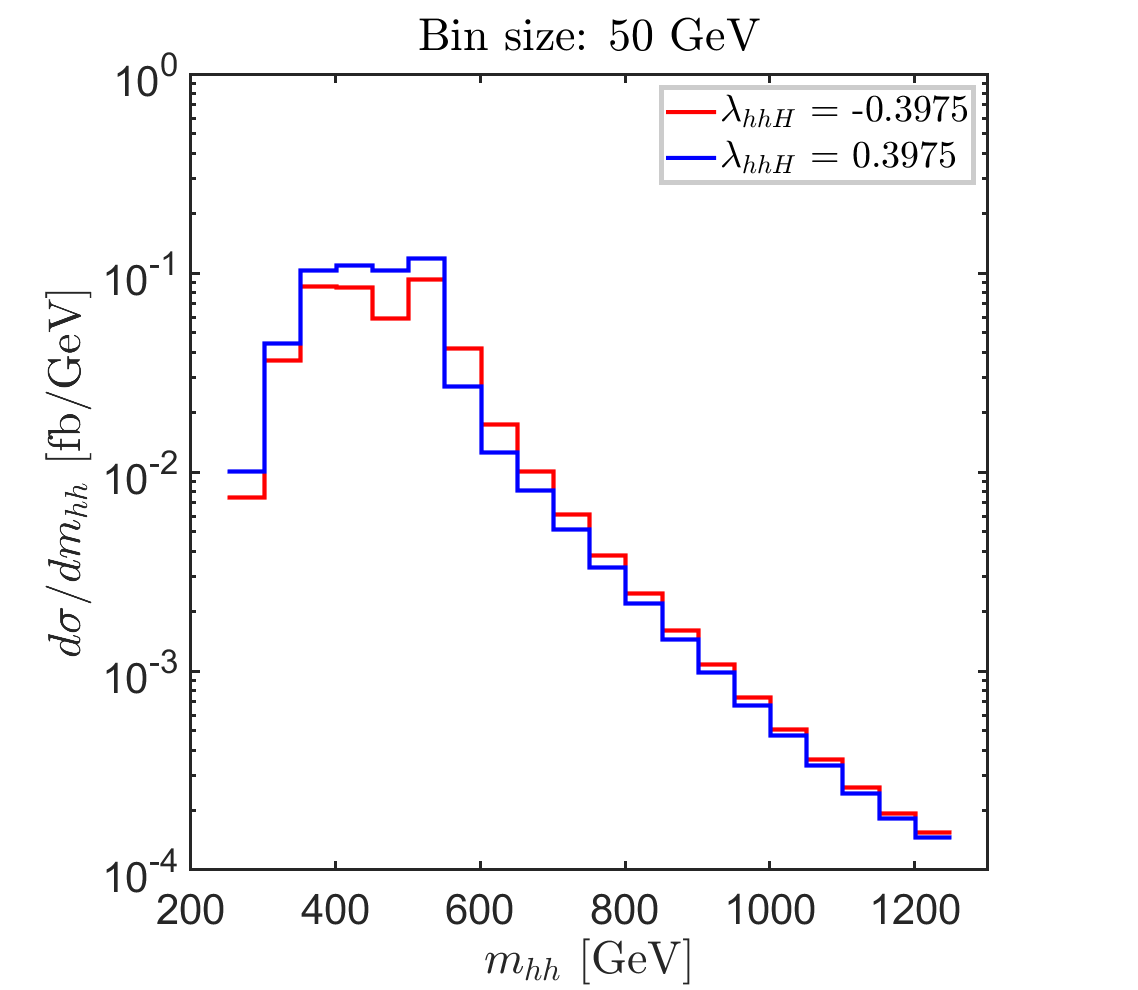}
  \end{center}
\vspace{-1em}
\caption{Invariant mass distributions for different signs of ($\lahhH
  \times \xi_H^t$) with a 15\% smearing applied. Results for different
  bin sizes: $10 \gev$ (upper left), $20 \gev$ (upper right), $40 \gev$
  (lower left) and $50 \gev$ (lower right).}
\label{fig:mhhsigns-smearedbin}
\end{figure}

The final step is the application of binning on top of a 15\% smearing, 
which is the most realistic experimental set up. The results are
shown in \reffi{fig:mhhsigns-smearedbin}. One can conclude that effectively
the dip-peak vs.\ peak-dip structure becomes unresolvable once both
smearing and  binning is applied.

\section{Conclusions}
\label{sec:conclusions}

In this work we have analyzed the impact of triple Higgs couplings
on the production cross section
of two $\sim 125 \gev$ Higgs bosons at the HL-LHC in the framework of the
2HDM. The first goal was to analyze the impact of \lahhh\ and \lahhH\ on
the di-Higgs production cross section. In a second step we analyzed the
potential sensitivity of the HL-LHC on the BSM THC \lahhH. This
sensitivity has been analyzed w.r.t.\ various experimental
uncertainties. 

We have chosen several observables that allow us to trace the impact of
the THCs. The first one is the Higgs pair production
cross section that is affected by the diagram containing this
coupling. Using the code \texttt{HPAIR} we 
have evaluated the cross sections of di-Higgs production including NLO
QCD corrections (in the heavy top limit) in the specific benchmark
planes and analyzed the  
particular impact of the THCs, the contribution of the heavy Higgs boson
in the propagator and its decay width.  

Differences of the di-Higgs production cross sections can originate from
a changed value of \lahhh, as well as from additional ``resonant''
contributions, given by the $s$-channel exchange of the heavy $\cp$-even
Higgs boson, $H$. In the analyzed benchmark planes we have found both
types of effects, which can yield a strong enhancement of the di-Higgs
production cross section w.r.t.\ its SM result. While these results
indicate that the effect of the resonant $s$-channel $H$~contribution
may be visible, from the cross section alone it is not possible
to disentangle the two sources. 

In order to gain more direct access to \lahhH\ we analyzed the invariant
di-Higgs mass distributions, $d\sig/d\mhh$ as a function of \mhh. In a
toy example we analyzed the interference of the resonant $H$~exchange
with the non-resonant (continuum) diagrams. The interference term
changes sign for $Q = \mhh = \MH$, resulting in a peak-dip (or
dip-peak) structure of the \mhh\ distribution. We demonstrated that the
interference effect becomes smaller with the total decay width of the
$H$ and that the ``total width of the effect'', given by the
width of the peak at half of its maximum value, increases with
increasing \GaHtot.
Furthermore, the structure of the interference (peak-dip
vs.\ dip-peak) depends on the sign of ($\lahhH \times \xi_H^t$), which
enters as a prefactor of the interference contribution
($\xi_H^t$ denotes the top Yukawa coupling of the heavy $\cp$-even Higgs
boson).

The effects in the toy example were reproduced for the complete
\mhh\ distribution for several points in the before chosen benchmark
planes. We demonstrated the dependences of the size of the interference
effects on \lahhH\ and $\xi_H^t$ for large di-Higgs production cross
sections, as well as for selected points closer to the alignment limit
($\CBA \sim 0.1$) and further away from the alignment limit ($\CBA \sim 0.2$).
These results indicate that the HL-LHC may have sensitivity to see
effects of the BSM THC \lahhH. As a theoretical variable quantifying the
{\it relative} sensitivity to \lahhH, the variable $R$ was defined, see
\refeq{eq:R}. However, it is not meant as a determination of the
experimental \lahhH\ precision that requires a detailed experimental
analysis, which is beyond the scope of our paper

In order to further analyze a possible sensitivity, we applied various
experimental uncertainties on $R$. The first one is the smearing due to
the detector resolution, which realistically can go down to $\sim 15\%$,
but we also analyzed a potential (optimistic) improvement down to
10\%. It is found that from the dip-peak structure mostly a bump
survives, with a very small reduction due to the dip w.r.t.\ the
unsmeared result. The next effect is the binning of the result, mostly
given by the $b$-jet mass resolution from the reconstruction of the 
$h \to b\bar b$ decay of one of the Higgs bosons. A realistic value of a
$50 \gev$ bin size was compared to more optimistic sizes down to
$10 \gev$. While the smearing has a visible effect on $R$, the binning
hardly reduces its value. Similarly, the location of the bin, which
is partially arbitrary, has a smaller impact on $R$ once we take into
account the finite resolution of the detector (smearing).
As a last step we showed 
that the sign($\lahhH \times \xi_H^t$) does not leave any measurable
imprint on the \mhh\ distributions, once the experimental uncertainties
are taken into account.

We conclude that, depending on the values of the underlying Lagrangian
parameters, a sizable resonant $H$ contribution to the di-Higgs
production cross section can leave possibly visible effects in the
\mhh\ distribution. This would pave the way for a first determination of
a BSM THC, a step that is crucial for the reconstruction of the Higgs
potential of the underlying BSM model.


\subsection*{Acknoledgements}

We thank D.~Azevedo, 
M.J.~Herrero,
J.~Schaarschmidt,
M.~Spira
and
G.~Weiglein
for helpful discussions.
The work of F.A.\ and S.H.\ has received financial support from the
grant PID2019-110058GB-C21 funded by
MCIN/AEI/10.13039/501100011033 and by ``ERDF A way of making Europe".
MEINCOP Spain under contract PID2019-110058GB-C21.
and in part by
by the grant IFT Centro de Excelencia Severo Ochoa CEX2020-001007-S
funded by MCIN/AEI/10.13039/501100011033.
The work F.A.\ has furthermore received financial support from
the Spanish ``Agencia Estatal de Investigaci\'on'' (AEI) and the EU
``Fondo Europeo de Desarrollo Regional'' (FEDER) 
through the project PID2019-108892RB-I00/AEI/10.13039/501100011033, 
from the European Union's Horizon 2020 research and innovation
programme under the Marie Sklodowska-Curie grant agreement No 674896 and
No 860881-HIDDeN; and from the FPU grant with code FPU18/06634.
The work of M.M.\ has been supported by the BMBF-Project 05H21VKCCA.
K.R.\ acknowledges support by the Deutsche Forschungsgemeinschaft
(DFG, German Research Foundation) under
Germany's Excellence Strategy -- EXC 2121 ``Quantum Universe'' --
390833306. This work has been partially funded by the Deutsche
Forschungsgemeinschaft (DFG, German Research Foundation) - 491245950.



\end{document}